\documentclass[fleqn,usenatbib]{mnras}

\usepackage{psfrag}
\usepackage{hyperref}
\usepackage{gensymb}
\usepackage{xcolor}
\usepackage{tabularx}
\usepackage{lipsum}
\usepackage[normalem]{ulem}
\usepackage[labelfont=bf, labelsep=period]{caption}
\usepackage{subcaption}
\usepackage{mathtools}
\usepackage{balance}
\usepackage{adjustbox}
\usepackage{float} 
\usepackage{placeins}
\usepackage{caption}
\usepackage{stfloats}
\usepackage{pifont}

\usepackage{enumitem}

\usepackage[switch,displaymath,mathlines]{lineno}

\usepackage{newtxtext,newtxmath}
\usepackage[T1]{fontenc}
\usepackage{amsmath}
\usepackage{graphicx}

\title{Multi-Clustering Needlet ILC for CMB $B$-mode component separation}
\author[A. Carones et al.]{Alessandro Carones$^{1,2}$\thanks{E-mail: alessandro.carones@roma2.infn.it}, 
Marina Migliaccio$^{1,2}$,
Giuseppe Puglisi$^{1,2}$,
Carlo Baccigalupi$^{4,5,6}$, 
\newauthor
Domenico Marinucci$^{3}$, Nicola Vittorio$^{1,2}$, Davide Poletti \vspace{+0.2 cm}
\newauthor
\hspace{+6 cm} for the LiteBIRD collaboration \vspace{+0.5 cm}\\
$^{1}$Dipartimento di Fisica, Universit\`a di Roma ``Tor~Vergata'', via della Ricerca Scientifica 1, I-00133, Roma, Italy \\
$^{2}$Sezione INFN Roma~2, via della Ricerca Scientifica 1, I-00133, Roma, Italy \\
$^{3}$Dipartimento di Matematica, Universita’ di Roma Tor Vergata, via della Ricerca Scientifica 1, I-00133, Roma, Italy
\\
$^{4}$International School for Advanced Studies (SISSA), Via Bonomea 265, 34136, Trieste, Italy\\
$^{5}$INFN Sezione di Trieste, Via Valerio 2, 34127 Trieste, Italy\\
$^{6}$IFPU, Via Beirut, 2, 34151 Grignano, Trieste, Italy
}

\date{Accepted XXX. Received YYY; in original form ZZZ}

\pubyear{2023}

\begin{document} 
\label{firstpage}
\pagerange{\pageref{firstpage}--\pageref{lastpage}}
\maketitle

\begin{abstract}
The Cosmic Microwave Background (CMB) primordial $B$-mode signal is predicted to be much lower than the polarized Galactic emission (foregrounds) in any region of the sky pointing to the need for sophisticated component separation methods. Among them, the blind Needlet Internal Linear Combination (NILC) has great relevance given our current poor knowledge of the $B$-mode foregrounds. However, the expected level of spatial variability of the foreground spectral properties complicates the NILC subtraction of the Galactic contamination. We therefore propose a novel extension of the NILC approach, the Multi-Clustering NILC (MC-NILC), which performs NILC variance minimization on separate regions of the sky (clusters) properly chosen to have similar spectral properties of the $B$-mode Galactic emission within them. Clusters are identified thresholding either the ratio of simulated foregrounds-only $B$ modes (ideal case) or the one of cleaned templates of Galactic emission obtained from realistic simulations. In this work we present an application of MC-NILC to the future \textit{LiteBIRD} satellite, which targets the observation of both reionization and recombination peaks of the primordial $B$-mode angular power spectrum with a total error on the tensor-to-scalar ratio $\delta r < 0.001$. We show that MC-NILC provides a CMB solution with residual foreground and noise contamination that is significantly lower than the NILC one and the primordial signal targeted by \textit{LiteBIRD} at all angular scales for the ideal case and at the reionization peak for a realistic ratio. Thus, MC-NILC will represent a powerful method to mitigate $B$-mode foregrounds for future CMB polarization experiments.

\end{abstract}


\section{Introduction}

Observations of the cosmic microwave background (CMB) temperature and polarization anisotropies led to the establishment of a cosmological concordance scenario, the $\Lambda$CDM model, with very tight constraints on its parameters (see, e.g., Boomerang:~\citealt{2006ApJ...647..799M}; \textit{WMAP}:~\citealt{2013ApJS..208...19H}; \textit{Planck}:~\citealt{2018arXiv180706209P}). 
CMB polarization is usually decomposed into the so-called $E$ and $B$ modes, of even and odd parity respectively, on the sky \citep{1997PhRvL..78.2058K, 1997PhRvD..55.1830Z}. Of these, CMB $B$ modes are thought to be generated by two independent processes. On small angular scales (typically a few arcminutes), they are mostly sourced by  primordial $E$ modes distorted into $B$ via weak gravitational lensing of the intervening  large scale structures along the line of sight between the last scattering surface and the observer. These are commonly known as \emph{lensing $B$ modes} \citep{1998PhRvD..58b3003Z}. The existence of this signal has been confirmed by  several  experiments, although detections are still with a modest signal-to-noise ratio or on very small patches of sky \citep{2013PhRvL.111n1301H, 2014PhRvL.113b1301A, 2016ApJ...833..228B, 2015ApJ...807..151K, 2017ApJ...848..121P,2020A&A...641A...8P}. On the other hand, at large angular scales, $B$ modes are expected to originate from tensor perturbations (primordial gravitational waves) generated in the very early Universe in a phase of cosmic inflation \citep{1997PhRvL..78.2058K}. Their amplitude is set by the tensor-to-scalar ratio $r$, which defines the amplitude of primordial tensor modes over the ones of initial density perturbations. To date, primordial $B$ modes have not been detected and \cite{2022PhRvD.105h3524T} provides the latest upper limit on $r \leq 0.032$ at $95\,\%$ CL. Their detection would represent a powerful test of cosmic inflation and a proxy to discriminate among different inflationary models. \\
The main features in the $BB$ tensor angular power spectrum are the reionization bump at very large angular scales ($\ell \lesssim 10$), 
which is associated to the integral of the linear polarization generated by quadrupolar anisotropy from the last scattering surface as seen by each free electron after reionization, and the recombination bump at smaller scales ($\ell \sim 80$), which is the imprint of the primordial gravitational waves on Last Scattering Surface physics. \\
Many experiments have been designed to  observe $B$-mode polarization, either from the ground:  %\textcolor{red}{[citations]}
POLARBEAR \citep{2010SPIE.7741E..1EA}, QUBIC \citep{2011APh....34..705Q}, BICEP \citep{2016JLTP..184..765W}, Keck-Array \citep{2003SPIE.4843..284K}, LSPE-STRIP \citep{2012SPIE.8446E..7CB}, ACT \citep{2020JCAP...12..047A}, SPT \citep{2020PhRvD.101l2003S}, Simons Observatory \citep{2019JCAP...02..056A}, CMB-S4 \citep{2022arXiv220308024A}; 
from balloons: %\textcolor{red}{[citations]}
SPIDER \citep{2010SPIE.7741E..1NF}, and LSPE-SWIPE \citep{2012SPIE.8452E..3FD}; 
or from space: %\textcolor{red}{[citations]}.
\textit{LiteBIRD} \citep{2014JLTP..176..733M} and \textit{PICO} \citep{2019arXiv190210541H}. Ground-based and balloon-borne missions will essentially be targeted to  observe  mainly  the recombination peak. Thus a  satellite mission, given its full-sky coverage, is crucial in  observing both the reionization and recombination bumps to possibly have a clearer detection of the primordial B-signal. \\
The development of sophisticated component separation methods has emerged as one of the most important aspects of the data analysis for all these experiments. The reason being that the quest for primordial $B$ modes is made much more difficult by the presence of instrumental noise and polarized foregrounds, especially Galactic thermal dust and synchrotron emission. These methods are usually divided in blind, parametric, and template removal techniques. Blind methods, such as ILC (\citealt{2003ApJS..148...97B}, \citealt{2003PhRvD..68l3523T}), NILC \citep{2009A&A...493..835D}, FastICA \citep{2002MNRAS.334...53M}, and SMICA \citep{2003MNRAS.346.1089D}, usually linearly combine the multi-frequency sky maps in such a way as to minimize some meaningful statistical quantity of the final map. Parametric methods, as Commander \citep{2008ApJ...676...10E}, FGBuster \citep{2009MNRAS.392..216S}, FastMEM \citep{2002MNRAS.336...97S} explicitly model the foreground frequency properties by means of a set of parameters which are fitted to the data. Such methods provide an easy way to characterise and propagate foreground residual errors, but their effectiveness depends on how reliable is the adopted model. Finally, template removal algorithms, like SEVEM \citep{2003MNRAS.345.1101M}, try to construct internal foreground templates using multi-frequency observations. \\
In this paper, we focus on the so-called Needlet Internal Linear Combination (NILC) algorithm \citep{2009A&A...493..835D}, which has great relevance in the context of CMB data-analysis because it performs noise and foreground subtraction with minimal prior information on their properties and, hence, it is not prone to systematic errors due to mis-modelling. This method will, then, play a key-role in the analysis of future missions' data, given our still poor knowledge of polarized foreground properties on the full sky. One possible drawback of such approach is that foreground residuals estimation usually relies on Monte-Carlo simulations or other bootstrapping techniques. Recently, an extension of the method has been proposed, the constrained moment Internal Linear Combination (cMILC), where the weights estimation is constrained to project out some moments of the foreground emission \citep{2021MNRAS.503.2478R}. 
The standard NILC technique produces generally too high foreground residuals given the targets of future CMB satellite missions \citep{2021MNRAS.503.2478R}. The new cMILC technique, instead, reduces the Galactic contamination in the CMB solution but at the price of an increase in the reconstruction noise. Thus, it would be desirable to build a pipeline which lowers the foreground residuals' amplitude without suffering of a noise penalty. \\
In this paper we develop a new NILC approach, the \textit{Multi-Clustering} NILC (MC-NILC), to be applied to multi-frequency $B$-mode datasets, where the standard NILC algorithm is improved by performing the foreground subtraction separately on different regions of the sky (\emph{clusters}) characterised by similar spectral properties of the $B$-mode foregrounds. Forecasts of the performance of $B$-mode foreground cleaning methods for space mission proposals are already available in the literature (see \textit{e.g.} \citealt{2009AIPC.1141..222D,2018JCAP...04..023R,2022MNRAS.514.3002A,2022arXiv221114342A,2022arXiv220202773L}).
In this work, we consider an application to the future CMB mission \textit{LiteBIRD} (the Lite (Light) satellite for the study of $B$-mode polarization and Inflation from cosmic background Radiation Detection), selected by the Japan Aerospace Exploration Agency (JAXA) in May 2019 as a strategic large-class (L-class) mission, with an expected launch in the late 2020s. \textit{LiteBIRD} will survey the microwave sky in $15$ frequency bands between $34$ and $448$ GHz at the Sun-Earth Lagrangian point L2 with an unprecedented total sensitivity of $2.2$ $\mu$K-arcmin \citep{2020SPIE11443E..2FH}. \\
The paper is organized as follows. We present the simulated multi-frequency \textit{LiteBIRD} dataset in Sect. \ref{Sec:sims}; in Sect. \ref{sec:methods} we detail the different procedures employed to perform a NILC or MC-NILC minimization and to assess their performance; in Sect. \ref{sec:results} we discuss the results obtained by applying the MC-NILC pipeline to the \textit{LiteBIRD} $B$-mode dataset; finally, in Sect. \ref{sec:concs} we report our conclusions.

\section{LiteBIRD simulated data}
\label{Sec:sims}
The dataset considered in this work includes the frequency maps of the \textit{LiteBIRD} satellite, simulated according to the specifications reported in Table $13$ in \citealt{2022arXiv220202773L}. The maps at the different frequencies are obtained by the co-addition of CMB, white Gaussian instrumental noise and Galactic polarized foregrounds (thermal dust and synchrotron emission). \\
The CMB maps are generated with the Healpix Python package\footnote{\url{https://github.com/healpy/healpy}} from the angular power spectra of the best-fit \textit{Planck} 2018 parameters \citep{2020A&A...641A...6P} with tensor-to-scalar ratio $r=0$, including lensing. 
Instrumental noise is simulated as white and isotropic Gaussian realisations with standard deviations in each pixel corresponding to the polarization sensitivities reported in \citealt{2022arXiv220202773L}. \\
Galactic emission is simulated with the \textit{PySM} Python package\footnote{\url{https://github.com/healpy/pysm}} \citep{2017MNRAS.469.2821T}. Synchrotron polarized emission is modelled with a power law \citep{Rybicki}:
\begin{equation}
	X_{\textrm{s}}(\hat{n},\nu) = X_{\textrm{s}}(\hat{n},\nu_{0}) \cdot  \Bigg(\frac{\nu}{\nu_{0}} \Bigg)^{\beta_{\textrm{s}}(\hat{n})} 
	\label{eq:sync}
\end{equation}
where $X=\{\textit{Q},\textit{U} \}$ are the Stokes parameters, $\beta_{\textrm{s}}$ is the spectral index, $\hat{n}$ the position in the sky, and $\nu$ the considered frequency. $X_{\textrm{s}}(\hat{n},\nu_{0})$ represents the synchrotron template at a reference frequency $\nu_{0}$. 
Thermal dust emission is modelled with a modified black-body (MBB):
\begin{equation}
	X_{\textrm{d}}(\hat{n},\nu) = X_{\textrm{d}}(\hat{n},\nu_{0}) \cdot  \Bigg(\frac{\nu}{\nu_{0}} \Bigg)^{\beta_{\textrm{d}}(\hat{n})} \cdot \frac{B_{\nu}(T_{\textrm{d}}(\hat{n}))}{B_{\nu_{0}}(T_{\textrm{d}}(\hat{n}))},
	\label{eq:dust}
\end{equation}
where $B_{\nu}(T)$ is the black-body spectrum, $\beta_{\textrm{d}}$ is the dust spectral index, $T_{\textrm{d}}$ is the dust temperature, and $X_{d}(\hat{n},\nu_{0})$ represents the dust template at a reference frequency $\nu_{0}$. \\
In most of the paper the Galactic emission is simulated according to the PySM \texttt{d1s1} model, in which the values of synchrotron and dust parameters, namely $\beta_{\textrm{s}}, \beta_{\textrm{d}}, T_{\textrm{d}}$,  depend on the position in the sky. The thermal dust template is the estimated dust emission at 353 GHz in polarization from the Planck-2015 analysis, smoothed with a Gaussian kernel of full width at half maximum (FWHM) equal to $2.6^\circ$ and with small scales added via the procedure described in \citealt{2017MNRAS.469.2821T}. The dust temperature and spectral index maps are obtained from the \textit{Planck} data using the Commander pipeline \citep{2016A&A...594A..10P}. The synchrotron template, instead, is the \textit{WMAP} 9-year 23-GHz Q/U map \citep{2013ApJS..208...20B}, smoothed with a Gaussian kernel of FWHM $5^\circ$ with small scales added as above. The synchrotron spectral index map was derived using a combination of the Haslam 408-MHz data and \textit{WMAP} 23-GHz 7-year data \citep{2008A&A...490.1093M}. \\
For some comparisons within the analysis, we consider alternative models of the thermal dust and synchrotron emission available in the PySM package which are conventionally labelled, respectively, with $\texttt{d}$ and $\texttt{s}$ followed by a number: 
\begin{itemize}
    \item \texttt{d0}, a MBB with $\beta_{\textrm{d}}=1.54$ and $T_{\textrm{d}}=20 K$ constant across the sky
    \item \texttt{d4}, a dust model composed of two MBBs with constant spectral indices and anisotropic temperatures according to \citealt{1999ApJ...524..867F}
    \item \texttt{d5}, a model presented in  \citealt{2017ApJ...836..179H}.
    \item \texttt{d6}, a MBB with frequency decorrelation, where the impact of averaging over spatially varying dust spectral indices both unresolved and along the line of sight is properly modelled
    \item \texttt{d7}, extension of \texttt{d5} which takes into account iron inclusions in the dust grain composition
    \item \texttt{d12}, a 3D model of polarized dust emission with $6$ layers, based on the work in \citet{2018MNRAS.476.1310M}; each layer has different templates, spectral index, and dust temperature
    \item \texttt{s0}, a power law with $\beta_{\textrm{s}}=-3,$ constant across the sky
    \item \texttt{s3}, a power law with spatially varying $\beta_{\textrm{s}}$ and a constant curvature term. 
\end{itemize}
While \texttt{d0} and \texttt{s0} are a too simplistic view of the Galaxy, all the other models are in agreement with current \textit{Planck} data and represent possible alternatives to $\texttt{d1}$ and $\texttt{s1}$ modelling of the Galactic polarized emission.  Most of the above models assume a single emission law for thermal dust and synchrotron. However, we have evidences that such approximation may not hold for precise modelling of foreground observations in $B$ modes (see \textit{e.g.} \citealt{2015MNRAS.451L..90T,2018MNRAS.476.1310M}). Models, like the \texttt{d12}, account for such expected higher complexity of the dust emission by assuming different layers of emitters with distinct physical conditions. \\
All the components included in the simulations are smoothed with the beam of the \textit{LiteBIRD} frequency channel with the lowest angular resolution ($70.5'$ FWHM) and, then, combined together. We have verified that performing the same analysis even bringing all frequency maps to the resolution of a \textit{LiteBIRD} CMB channel ($23.7'$ FWHM) we obtain analogous results both in terms of foreground subtraction and CMB reconstruction (especially at the recombination peak).\\
All the maps are generated adopting the HEALPix pixelisation scheme \citep{2005ApJ...622..759G} with $N_{\textrm{side}}=64$ to fasten the analysis and because we are interested in evaluating the performance of the different methods on the largest angular scales, which are the ones where tensor modes are expected to be detected. Analogous results are obtained with $N_{\textrm{side}}=128$.
%(see Appendix \ref{sec:res_128}).
However, the pipeline is completely general and can be easily applied to maps with any spatial resolution without any significant additional computational cost. The $B$-mode maps needed for our analysis are obtained by a full-sky spin-2 harmonic transform of Q and U simulated data at the different frequencies.

\section{Methodology}
\label{sec:methods}
In this section we present and discuss:
\begin{itemize}
   
    \item the NILC method and the problem of the bias of the CMB reconstruction in Sect. \ref{Sec:NILC}
    \item the MC-NILC pipeline in Sect. \ref{sec:clusters}
    \item an optimised method to obtain cleaned templates of Galactic emission with a model-independent approach in Sect. \ref{sec:MCGNILC}
    \item the adopted masking strategies to avoid the most contaminated regions of the sky in Sect. \ref{sec:masks}
    \item the estimation of the effective tensor-to-scalar ratio associated to the foreground residuals and to the CMB reconstruction bias in Sect. \ref{sec:like}
    
\end{itemize}
All the obtained results shown in this paper are primarily evaluated through the use of the estimator of the angular power spectrum, which indicates the rotationally invariant variance of the harmonic coefficients of a map: 
\begin{equation}
    \hat{C}_{\ell}=\frac{1}{2\ell +1}\sum_{m=-\ell}^{m=\ell}\Big|a_{\ell m}\Big|^{2}.
\label{eq:Cl}
\end{equation}
It is customary to report an analogous quantity: $\hat{D}_{\ell}=\frac{\ell(\ell+1)}{2\pi}\hat{C}_{\ell}$, which would be constant on large scales for a scale invariant primordial perturbations density spectrum. \\
We estimate angular power spectra directly on $B$-mode maps obtained from full-sky simulated $Q$ and $U$ data, thus $B$-mode purification is not needed. Angular power spectra of $B$-mode maps have been estimated according to the MASTER formalism \citep{MASTER, MASTER2} using the \emph{pymaster} Python package\footnote{\url{https://namaster.readthedocs.io/en/latest/}}, which extends the estimator presented in Eq. \ref{eq:Cl} taking into account the effects of the masking and of the beam convolution (see \citealt{2019MNRAS.484.4127A}).

\subsection{NILC}
\label{Sec:NILC}
NILC is a blind component separation method which performs a linear combination of several frequency maps, filtered with a particular wavelet system (\emph{Needlets}), which minimizes the variance of the final map \citep{2009A&A...493..835D}.
Needlet filtering guarantees simultaneous localisation in harmonic and pixel space; it has been introduced in the statistical literature by \citealt{doi:10.1137/040614359} and firstly applied to CMB data by \citealt{2006PhRvD..74d3524P}. 
Needlets encode very remarkable properties \citep{2008MNRAS.383..539M} as : i) they do  not rely on the tangent plane approach, which implements a local flat sky approximation, ii) they are described by a simple reconstruction formula, where the same needlet functions appear both in the direct and the inverse transform, iii) they are defined in  concentrated regions in pixel space, because the window function goes to zero faster than any polynomial,
iv) they have exact harmonic localisation, set by the input parameter $B$, on a finite range of multipoles, v)
the coefficients are asymptotically uncorrelated   at any fixed angular distance, when the needlet scale $j$ increases. \\
\begin{figure}
	\centering
    \includegraphics[width=0.48\textwidth]{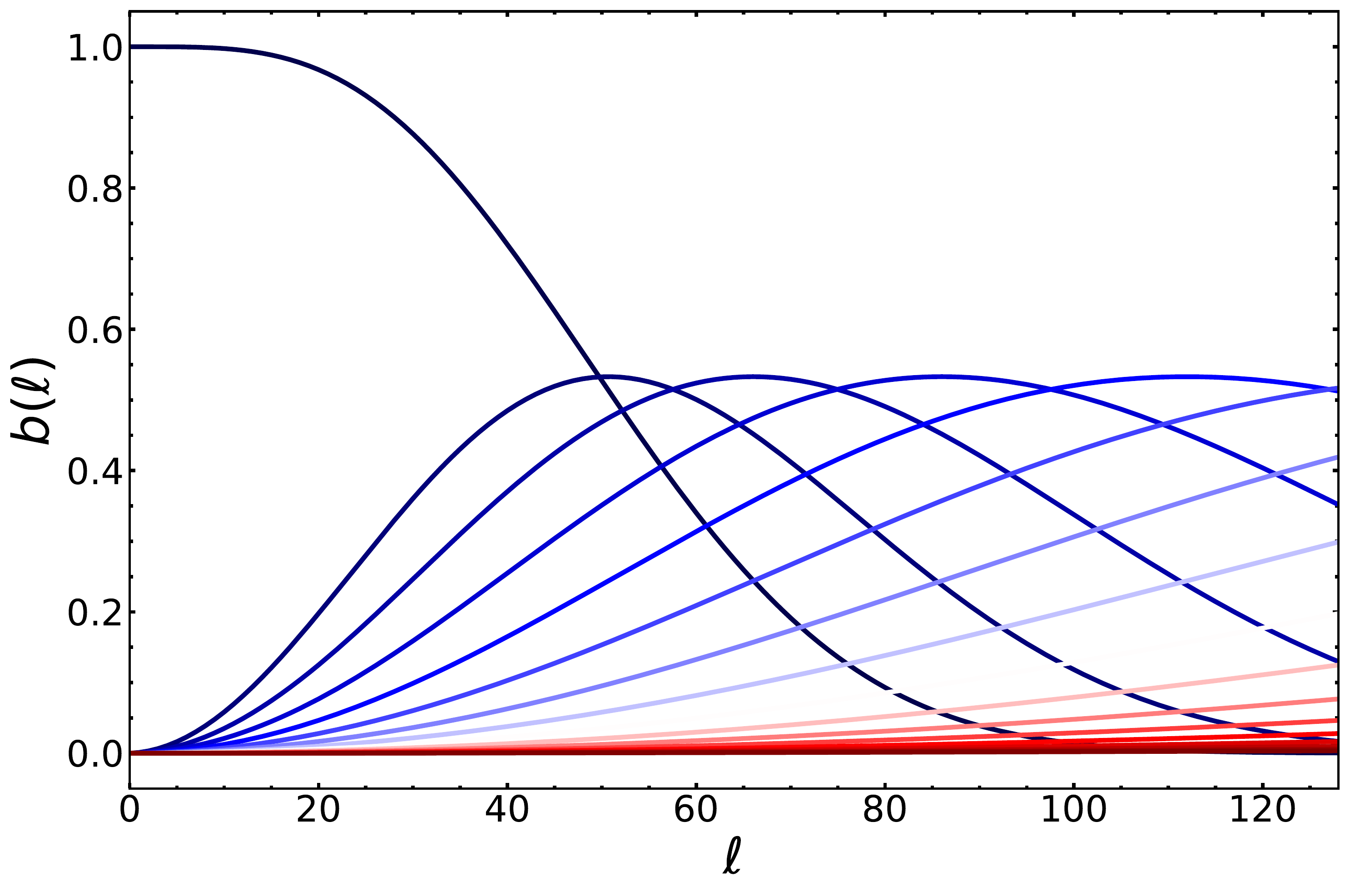} 
	\caption{Representation of Mexican needlet bands with $B=1.3$ in the harmonic domain. The first fourteen bands have been merged together according to Eq. \ref{eq:merge}, while the other bands are left unchanged. }
	\label{fig:bands}
\end{figure}
In practice, the needlet coefficients of an input $B$-mode map at frequency $i$, $\beta_{j}^{i}$, are obtained by filtering its harmonic coefficients $a_{\ell m}^{i}$ with a weighting function $b_{j}(\ell)$ which selects modes at different angular scales for each needlet scale $j$:
\begin{equation}
\beta_{j}^{i}(\hat{\gamma})
=\sum_{\substack{\ell,m}} (a_{\ell m}^{i}\cdot b_{j}(\ell))\cdotp Y_{\ell m}(\hat{\gamma}),
%\label{eq:Psi_nl}
\label{eq:needlets_map}
\end{equation}
where $\hat{\gamma}$ is a direction in the sky. This procedure in harmonic space is equivalent to performing a convolution of the map in real domain. \\
The shape of the needlet bands is defined by the choice of the harmonic function $b$ whose width is set by the parameter $B$: lower values of $B$ correspond to a tighter localisation in harmonic space (less multipoles entering into any needlet coefficient), whereas larger values ensure wider harmonic bands.  \\
In this work we adopt as baseline % the \emph{standard needlet} construction (\citealt{doi:10.1137/040614359}) and 
the \emph{Mexican needlet} construction \citep{2008arXiv0811.4440G} with $B=1.3$ (see Fig. \ref{fig:bands}  for their representation in harmonic space). However, we have verified that similar results can be obtained using even \emph{standard needlets} \citep{doi:10.1137/040614359} with $B=2$. The needlet filters have been obtained through the \emph{MTNeedlets}\footnote{\url{https://javicarron.github.io/mtneedlet}} Python package. \\
In the NILC method, the needlet coefficients of input multi-frequency data are linearly combined in such a way to obtain a minimum variance map $\beta_{j}^{\textrm{NILC}}$ on each scale $j$:
\begin{equation}
\beta_{j}^{\textrm{NILC}}(\hat{\gamma}) = \sum_{i=1}^{N_{\nu}}w_{i}^{j}(\hat{\gamma})\cdot \beta_{j}^{i}(\hat{\gamma})=\sum_{\substack{\ell,m}} a_{\ell m,j}^{\textrm{NILC}} \cdotp Y_{\ell m}(\hat{\gamma}),
\label{eq:NILC}
\end{equation}
where $i$ runs over the frequency channels of the experiment and $a_{\ell m,j}^{\textrm{NILC}}$ are the harmonic coefficients of $\beta_{j}^{\textrm{NILC}}$. The pixel-dependent weights which minimize the output variance are estimated through the following formula:
\begin{equation}
w_{i}^{j}(\hat{\gamma})=\frac{\sum_{k}C_{ik}^{j}(\hat{\gamma})^{-1}}{\sum_{ik}C_{ik}^{j}(\hat{\gamma})^{-1}},
\label{eq:NILC_weights} 
\end{equation}
where the covariance matrix $C_{ik}^{j}(\hat{\gamma})=\langle \beta_{j}^{i}\cdot \beta_{j}^{k} \rangle$ for scale $j$ at pixel $\hat{\gamma}$  is estimated  as  the  average  of  the  product  of  computed needlet  coefficients  over  some  space  domain $\mathcal{D}$.  This domain is usually chosen to be a circularly symmetric top-hat or a Gaussian window function centred at $\hat{\gamma}$, whose width varies with the considered needlet scale $j$. The latter is the adopted approach in this work. \\
The final NILC map is then reconstructed filtering again the harmonic coefficients $a_{\ell m,j}^{\textrm{NILC}}$ in Eq. \ref{eq:NILC} with $b_{j}(\ell)$ and summing them all:
\begin{equation}
X_{\textrm{NILC}}(\hat{\gamma})=\sum_{\substack{\ell,m}} a_{\ell m}^{\textrm{NILC}} \cdotp Y_{\ell m}(\hat{\gamma})
=\sum_{\substack{\ell,m}}\Bigg(\sum_{j} a_{\ell m,j}^{\textrm{NILC}}\cdot b_{j}(\ell)\Bigg)\cdotp Y_{\ell m}(\hat{\gamma}).
\label{eq:needlets_alm_final}
\end{equation}
Minimizing the variance (and hence the contamination) separately on different needlet scales leads to a more effective cleaning with respect to a simple minimization in real space. Indeed, on large scales, diffuse Galactic foregrounds dominate over the other components and are better removed in NILC. The same is expected to happen for the instrumental noise on smaller scales (larger multipoles). \\
The NILC solution can be statistically interpreted as a local maximization of the likelihood function in needlet space to recover the CMB signal. Indeed, in each pixel our multi-frequency set of needlet coefficients $y$ can be seen as the co-addition of the signal of interest $\beta$ (the CMB) with a known spectral behaviour ($X$) and the contamination of foregrounds and instrumental noise  $\epsilon$, which here for simplicity are assumed to be Gaussian:
\begin{equation}
    y = X\ \beta + \epsilon.
\end{equation}
In this case, the log-likelihood function can be written as:
\begin{equation}
    \Lambda \propto - (y - X\ \beta)^{\textrm{T}} \Omega^{-1} (y - X\ \beta),
\label{eq:like_Dom}
\end{equation}
where $\Omega=E(\epsilon^{\textrm{T}} \epsilon)$ is the ensemble covariance of the noise component. Given that $\Omega^{1/2}\Omega^{1/2}=\Omega$, Eq. \ref{eq:like_Dom} can be recast in the following way:
\begin{equation}
    \Lambda \propto - (\Tilde{y} - \Tilde{X}\ \beta)^{\textrm{T}} (\Tilde{y} - \Tilde{X}\ \beta),
\label{eq:like_Dom_1}
\end{equation}
where $\Tilde{y}=\Omega^{-1/2}y$ and $\Tilde{X}=\Omega^{-1/2}X$. The solution $\hat{\beta}$ which maximizes $\Lambda$ can then be easily computed:
\begin{equation}
    \hat{\beta} = (\Tilde{X}^{\textrm{T}} \Tilde{X})^{-1} \Tilde{X}^{\textrm{T}} \Tilde{y} = (X^{\textrm{T}}\Omega^{-1} X)^{-1} X^{\textrm{T}} \Omega^{-1} y,
\label{eq:like_Dom_2}
\end{equation}
which corresponds exactly to the NILC CMB solution in needlet space which is obtained employing Eq. \ref{eq:NILC_weights} and substituting the ensemble covariance $\Omega$ with the empirical one $C_{ik}^{j}$ which includes both noise and signal. \\
The estimation of this covariance matrix is performed locally in needlet space around each pixel due to the choice of a proper Gaussian window $\mathcal{D}$. However, the choice of such domain does not account for the foreground spectral variability within it. Thus, we can consider the NILC minimization as local in needlet space and global in the space of foreground spectral parameters. \\
One of the main issues that need to be taken into account when applying the NILC method is the generation of empirical correlations between the CMB signal and the residual contaminants \citep{2009A&A...493..835D} caused by the departure of
the empirical correlation matrix $C_{ik}$ from its ensemble average because of the
finite size of the sample over which it is estimated. This leads to a loss of CMB power especially at low multipoles (where fewer modes are sampled). \\
In practice, after the application of a component separation method on the $B$-mode data, we would like to extract cosmological information from the angular power spectrum ($C_{\ell}^{\textrm{out}}$) of the $BB$ cleaned map. This quantity is composed of several terms: 
\begin{equation}
    C_{\ell}^{\textrm{out}} = C_{\ell}^{\textrm{cmb}} + C_{\ell}^{\textrm{fgds}} + C_{\ell}^{\textrm{noi}} + 2\cdot C_{\ell}^{\textrm{c-f}} + 2\cdot C_{\ell}^{\textrm{c-n}} + 2\cdot C_{\ell}^{\textrm{n-f}},
\label{eq:spectra_out}
\end{equation}
where $C_{\ell}^{\textrm{cmb}},\ C_{\ell}^{\textrm{fgds}}$, and $C_{\ell}^{\textrm{noi}}$ are the angular power spectra of CMB, foreground, and noise residuals, while the other terms represent the corresponding correlations among these components. One can correct for the noise bias term, $C_{\ell}^{\textrm{noi}}$, either estimating it from Monte-Carlo simulations or computing the angular power spectra by correlating maps of independent data splits. \\
The foreground contamination ($C_{\ell}^{\textrm{fgds}}$), on the other hand, can be marginalised at the level of the likelihood (hoping to be as low as possible). Therefore our final estimate of the angular power spectrum of the NILC CMB solution results to be:
\begin{equation}
    \hat{C}_{\ell}^{\textrm{out}} = C_{\ell}^{\textrm{cmb}} + 2\cdot C_{\ell}^{\textrm{c-f}} + 2\cdot C_{\ell}^{\textrm{c-n}} + 2\cdot C_{\ell}^{\textrm{n-f}}
\label{eq:bias}
\end{equation}
If NILC is properly implemented, the correlation terms should be very small and the CMB well reconstructed. Taking advantage of simulations, we checked the goodness of the CMB angular power spectrum reconstruction by comparing $ \hat{C}_{\ell}^{\textrm{out}} = C_{\ell}^{\textrm{out}} - C_{\ell}^{\textrm{fgds}} - C_{\ell}^{\textrm{noi}}$ with the input $C_{\ell}^{\textrm{cmb}}$. \\
In order to include more modes in the minimization process at large angular scales, thus reducing the statistical uncertainty of the covariance matrix estimation of Eq. \ref{eq:NILC_weights} and the negative bias on the CMB reconstruction previously described, we optimise the employed set of needlet filters by adding together several bands at low multipoles with the following procedure:
\begin{equation}
b^{\textrm{new}}(\ell) =\sqrt{\sum_{j=j_{min}}^{j_{max}}b_{j}^{2}(\ell)}.
\label{eq:merge}
\end{equation}
In this analysis, the first fourteen Mexican bands have been merged, while filters on smaller scales are left unchanged. Such configuration of needlet bands is shown in Fig. \ref{fig:bands}.

\subsection{Clusters and MC-NILC}
\label{sec:clusters}
In this work we aim at performing the NILC variance minimization independently on regions of the sky where polarized $B$-mode Galactic emission presents similar spectral properties: \emph{MC-NILC} (Multi-Clustering NILC).
Throughout all the paper, we will refer to these regions identically as either clusters or patches. \\
The implementation of the MC-NILC pipeline requires the analysis of two separate aspects:
\begin{enumerate}[label=\alph*)]
    \item the identification of a blind tracer of the spectral properties of Galactic foregrounds in $B$ modes
    \item the clustering technique to adopt to assign pixels with close values of this tracer to the same patch.
\end{enumerate}
Both of them are discussed in the following sections.
\subsubsection{Tracer of foreground spectral properties}
\label{sec:GNILC&cPILC}
In order to estimate the spatial distribution of the dust and synchrotron spectral properties in $B$ modes with a blind approach, we construct the ratio of $B$-mode maps at two separate frequencies. Such frequency maps are chosen in such a way that dust and synchrotron frequency dependence is encompassed through the whole frequency coverage of an experiment. \\
The selection of the frequencies to use in the construction of the ratio can be calibrated with realistic simulations of the considered experiment. 
As it will be detailed in Sect. \ref{sec:results}, the best choice of frequencies for the ratio is a template of foreground $B$ modes at high-frequency, such as $402$ or $337$ GHz, and one at a 'CMB' frequency channel, as $119$ GHz. \\
In first approximation, such a ratio can be identified with the map of $(\nu_1/\nu_2)\,^{\beta_{\textrm{eff}}^{\textrm{B}}}$, where $\nu_1$ and $\nu_2$ are the chosen frequencies and $\beta_{\textrm{eff}}^{\textrm{B}}$ is an effective spectral index of $B$-mode foreground emission if modelled with a power law. Such an approximation is expected to be quite effective in describing both synchrotron \texttt{s1} and thermal dust \texttt{d1} emission in the \textit{LiteBIRD} frequency range which is almost insensitive to the dust temperature $T_{\textrm{d}}$ and where both components can be approximately modelled with power laws. It would be interesting in the future to assess the performance of MC-NILC with this blind tracer for CMB missions observing at higher frequencies than \textit{LiteBIRD}, such as \textit{PICO} \citep{2019arXiv190210541H}, since the dust temperature spatial variations will not be degenerate anymore with the spectral index ones. \\
In Appendix \ref{app:tracers}, we show more in detail that the ratio introduced above traces the combination of an effective spectral index of dust $B$ modes and of an emission ratio between synchrotron and dust components at a central frequency channel. Moreover, we prove that a single ratio is able to better blindly trace the spectral properties of the $B$-mode foregrounds than two distinct ratios: one at low (for synchrotron) and one at high frequencies (for dust). 
The choice of employing $B$-mode maps in the ratio, instead of considering the spectral indices of Q and U or some other combinations of the Stokes parameters, is mainly due to the fact that the spatial distribution of Q and U spectral properties is completely different with respect to the one of $B$ modes, given the non-locality of the QU-to-B transformation. \\
Pixels with close values of the chosen ratio are then grouped together in separate patches where the variance of the output solution is minimized independently for each needlet scale. Two different cases are considered:
\begin{enumerate}
    \item an \emph{ideal} one, where a different ratio is built for each needlet scale $j$ employing 'noiseless' foreground needlet coefficients at the two different frequencies,
    \item a \emph{realistic} one, where a unique ratio is used for all needlet scales constructed with templates (estimated from input data) of foreground $B$ modes at the two different frequencies filtered with the first needlet band $b_{0}$ to avoid the more significant contamination of CMB and noise at small angular scales.
\end{enumerate}
The first assessment of the performance of MC-NILC is performed in the ideal case in Sect. \ref{sec:ideal_app}, while the results for the realistic one are shown in Sect. \ref{sec:real_appr}. \\
Given the current uncertainties in the foreground emission models, it is crucial to consider a \emph{realistic} approach where templates of Galactic $B$-mode emission are obtained directly from multi-frequency data with a reduced amount of CMB and instrumental noise contamination. To achieve this goal, we employ the Generalised Needlet Internal Linear Combination (GNILC, \citealt{2011MNRAS.418..467R}).
GNILC is a multi-dimensional generalisation of
the standard NILC method, where contaminating components (as CMB and noise) can be de-projected from a multi-frequency dataset with some a-priori information of their statistical properties. This method has been successfully employed in the past to reconstruct the intensity of the Galactic emission at the \textit{Planck} frequencies \citep{2016A&A...596A.109P}. It provides a set of foreground templates (at the frequencies of the employed dataset) which includes all the independent Galactic modes whose power is larger with respect to the one of CMB and instrumental noise. However these maps are still affected by some residual CMB and noise contamination. \\
In this work, the domain selection of the GNILC method has been optimised according to the same MC-NILC principle. The weights are separately computed in different patches of a sky partition which accounts for the spatial variation of the spectral properties of the $B$-mode foregrounds. Details of this new implementation, Multi-Clustering GNILC (\emph{MC-GNILC}), can be found in Sect. \ref{sec:MCGNILC}. 
In such realistic approach, the $B$-mode foreground templates, both at high frequency and at $119$ GHz, are obtained in a fully model-independent way through a MC-GNILC run on the \textit{LiteBIRD} $B$-mode dataset. \\
A $B$-mode field at any frequency fluctuates between positive and negative values being in some pixels close to zero. In these pixels we expect low values of foreground needlet coefficients at both the frequencies adopted in the ratio and, indeed, we have verified that no significant numerical instabilities arise in the ideal case. Vice versa, in the realistic approach they are affected by CMB and noise contamination and are likely to be assigned to a wrong cluster. However, these points with low values of the $B$-mode foreground emission are expected to less affect the residual contamination in the final MC-NILC CMB solution and, thus, this mis-modelling of the partition should not significantly impact the performance of the component separation.

\begin{figure*}
\centering
	\includegraphics[width=0.49\textwidth]{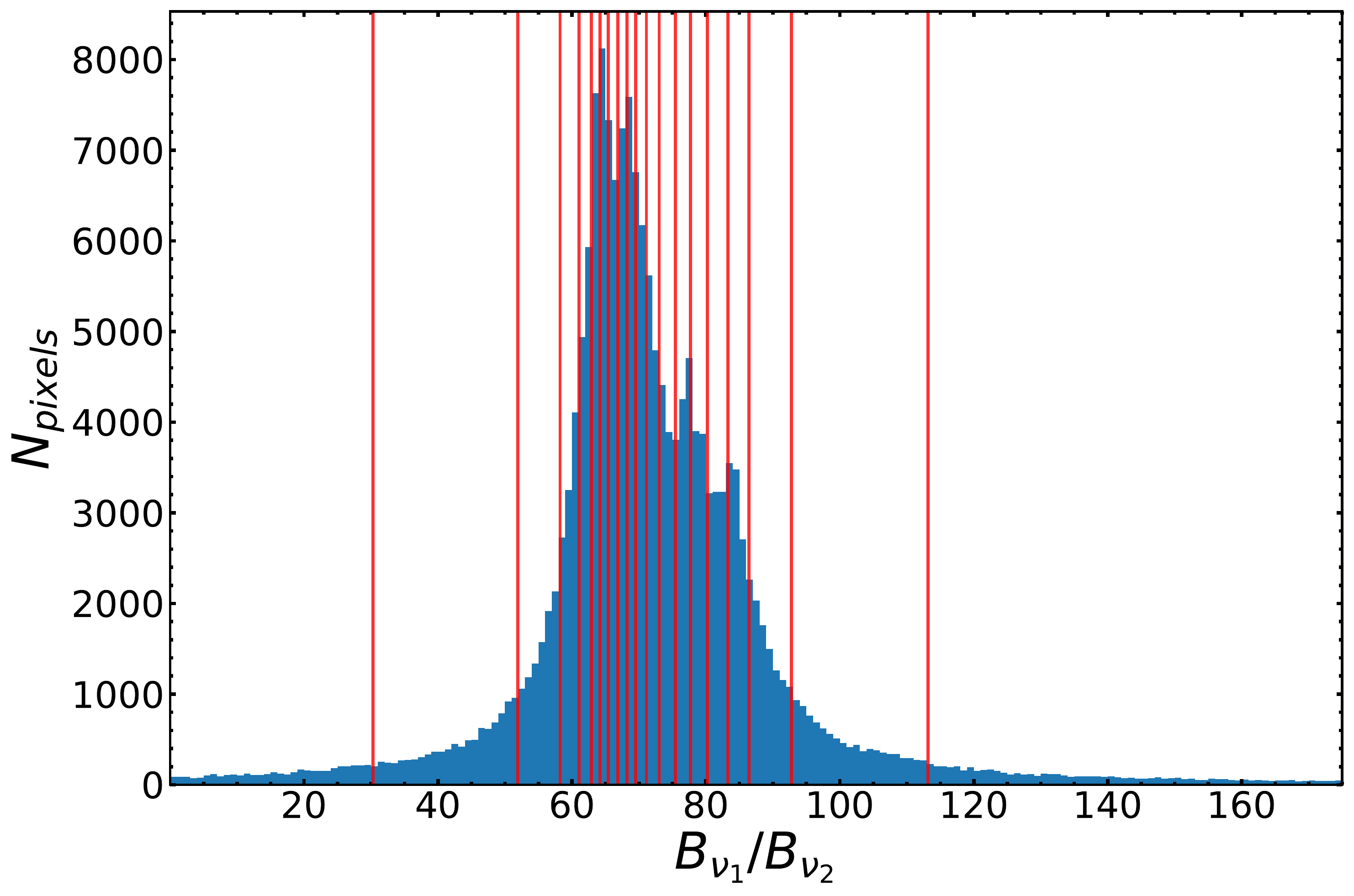}
	\hspace{0.2 cm}
	\includegraphics[width=0.49\textwidth]{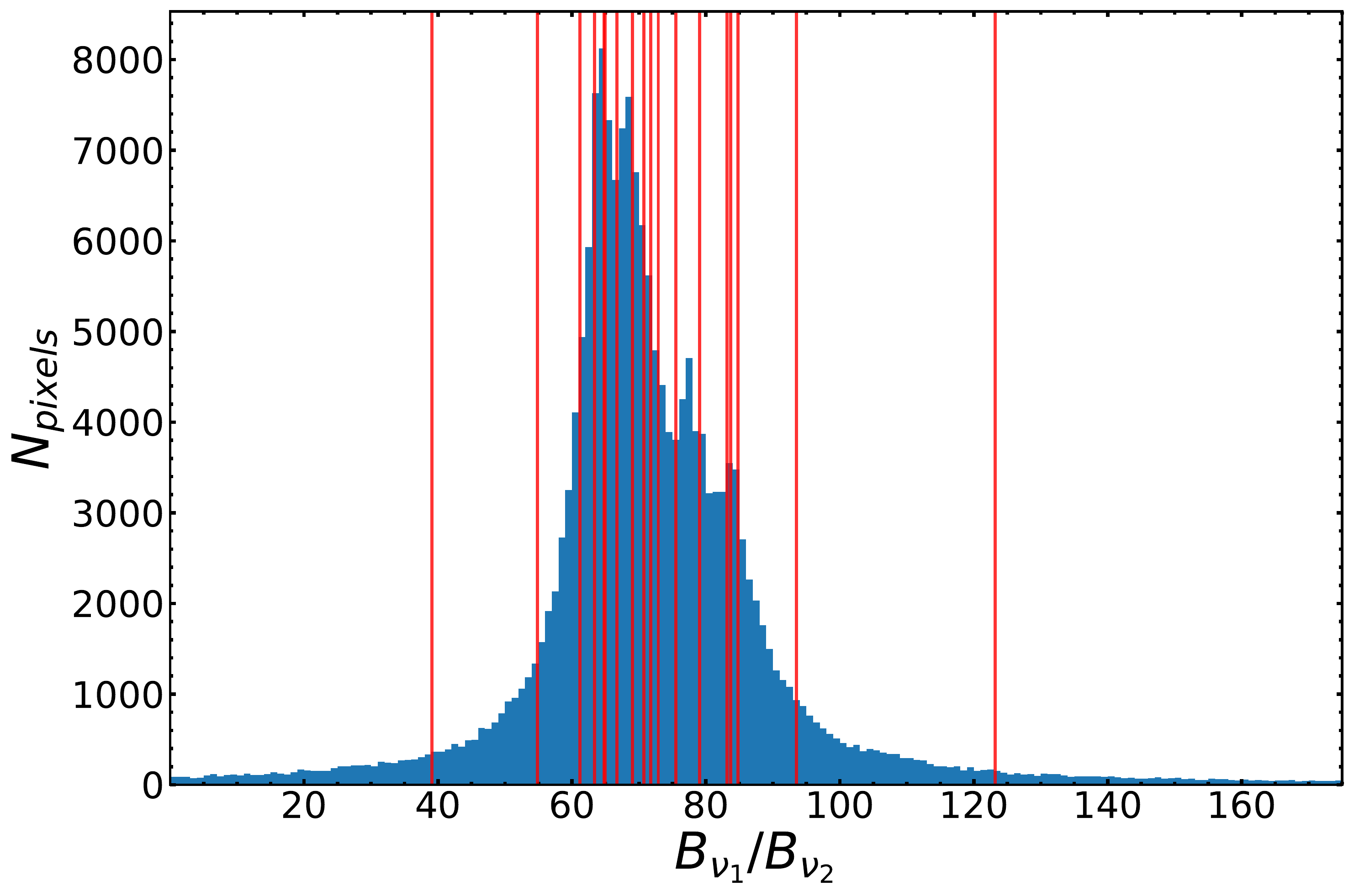}
	\caption{Representation of the CEA-thresholding (left) and of one realisation of RP thresholding (right) of the histogram of the ratio of $B$-mode \texttt{d1s1} foreground emission at $402$ and $119$ GHz, filtered with the first Mexican needlet band shown in Fig. \ref{fig:bands}. Here the the number of patches is $K=20$ just for visualisation purposes.}
\label{fig:histograms}
\end{figure*}
\subsubsection{Clustering approaches}
\label{sec:clustering}
Once a tracer of the $B$-mode foreground spectral properties has been built, we partition the sky in different patches which collect pixels with close values of such quantity. In practice, the clusters are obtained by thresholding the histogram of the considered ratio. The thresholding is performed following two separate approaches:
\begin{itemize}
    \item \emph{CEA} (Clusters of Equal Area): a unique partition of the sky is obtained where the different clusters have all the same area and collect sub-sets of $N_{\textrm{p}}/K$ pixels with increasing values of the ratio, where $N_{\textrm{p}}$ is the total number of pixels and $K$ the chosen amount of clusters;
    \item \emph{RP} (Random Partitions): we generate several different partitions of the sky where in each partition each cluster collects a random amount of pixels.
\end{itemize}
The CEA thresholding and one realisation of the RP thresholding are shown in Fig. \ref{fig:histograms}. \\ 
Two MC-NILC pipelines have been developed exploiting the two different clustering techniques: \emph{CEA-MCNILC} and \emph{RP-MCNILC}. In CEA-MCNILC, the CMB solution with minimum variance is estimated independently within each patch. In RP-MCNILC, instead, we generate $K_{\textrm{p}}=50$ different partitions of the sky and for each partition MC-NILC is applied obtaining $K_{\textrm{p}}$ different full-sky CMB solutions. The final RP-MCNILC CMB map is estimated computing the average among all these different solutions. The number of partitions $K_{\textrm{p}}$ is chosen by estimating when a convergence in the variance of the output RP-MCNILC solution is reached varying $K_{\textrm{p}}$.
The rationale of applying the RP approach, already introduced in \citealt{2019JCAP...02..039K}, is that pixels with close values of the ratio are assigned to the same cluster with larger probability with respect to CEA. Throughout the paper, the CEA approach is adopted unless otherwise specified. \\
In general the number of clusters should be chosen taking into account two main factors:
\begin{itemize}
    \item $B$-mode foreground residuals,
    \item the bias in the  reconstructed CMB power spectrum.
\end{itemize}
We remark here, that the former criterion can be intuitively associated to a sort of \emph{under-partition}  measure, as the foreground residuals decrease with increasing  number  of clusters $K$. Vice versa, the latter acts as an \emph{over-partition} measure, with bias increasing with $K$. \\
We have seen in Eq. \ref{eq:like_Dom_2} that NILC solution corresponds to the CMB estimate which maximizes a likelihood which is local in needlet domain and global in the space of foreground spectral indices, assuming that Galactic emission follows a Gaussian distribution. MC-NILC, instead, can be interpreted statistically as a local maximization of the likelihood in the space of foreground spectral indices. \\
The minimization of the variance in several different patches in the sky can introduce a significant bias in the CMB reconstruction. Thus, to alleviate such phenomenon, we adopt a technique presented in Coulton et al. in prep.: the MC-NILC covariance in each pixel is computed considering all the other pixels in the same cluster but excluding the pixel itself and a circular region around it. This procedure highly reduces the bias, but it does not completely eliminate it; therefore the merging of first needlet bands of Eq. \ref{eq:merge} is still required.

\subsection{MC-GNILC}
\label{sec:MCGNILC}
\begin{figure}
	\centering
    \includegraphics[width=0.475\textwidth]{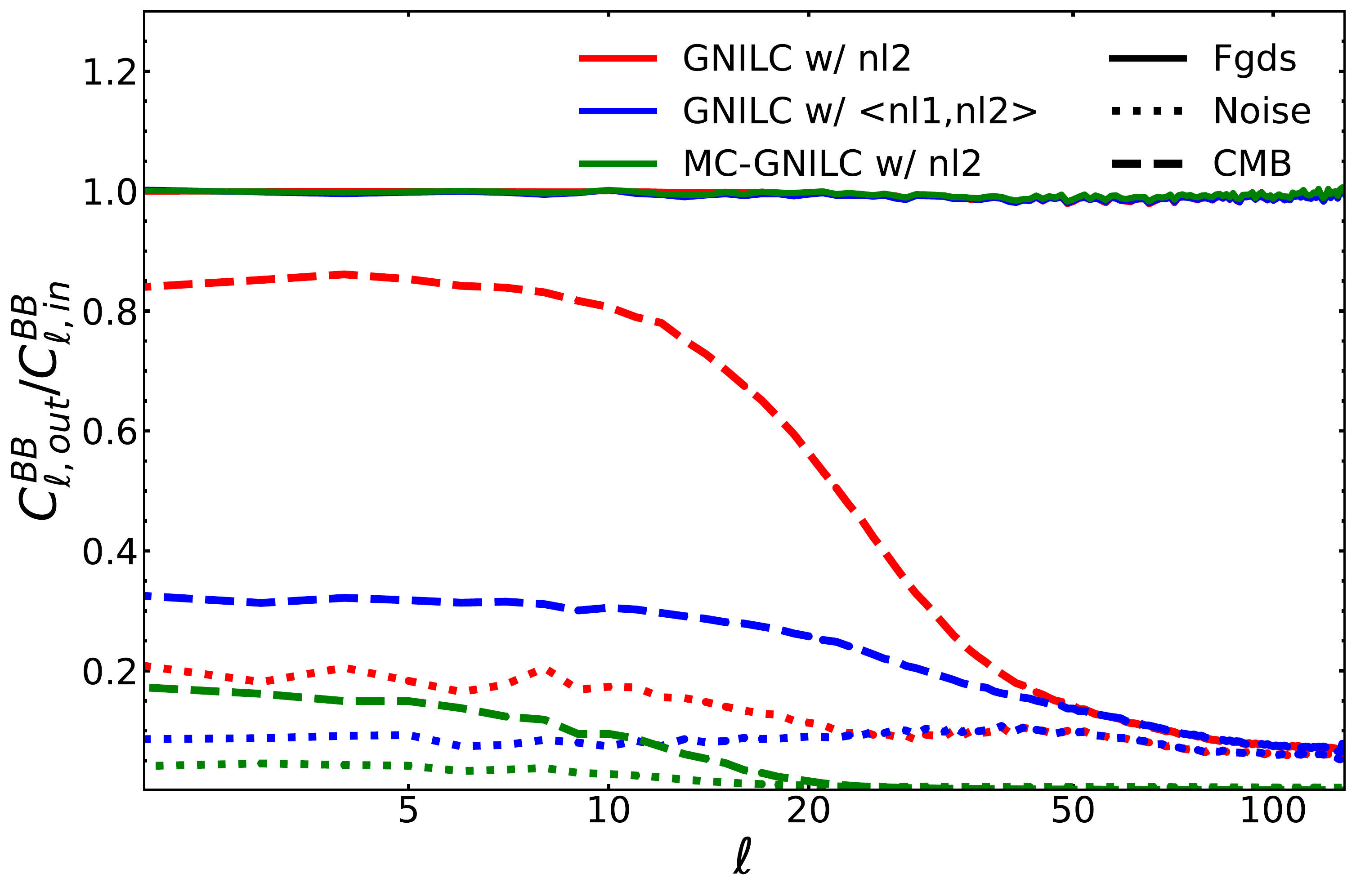} 
	\caption{Ratio of average angular power spectra over $200$ simulations of output foreground (solid lines), noise (dotted), and CMB (dashed) components and input ones at $119$ GHz. Three different applications are shown: GNILC with $\textit{nl2}$ needlet configuration (red), the average of GNILC solutions with $\textit{nl1}$ and $\textit{nl2}$ (blue), MC-GNILC with $\textit{nl2}$ (green). The angular power spectra are computed employing the \textit{Planck} \textit{GAL60} Galactic mask  with $f_{\textrm{sky}}=60\,\%$. The PySM \texttt{d1s1} model is assumed for the Galactic emission. }
	\label{fig:cls_gnilc}
\end{figure}

GNILC allows to obtain cleaned templates of the Galactic emission from microwave data with a fully model--independent approach. It consists of linearly combining the input $N_{\nu}$ maps of Eq. \ref{eq:needlets_map} at each needlet scale $j$:
\begin{equation}
    f_{j}=W \beta_{j},
\label{eq:gnilc_f}
\end{equation}
where $f_{j}$ is the set of $N_{\nu}$ reconstructed templates and $W$ is designed to offer a unit response to the Galactic foreground emission while minimising the total variance of $f_{j}$. As reported in \citealt{2016A&A...596A.109P}, such weights can be exactly derived:
\begin{equation}
    W=F(F^TC^{-1}F)^{-1}F^TC^{-1},
\label{eq:w_gnilc}
\end{equation}
with $C$ the empirical covariance matrix of Eq. \ref{eq:NILC_weights} and $F$ a $N_{\nu} \times m$ mixing matrix which links the real foreground emission, $f_{j}$ maps,  to a set of $m$ independent (non-physical) Galactic templates, $t_{j}$:
\begin{equation}
    f_{j}=F t_{j}.
    \label{eq:Ft}
\end{equation}
The Galactic signal mixing matrix $F$ is explicitly estimated as:
\begin{equation}
    F=R_{N}^{1/2} U_s,
\end{equation}
where $R_{N}$ is the covariance of the \emph{nuisance} terms and $U_s$ is the set of eigenvectors of $R_{N}^{-1/2}CR_{N}^{-1/2}$ with eigenvalues greater than unity. Therefore, linearly combining the input maps with the GNILC weights $W$ permits to reconstruct all the independent foreground modes whose power is larger with respect to the nuisance components (CMB and instrumental noise). The Galactic emission in pixel space is then reconstructed at each frequency through an inverse needlet transform of the needlet maps $f_{j}$. \\
\begin{figure}
	\centering
    \includegraphics[width=0.475\textwidth]{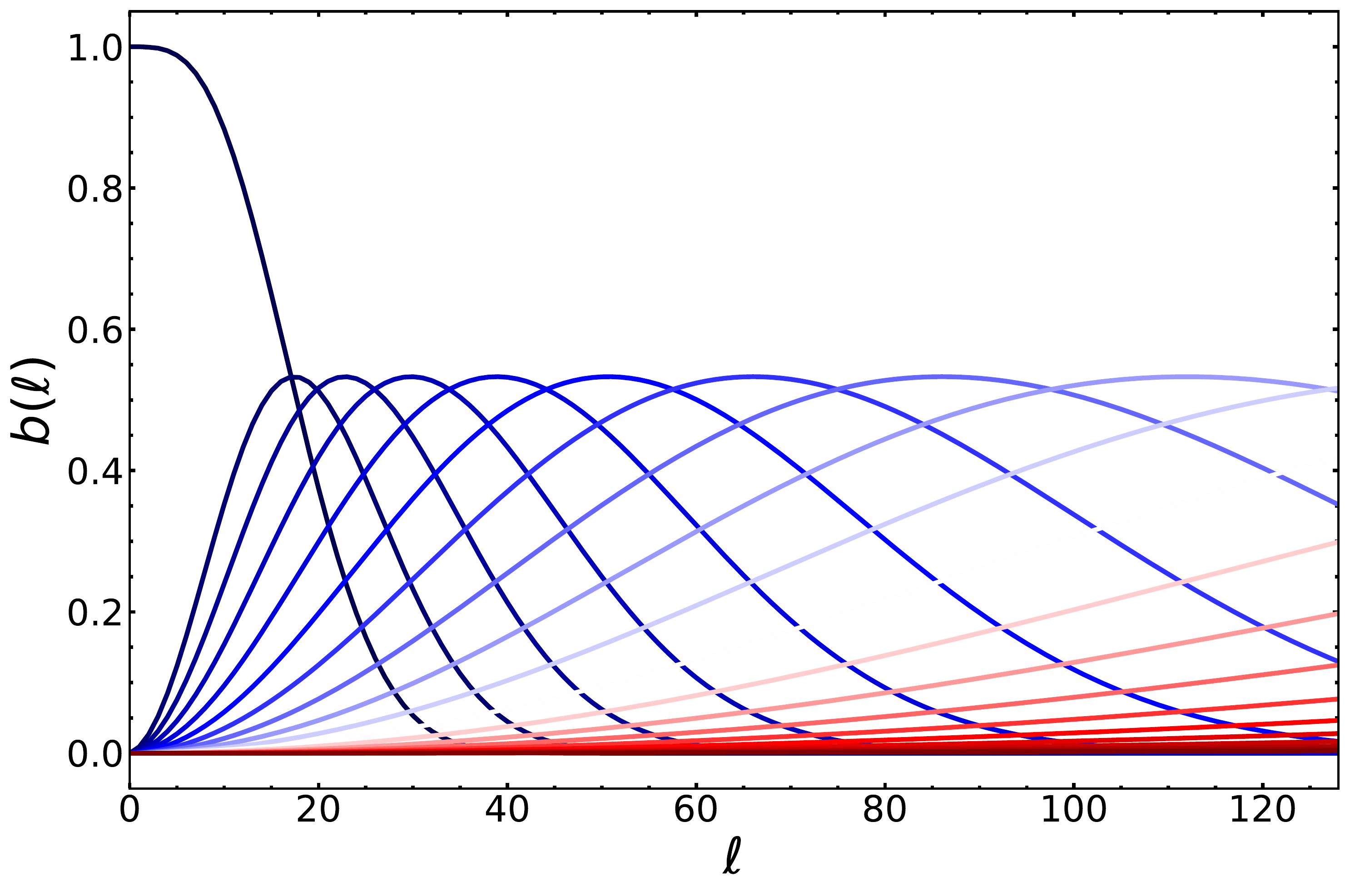} 
	\caption{Representation of Mexican needlet bands with $B=1.3$ in the harmonic domain. The first ten bands have been merged together according to Eq. \ref{eq:merge}, while the other bands are left unchanged. }
	\label{fig:bands_1}
\end{figure}
In GNILC, the covariance matrix $C$ is locally estimated by averaging the product of needlet coefficients within gaussianly shaped domains. In this work, we present an optimisation of the domain selection, \emph{MC-GNILC} (Multi-Clustering GNILC), analogous to that implemented in MC-NILC. In MC-GNILC, $C$ is computed in each pixel taking into account the affinity of the spectral properties of $B$-mode foregrounds instead of the spatial locality. The covariance matrix and, thus, the GNILC weights are separately estimated in different patches obtained by thresholding a ratio between $B$-mode foreground templates at two different frequencies, as described in Sect. \ref{sec:clustering}. In this case, such templates are obtained by applying the standard GNILC method. \\
The MC-GNILC method is applied by adopting the needlet configuration (hereafter $\textit{nl2}$) shown in Fig. \ref{fig:bands_1}. It differs from that employed for MC-NILC and shown in Fig. \ref{fig:bands}  (hereafter $\textit{nl1}$), because, in this case, the first needlet band is obtained by merging the first ten bands (instead of fourteen) with Eq. \ref{eq:merge}. Therefore, $\textit{nl2}$ configuration has a larger number of bands at low multipoles and, thus, if compared to $\textit{nl1}$, permits to better reconstruct the foreground emission on the largest angular scales at the price of an enhanced CMB and noise contamination in the reconstructed maps. \\
The specific MC-GNILC pipeline which has been validated on \textit{LiteBIRD} $B$-mode simulated dataset can be summarised as follows:
\begin{itemize}
    \item applying GNILC with the $\textit{nl1}$ needlet configuration shown in Fig. \ref{fig:bands};
    \item applying GNILC with the $\textit{nl2}$ needlet configuration shown in Fig. \ref{fig:bands_1};
    \item the final GNILC Galactic templates are obtained by averaging the solutions with $\textit{nl1}$ and $\textit{nl2}$ configurations;
    \item at each needlet scale, a ratio is built from the corresponding needlet GNILC templates at $337$ and $119$ GHz and the RP approach is adopted to generate sky partitions (see Sect. \ref{sec:clustering}), each with $50$ patches;
    \item MC-GNILC is applied for the different random partitions and the final Galactic templates are obtained by averaging among the different solutions.
\end{itemize}
The average of the GNILC templates derived with the $\textit{nl1}$ and $\textit{nl2}$ needlet configurations allows to lower the CMB and noise contamination with respect to the $\textit{nl2}$ GNILC solution while preserving some of the Galactic signal that is de-projected in the $\textit{nl1}$ GNILC template. \\
In Fig. \ref{fig:cls_gnilc}, we show the ratio between the angular power spectra of foreground, CMB, and noise components in output Galactic templates and those of input at $119$ GHz for three different cases: GNILC with $\textit{nl2}$ configuration, the average of GNILC templates with $\textit{nl1}$ and $\textit{nl2}$, MC-GNILC with $\textit{nl2}$. These angular power spectra are computed adopting the publicly released \textit{Planck} \textit{GAL60} Galactic mask \footnote{\url{https://pla.esac.esa.int}}. It is possible to observe that MC-GNILC leads to a better CMB and noise subtraction without paying much penalty in Galactic signal loss. These results are obtained assuming the \texttt{d1s1} model for the foreground emission, but analogous outcomes have been found for the other models presented in Sect. \ref{Sec:sims}. The MC-GNILC Galactic templates are thus employed in the MC-NILC realistic approach of Sect. \ref{sec:real_appr}.

\subsection{Foregrounds masks}
\label{sec:masks}
In order to perform a cosmological analysis of a NILC or MC-NILC CMB solution, we have to mask the most contaminated regions to lower the foreground residuals in the output angular power spectrum. Throughout all the paper, masks are generated in analogy with what has been done in \citet{2022arXiv220202773L}. The Galactic plane is masked with the publicly released \textit{Planck} \textit{GAL60} mask, which retains a sky fraction of $60\,\%$. A further $10\,\%$ of the sky is removed by thresholding the map of the average foreground residuals after a smoothing with a beam of FWHM $3^\circ$. The obtained mask yields to a final sky fraction of $50\,\%$. \\
The above described masking procedure, hereafter \texttt{mask1}, cannot be applied on real data. However, as already stated in \citet{2022arXiv220202773L}, our knowledge about polarized $B$-mode foreground emission will be expanded in the near future thus leading us to the development of realistic and optimal methodologies to proper mask the most foreground contaminated regions across the sky by the time of the \textit{LiteBIRD} launch. \\
The choice of the \texttt{mask1} strategy allows us to fairly compare the MC-NILC performance in subtracting Galactic contamination with the results reported by the LiteBIRD Collaboration in \citet{2022arXiv220202773L}. However, in this work, we have even considered a more conservative procedure to mask the output CMB $B$-mode solution: \texttt{mask2}. We linearly combine the GNILC templates obtained with the $\textit{nl2}$ needlet configuration (see Sect. \ref{sec:MCGNILC}) with the MC-NILC weights. This procedure enables us to obtain an estimate of the output foreground residuals, $\Tilde{B}_{\textrm{res}} = \Tilde{B}_{\textrm{fgds}} + \Tilde{B}_{\textrm{noi}} + \Tilde{B}_{\textrm{CMB}}$, with a reduced contamination of CMB ($\Tilde{B}_{\textrm{CMB}}$) and noise ($\Tilde{B}_{\textrm{noi}}$) with respect to their contribution in the MC-NILC CMB solution. The mask is then obtained by thresholding the local root mean square of such derived residuals map, estimated as $\sqrt{<(\Tilde{B}_{\textrm{fgds}} + \Tilde{B}_{\textrm{CMB}})^2>}$. In practice, the averaging is performed locally in each pixel by smoothing the $(\Tilde{B}_{\textrm{fgds}} + \Tilde{B}_{\textrm{CMB}})^2$ map with a Gaussian beam of FWHM=$3\degree$. The noise term $\Tilde{B}_{\textrm{noi}}$ can be safely neglected if the beam is sufficiently large and the above quantity is computed as the cross-product of two solutions obtained from data-splits with uncorrelated noise. In this case, we have adopted the GNILC templates instead of the MC-GNILC ones, because they retain a larger number of foreground modes and, thus, better allow to trace locally the spatial distribution of the residuals. At the same time, to lower the CMB contamination at the largest angular scales, the GNILC weights in the first needlet band are derived adding the further constraint of de-projecting some of the CMB signal: $W_{\textrm{GNILC}}\cdot a_{\textrm{CMB}}=0.7$, with $a_{\textrm{CMB}}$ the CMB spectral energy distribution (SED). With such more conservative masking strategy, the retained sky fraction is $40\%$. \\
A comparison of angular power spectra computed with both masking procedures is provided in Fig. \ref{fig:cls_diffmasks} for the realistic MC-NILC approach. We can notice that the amount of foreground residuals is comparable, although obtained with unequal observed sky fractions. Therefore, similar constraints to those reported in this paper would be found with \texttt{mask2} approach. However, we expect a reduction of the sensitivity on the tensor-to-scalar ratio due to the larger variance of the reconstructed power spectrum led by the lower observed sky fraction. 

\subsection{Estimation of the residual contamination and NILC bias}
\label{sec:like}
When NILC or MC-NILC are applied on simulated data, the foreground and noise residuals in the $B$-mode CMB solution can be estimated simply by linearly combining the input foreground and noise maps with the corresponding NILC or MC-NILC weights. Therefore, to assess the amount of residual contamination, it is possible to compare their angular power spectra with a $BB$ primordial signal of tensor modes corresponding to a value of $r$ comparable to the one targeted by \textit{LiteBIRD}. \\
Considering Eq. \ref{eq:spectra_out} and assuming that cross-correlations among components are negligible, the angular power spectrum of the output solution is contaminated by foreground ($C_{\ell}^{\textrm{fgds}}$) and noise ($C_{\ell}^{\textrm{noi}}$) residuals. The noise term can be corrected for either by using Monte-Carlo simulations or by analysing cross-angular power spectra of independent subsets of the data. Thus, the main bias to $C_{\ell}^{\textrm{out}}$ will be given by the Galactic contamination. \\
In order to assess the relevance of such bias, it is possible to perform a fit of an effective tensor-to-scalar ratio $r_{\textrm{fgds}}$ on binned foreground residuals average angular power spectrum, $C_{\ell_{\textrm{b}}}^{\textrm{fgds}}$, assuming a Gaussian likelihood \citep{2008PhRvD..77j3013H}:
\begin{equation}
    -2\log\mathcal{L}(r)=\sum_{\ell_{\textrm{b}},\ell'_{\textrm{b}}}\Big(C_{\ell_{\textrm{b}}}^{\textrm{fgds}}-rC_{\ell_{\textrm{b}}}^{r=1}\Big)M_{\ell_{\textrm{b}}\ell'_{\textrm{b}}}^{-1}\Big(C_{\ell'_{\textrm{b}}}^{\textrm{fgds}}-rC_{\ell'_{\textrm{b}}}^{r=1}\Big),
\label{eq:like}
\end{equation}
where $C_{\ell_{\textrm{b}}}^{r=1}$ is the primordial CMB $B$-mode power spectrum for a tensor-to-scalar ratio $r=1$, while $M_{\ell_{\textrm{b}}\ell'_{\textrm{b}}}^{-1}$ is the inverse of the covariance matrix of the binned output angular power spectrum and it has been estimated through Monte-Carlo simulations:
\begin{equation}
M_{\ell_{\textrm{b}}\ell'_{\textrm{b}}}=\mathrm{Cov}\Big(C_{\ell_{\textrm{b}}}^{\textrm{out}},C_{\ell'_{\textrm{b}}}^{\textrm{out}}\Big).
\label{eq:Mllb}
\end{equation}
Binning the angular power spectra makes them gaussianly distributed and mitigates the mode coupling generated by the use of a mask. In this work, we employ a constant binning scheme with $\Delta_{\ell}=10$. We have tested the robustness of the obtained constraints both varying the binning scheme and even considering an inverse-Wishart likelihood function. Eq. \ref{eq:Mllb} accounts for the cosmic variance of the lensing signal, the sample variance of the residual foreground power spectrum, together with that of the noise power spectrum and all their cross-terms. 
In this work we assume no de-lensing of the CMB $B$ modes in analogy with the analysis in \citealt{2022arXiv220202773L}. \\
A similar procedure is performed to assess the relevance of the bias in the CMB reconstruction in the NILC and MC-NILC solutions. In this case, in Eq. \ref{eq:like}, $C_{\ell_{\textrm{b}}}^{\textrm{fgds}}$ is replaced by
$C_{\ell_{\textrm{b}}}^{\textrm{bias}} = C_{\ell_{\textrm{b}}}^{\textrm{out}} - C_{\ell_{\textrm{b}}}^{\textrm{noi}} - C_{\ell_{\textrm{b}}}^{\textrm{fgds}} - C_{\ell_{\textrm{b}}}^{\textrm{cmb}}$. \\
For all results, when a posterior distribution of $r_{\textrm{fgds}}$ is shown, if $r_{\textrm{fgds}}=0$ is within $2\sigma$, we report upper bounds at $68\ \%$ CL to be compared with the targeted sensitivity at equal significance of the LiteBIRD experiment: $r\sim 10^{-3}$.

\section{Results}
\label{sec:results}
\begin{figure*}
	\centering
	\includegraphics[width=0.48\textwidth]{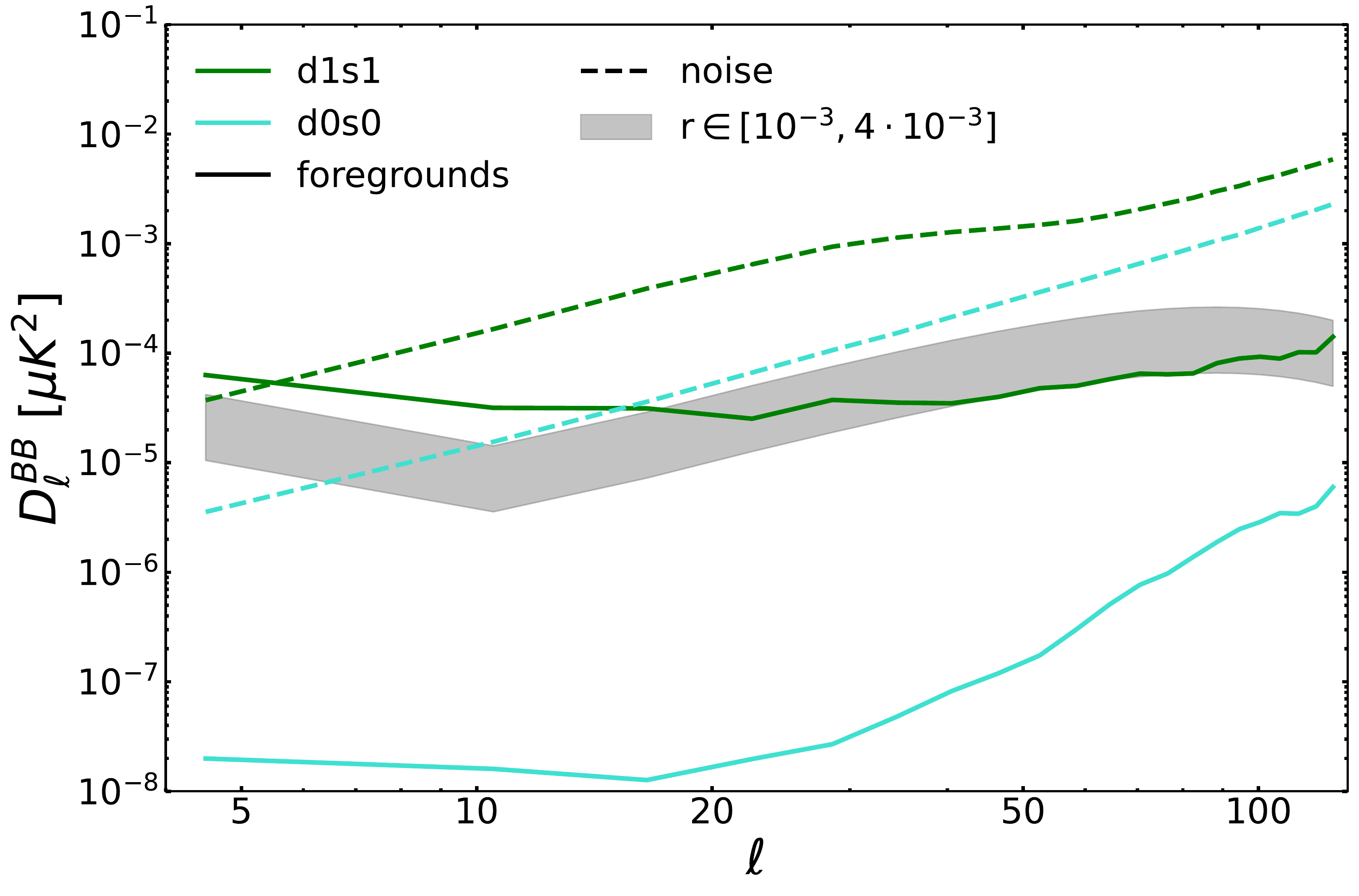} 
	\hspace{0.5 cm}
	\includegraphics[width=0.48\textwidth]{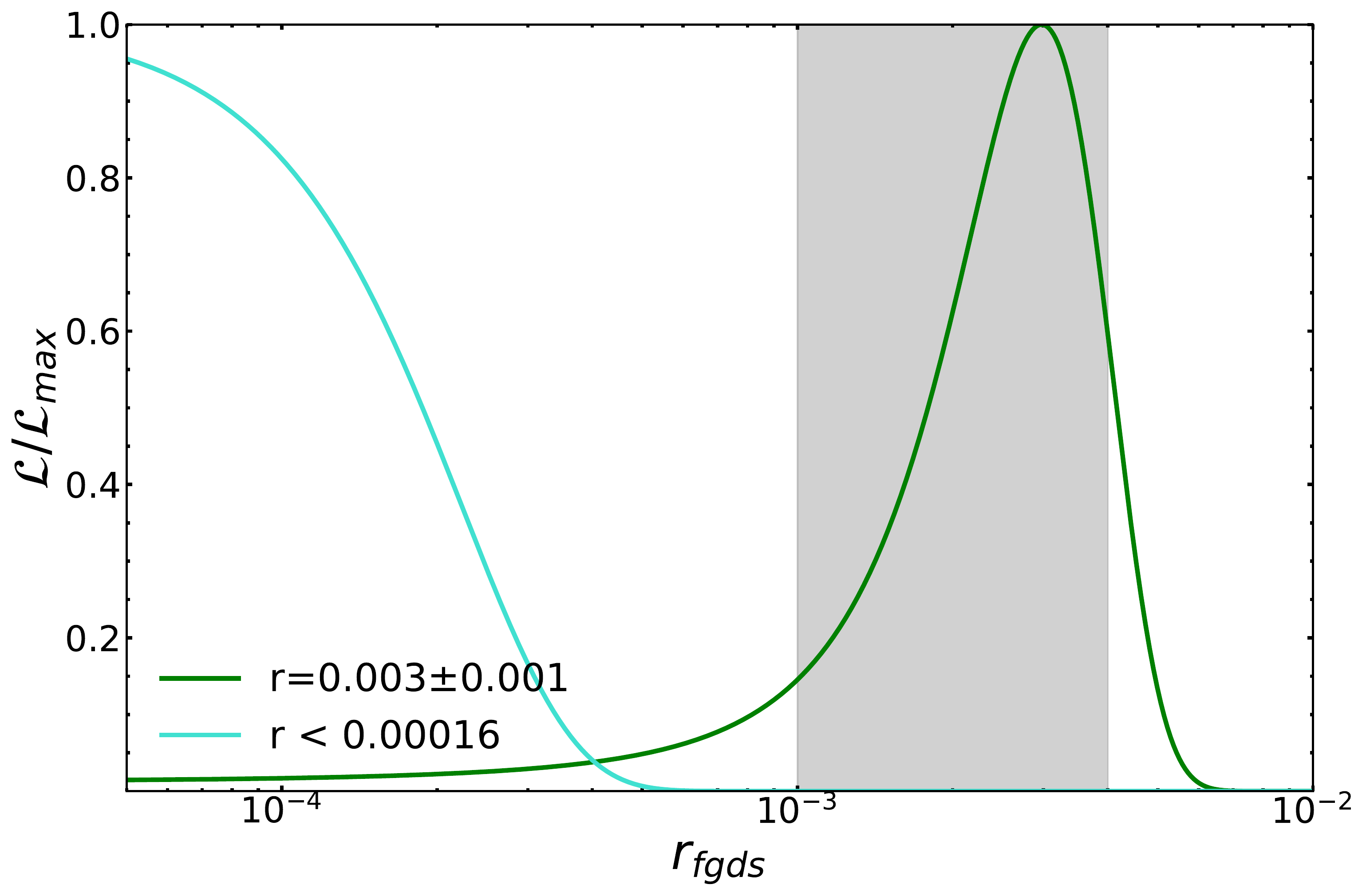} 
\caption{On the left: angular power spectra of foreground (solid) and noise (dashed) residuals when NILC is applied on $200$ \textit{LiteBIRD} different simulated datasets assuming the \texttt{d1s1} (green) or the \texttt{d0s0} (turquoise) Galactic sky models. The angular power spectra are computed employing masks obtained according to the first strategy (\texttt{mask1}) described in Sect. \ref{sec:masks} with $f_{\textrm{sky}}=50\,\%$. The adopted binning scheme is $\Delta\ell =6$ for visualisation purposes. On the right: the posterior distribution of an effective tensor-to-scalar ratio fitted on the foreground residuals for the two different cases. For the estimation of the posteriors a binning scheme of $\Delta\ell =10$ has been used as explained in Sect. \ref{sec:like}. The reported upper bounds and confidence intervals refer to the $68\%$ CL. The grey areas highlight the range of amplitudes of the primordial tensor signal targeted by \textit{LiteBIRD}: $r\in [0.001,0.004]$.}
\label{fig:NILC_litebird}
\end{figure*}

In this section, we assess the performance of NILC and MC-NILC application on the \textit{LiteBIRD} simulated dataset. 
All the results are obtained considering $200$ simulations (unless otherwise specified) of the \textit{LiteBIRD} $B$-mode multi-frequency maps including Galactic foregrounds, different realisations of CMB with only lensing and instrumental noise. \\
In the case of NILC application, we have considered both the \texttt{d0s0} and \texttt{d1s1} foreground models  described in Sect. \ref{Sec:sims}.
The average $BB$ angular power spectra of NILC foreground and instrumental noise residuals from the $200$ simulations are compared with the theoretical spectrum of the primordial cosmological tensor signal targeted by the \textit{LiteBIRD} mission ($r\sim 10^{-3}$) in Fig. \ref{fig:NILC_litebird}. These spectra are computed on maps masked with \texttt{mask1} procedure as described in Sect. \ref{sec:masks} with $f_{\textrm{sky}} =50\,\%$. 
Right panel of Fig. \ref{fig:NILC_litebird} highlights the effective tensor-to-scalar ratio associated to the Galactic contamination (computed according to Sec. \ref{sec:like}). \\
The plots  in  Fig. \ref{fig:NILC_litebird} indicate how NILC is capable to subtract most of the foreground contamination when the simple \texttt{d0s0} Galactic model with constant spectral parameters across the sky is assumed, leading to an upper bound on $r$ fitted on foreground residuals of $r_{\textrm{fgds}}\sim 1.6 \cdot 10^{-4}$ at $68\,\%$ CL. Vice versa, when we simulate foreground emission at the \textit{LiteBIRD} frequency channels with   
\texttt{d1s1} model, the residuals are larger both  at the reionization peak ($\ell\approx 5$) and at the recombination bump ($\ell\approx 80$) than the primordial tensor signal targeted by \textit{LiteBIRD}, leading to a clear bias at the level of $r_{\textrm{fgds}}\sim 3\cdot 10^{-3}$. \\
In order to successfully apply a blind method on \textit{LiteBIRD} data, then, the Galactic contamination, especially on the largest scales, has to be lowered with an alternative strategy. 
We thus apply MC-NILC. \\
We partition the sky into multiple patches driven by the spatial variability of the foreground SEDs in $B$-mode data following the procedure described in Sect. \ref{sec:clusters} and then the variance of the output solution is separately minimized within each patch.
Such patches are identified with a trade off in the size. They should be small enough to assume nearly zero  variability of the foregrounds and in the meantime large enough to avoid a significant bias in the CMB reconstruction. In fact, the NILC variance minimization on different regions can lead to such a bias whose amount is related to the number of modes sampled in each region and the number of employed frequency channels. \\
As detailed in Sect. \ref{sec:clusters}, the sky is partitioned considering the ratio of needlet $B$-mode maps at two different \textit{LiteBIRD} channels and following two different approaches: CEA and RP. If it is not specified, the former is adopted. 
In Sects. \ref{sec:ideal_app} and \ref{sec:real_appr} we will present the results of MC-NILC application when the ratio is estimated employing, respectively, input foregrounds-only needlet coefficients (\emph{ideal} case of Sect. \ref{sec:GNILC&cPILC}) and B-modes templates of Galactic emission obtained from input data (\emph{realistic} case of Sect. \ref{sec:GNILC&cPILC}).
\begin{figure} 
\centering
	\includegraphics[width=0.49\textwidth]{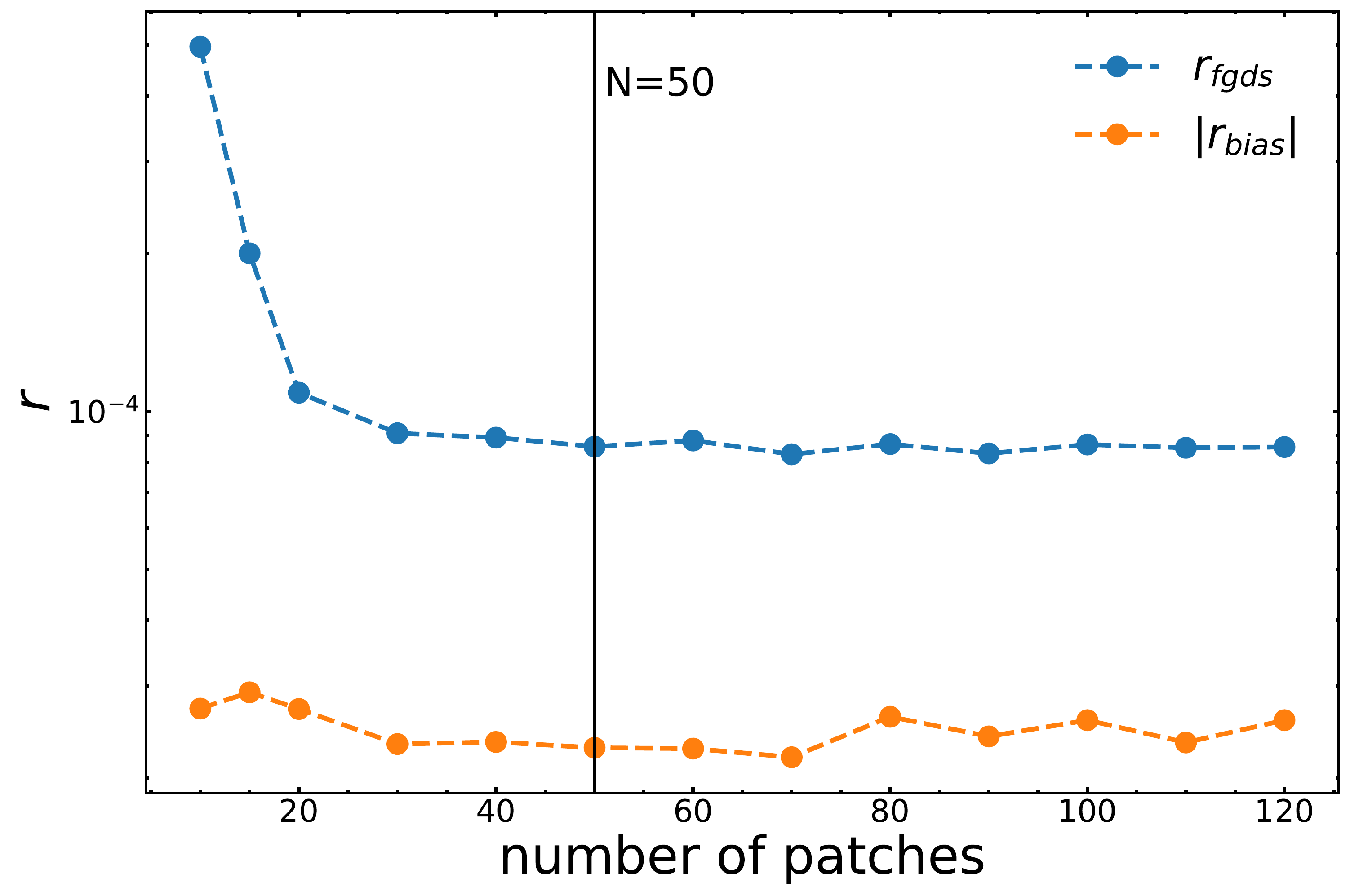}
	% \hspace{0.5 cm}
	%\includegraphics[width=0.48\textwidth]{figures_nersc/BB_res_NILCs_clusters_ratios_20or40new_mexB1.3_b0b15_wl01_LiteBIRD_fwhm1_fsky51_nside64.pdf}
    \caption{Left: trend of the mean effective tensor-to-scalar ratio fitted on the MC-NILC foreground residuals average power spectrum $C_{\ell}^{\textrm{fgds}}$ ($r_{\textrm{fgds}}$, in blue) and on the average bias $C_{\ell}^{\textrm{bias}}$ ($r_{\textrm{bias}}$ in orange) varying the number of clusters $K$ to partition the sky. The PySM \texttt{d1s1} model is assumed for the Galactic emission. The posterior distributions for the effective tensor-to-scalar ratios are obtained with a binning scheme of $\Delta\ell =10$.} 
	\label{fig:fgds_vs_bias}
\end{figure}
\subsection{Ideal approach}
\label{sec:ideal_app}
As a starting point, we use simulated foregrounds-only $B$-mode needlet coefficients to build the ratio (\emph{ideal} case). 
All the results shown in this section are obtained considering the PySM \texttt{d1s1} model for the Galactic emission, unless otherwise specified. \\
The optimal number of clusters $K$ to adopt for the application of MC-NILC is chosen by comparing  the obtained foreground residuals and the bias in the CMB reconstruction in terms of $r$ for different values of $K$. $r_{\textrm{fgds}}$ and $r_{\textrm{bias}}$ are estimated via Eq. \ref{eq:like} from, respectively, $B$-mode power spectra of foreground residuals, $C_{\ell}^{\textrm{fgds}}$, in the recovered MC-NILC CMB map and from the ones of the bias $C_{\ell}^{\textrm{bias}}$. The covariance in Eq. \ref{eq:like}, used to estimate the effective tensor-to-scalar ratios, is sourced by the input CMB lensing signal and the residuals obtained in each case. \\
The trend of average $r_{\textrm{fgds}}$ and $r_{\textrm{bias}}$ is shown in the left panel of Fig. \ref{fig:fgds_vs_bias}. It is possible to observe, as expected, that the amplitude of Galactic contamination decreases as the amount of patches increases up to a convergence which is reached at $K\approx 50$. The amplitude of foreground residuals indeed does not significantly decrease, at least for the largest angular scales, for $K > 50$ . \\
%(see right panel of Fig. \ref{fig:fgds_vs_bias}). \\
On the other hand, $r_{\textrm{bias}}$ is almost insensitive to $K$ and this is thanks to the choice of excluding the pixel and a circular region around it to compute the covariance referred to the pixel itself. If this procedure had not been performed, we would have observed an increase of $r_{\textrm{bias}}$ with $K$. Given the null impact of the number of patches on the CMB reconstruction bias, we adapt the number of clusters only to the trend of $r_{\textrm{fgds}}$ and, thus, we choose $K=50$ for the application of MC-NILC to \textit{LiteBIRD} simulated datasets performed in this work. \\
The comparison of $r_{\textrm{fgds}}$ and $r_{\textrm{bias}}$ permits to easily assess the optimal number of clusters when dealing with simulations. Having proven that the methodology is un-biased for a large range of plausible values of $K$, a similar analysis can be performed even with real data evaluating the trend of the total variance of the output CMB solution. \\
\begin{figure*} 
\centering
	\includegraphics[width=0.48\textwidth]{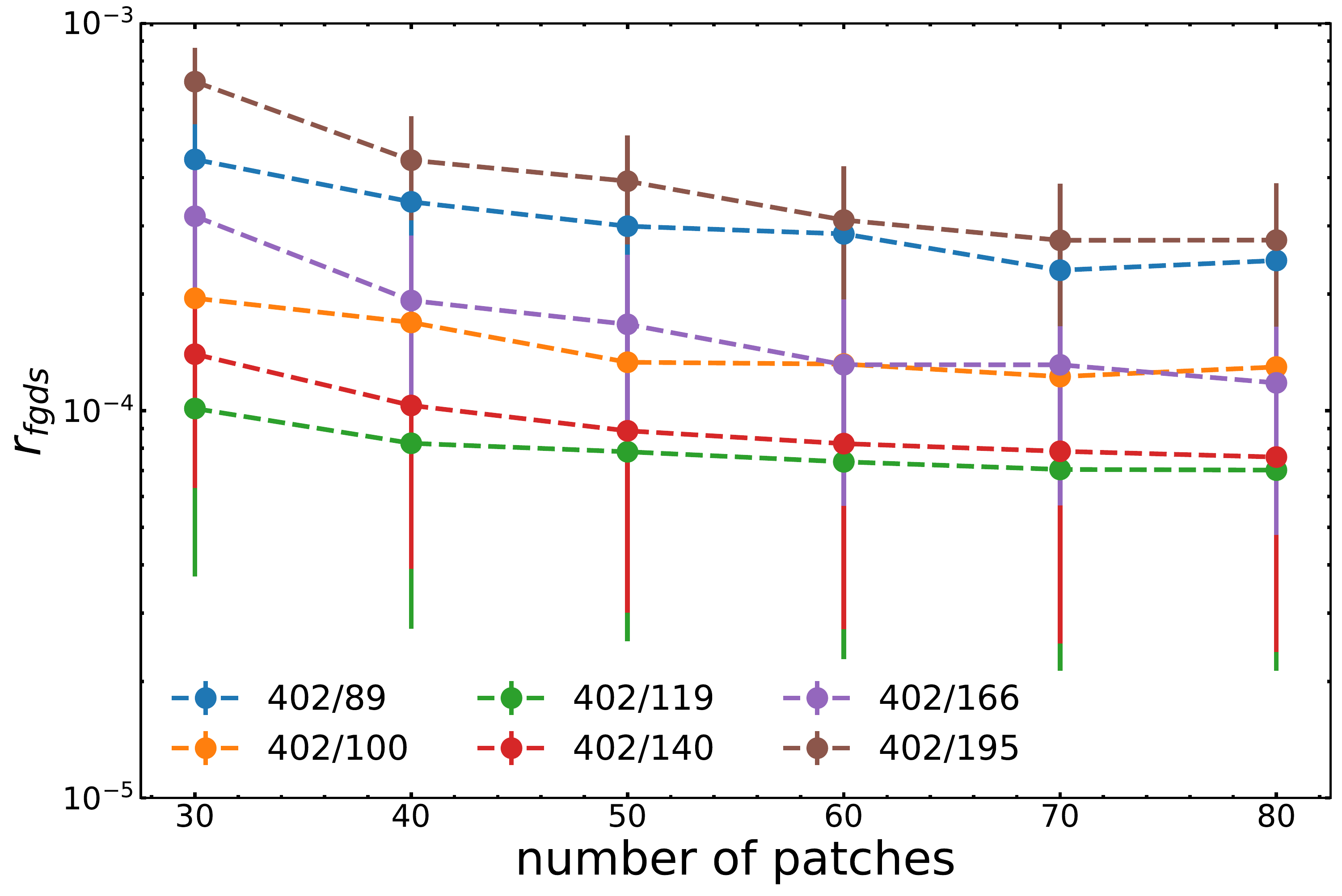}
	\hspace{0.5 cm}
    \includegraphics[width=0.48\textwidth]{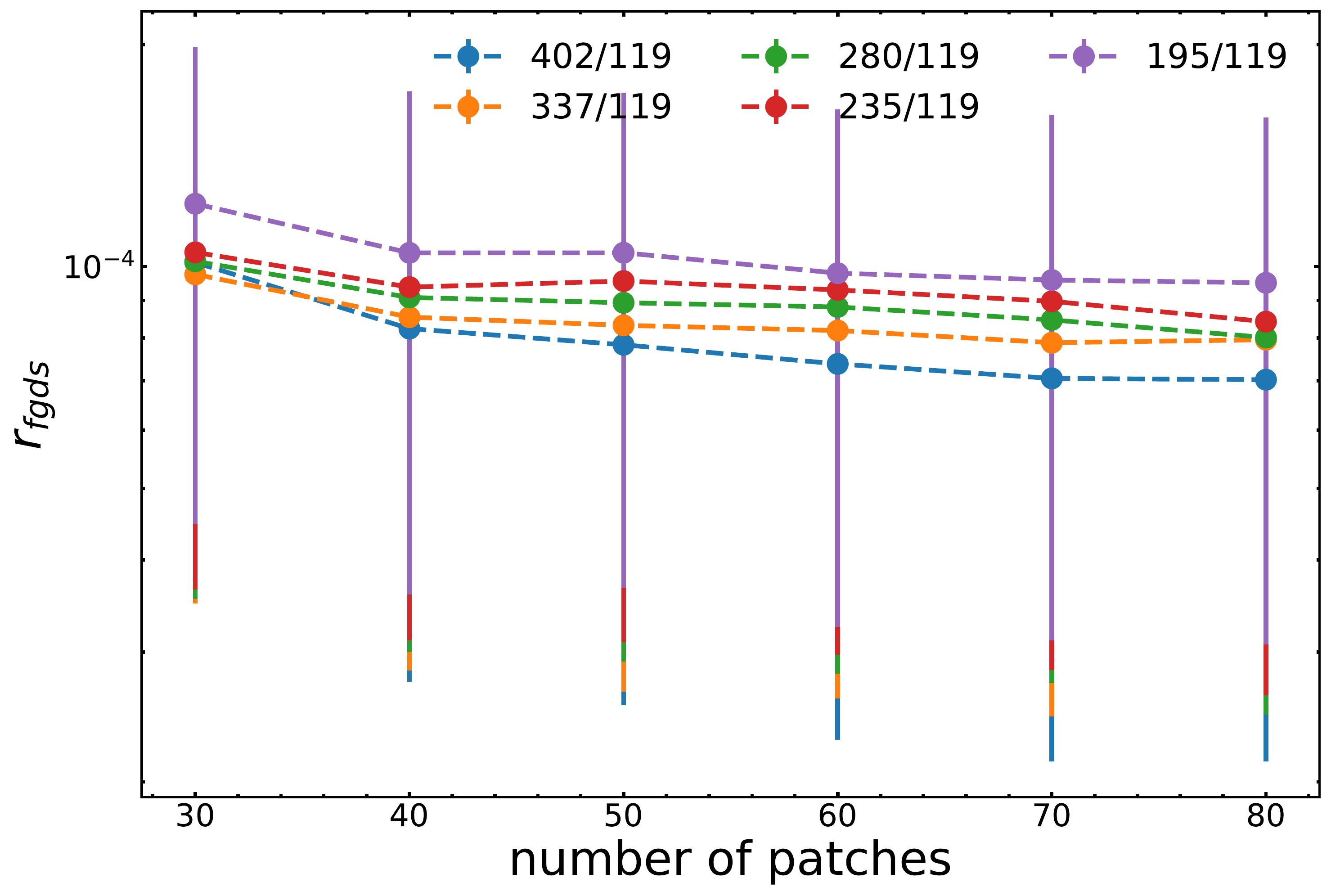}
    \caption{Tensor-to-scalar ratio fitted on the foreground residuals (see Eq. \ref{eq:like}) when CEA-MCNILC is applied on $200$ \textit{LiteBIRD} full simulations modelling the Galaxy with the \texttt{d1s1} sky and the number of patches ranges between $K=30$ and $K=70$. On the left: the map at the numerator of the ratio adopted as tracer of the spectral variability of the Galactic emission is kept fixed to be the needlet foreground $B$-mode coefficients at $402$ GHz for each needlet scale while the one at the denominator varies. On the right: the map at the denominator is kept fixed to be the needlet foreground $B$-mode emission at $119$ GHz while the frequency of the needlet $B$-mode foreground map at the numerator changes. A sky fraction of $f_{\textrm{sky}}=50\,\%$ is assumed. The posterior distributions of the effective tensor-to-scalar ratios are obtained with a binning scheme of $\Delta\ell =10$ for the angular power spectra  %to make the angular power spectrum gaussianly distributed 
    (see Sect. \ref{sec:like}). The error-bars in these plots show the $68\,\%$ CL.}
	\label{fig:num_and_den}
\end{figure*}
\begin{figure*}
\centering
	\includegraphics[width=0.495\textwidth]{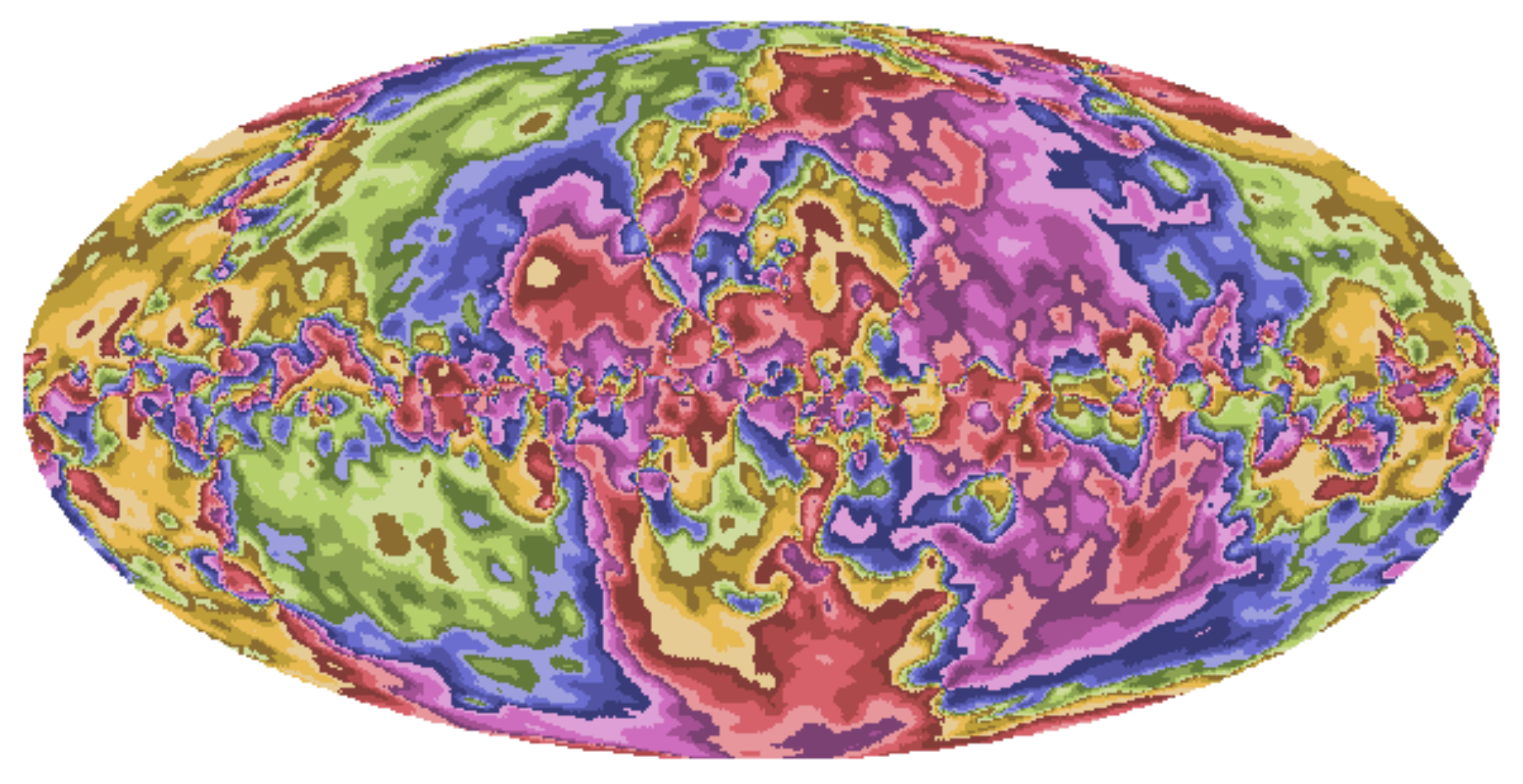}
	\includegraphics[width=0.495\textwidth]{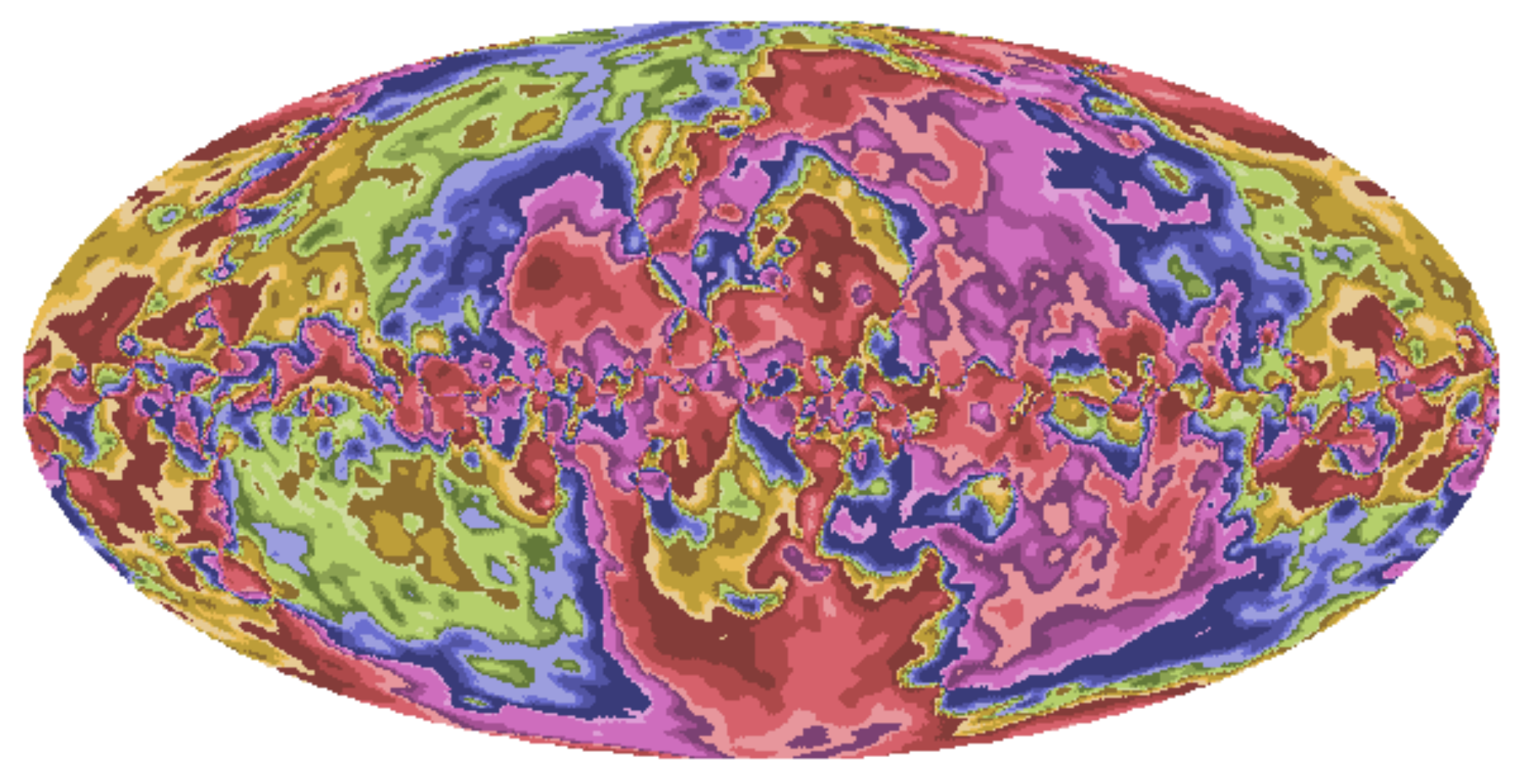}
	\caption{Partition of the sky in 20 patches with equal (CEA, left) and random (RP, right) values of the areas  obtained from the ratio of $B$-mode \texttt{d1s1} foreground emission at $402$ and $119$ GHz, filtered with the first Mexican needlet band shown in Fig. \ref{fig:bands}.}
\label{fig:clusters}
\end{figure*}
\begin{figure*}
	\centering
	\includegraphics[width=0.48\textwidth]{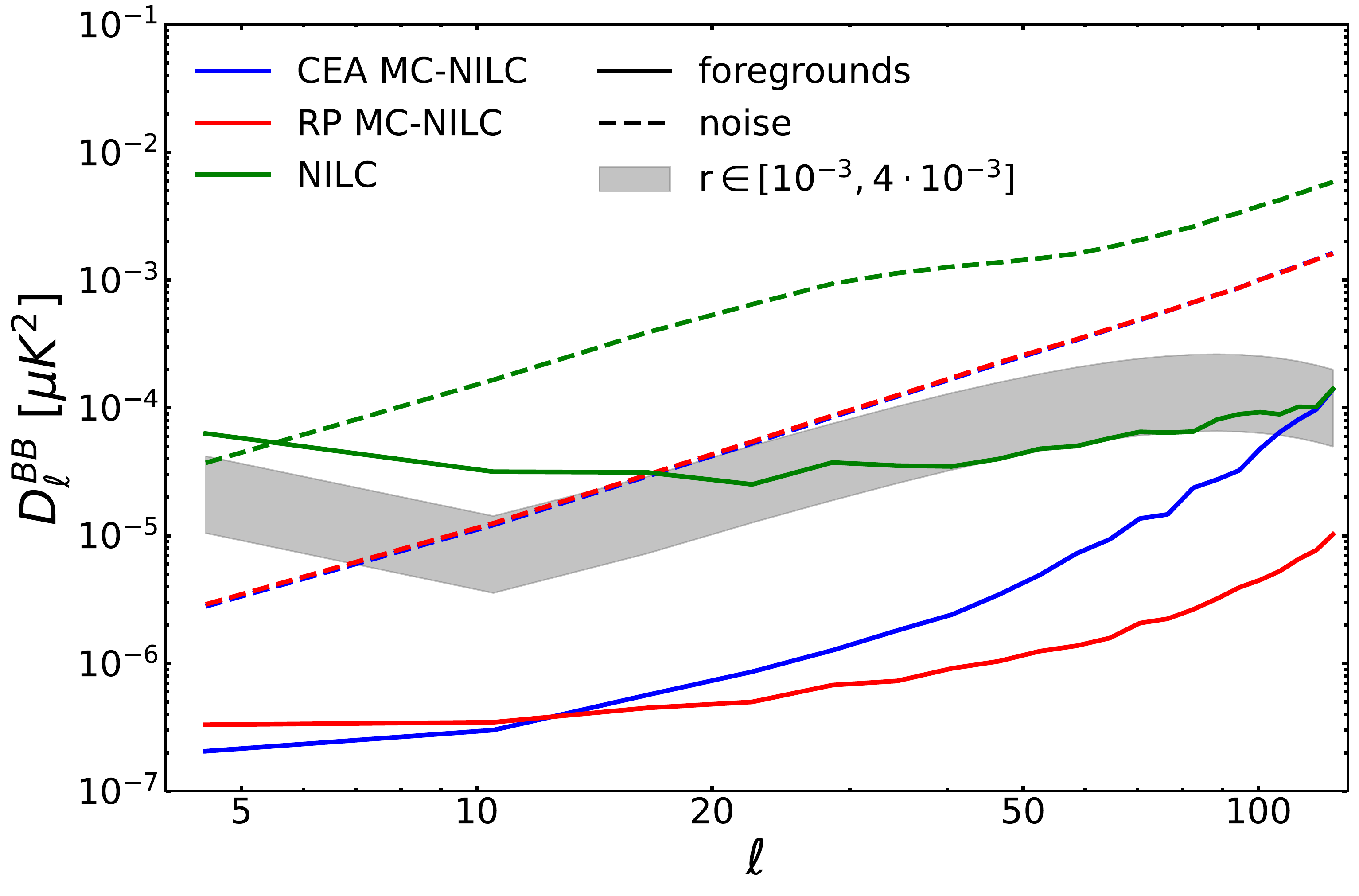}
	\hspace{0.5 cm}
	\includegraphics[width=0.48\textwidth]{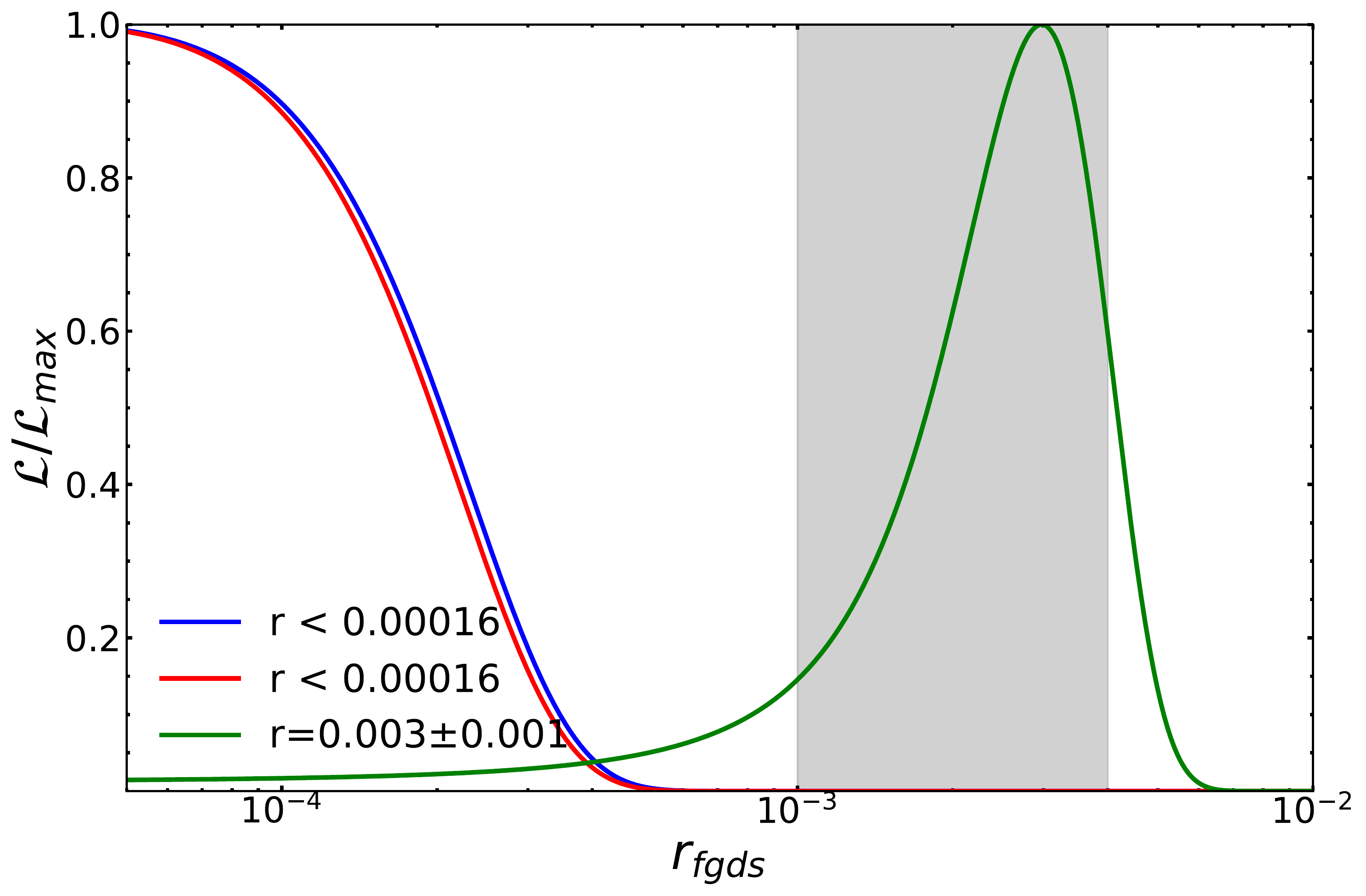} \\
	\hspace{-0.3 cm}
	\includegraphics[width=0.48\textwidth]{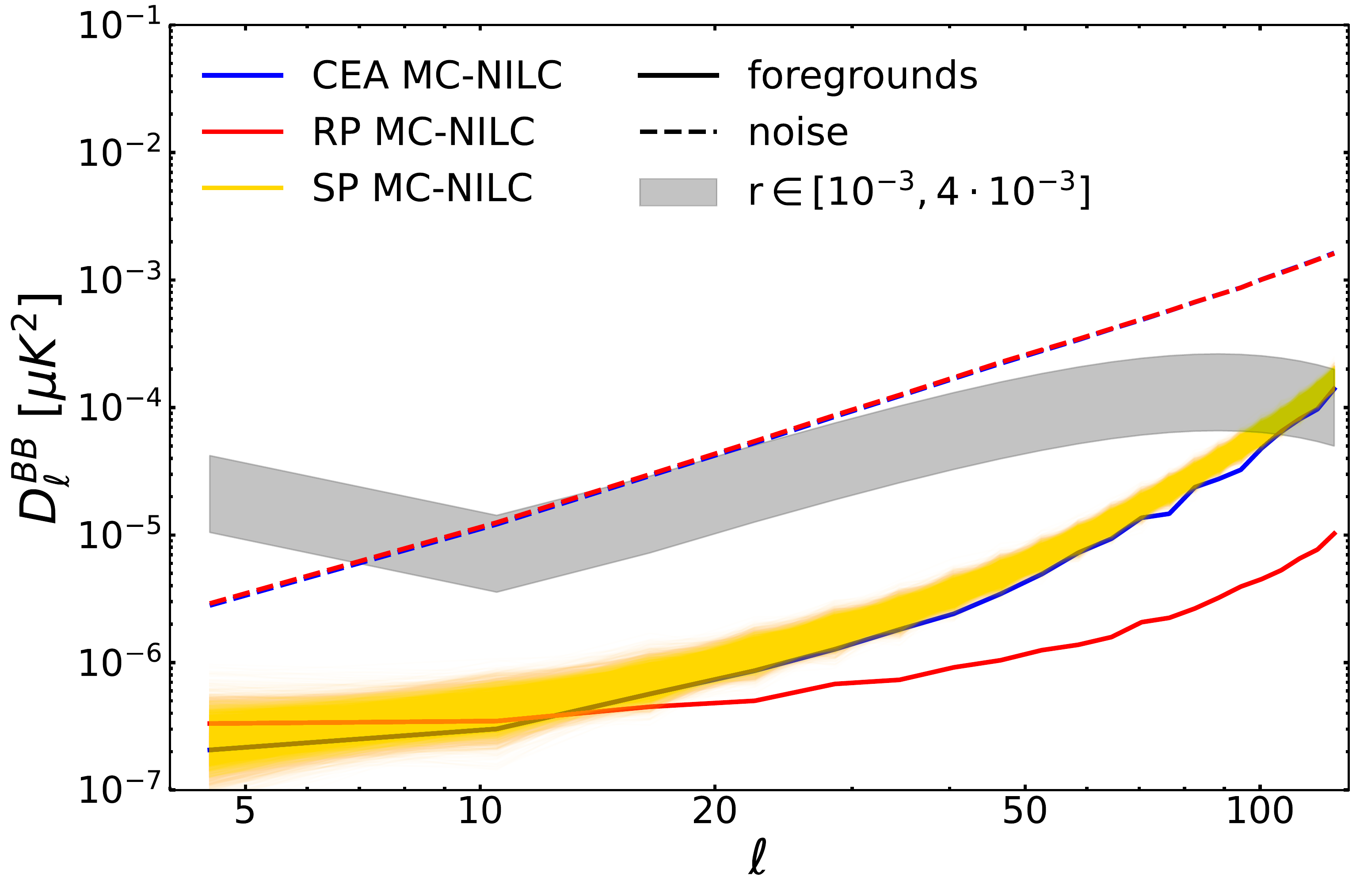} 
	\hspace{0.2 cm}
	\includegraphics[width=0.48\textwidth]{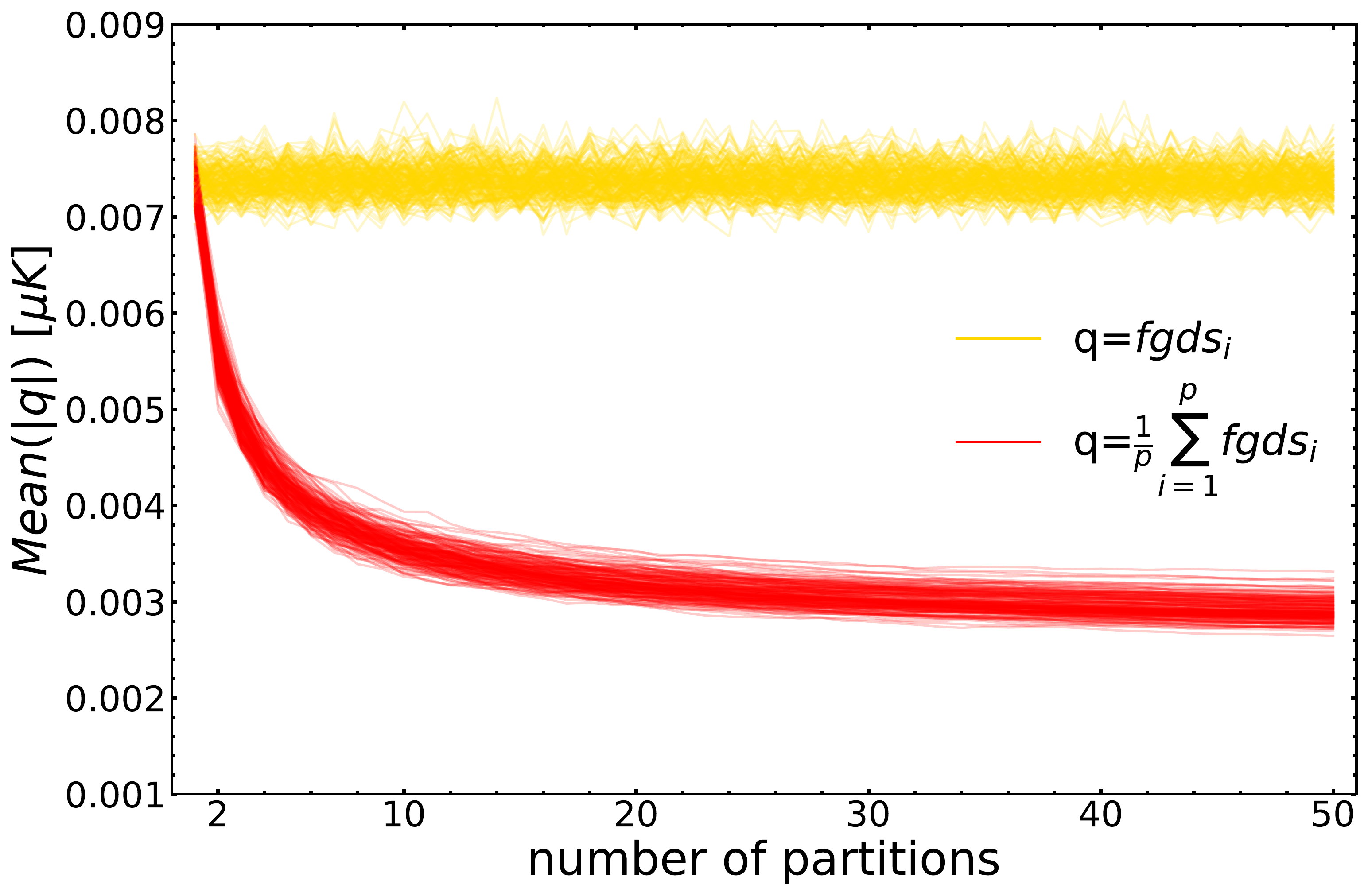} 
\caption{On the top left: average angular power spectra over $200$ CMB solutions of foreground (solid) and noise (dashed) residuals in three different cases: CEA-MCNILC (blue), RP-MCNILC (red), and NILC (green) when the Galaxy is modelled with the \texttt{d1s1} sky. CEA-MCNILC and RP-MCNILC present same amount of noise residuals. The angular power spectra are computed employing masks obtained according to the first strategy (\texttt{mask1}) described in Sect. \ref{sec:masks} with $f_{\textrm{sky}}=50\,\%$. On the top right: the posterior distribution of an effective tensor-to-scalar ratio fitted on the foreground residuals for the three previously mentioned different cases. For the estimation of the posteriors a binning scheme of $\Delta\ell =10$ has been used to make the angular power spectrum gaussianly distributed 
(see Sect. \ref{sec:like}). The reported upper bounds and confidence intervals refer to the $68\%$ CL. On the bottom left: foreground (solid) and noise (dashed) residuals when CEA-MCNILC (blue) and RP-MCNILC (red) are applied; in yellow the angular power spectra of the foreground residuals obtained applying MC-NILC minimization only in a specific random partition of RP-MCNILC without computing any mean among them (SP-MCNILC). In all previous plots the grey areas highlight the range of amplitudes of the primordial tensor signal targeted by \textit{LiteBIRD}: $r\in [0.001,0.004]$. On the bottom right: the trend of the mean of the absolute values of the map $q$ outside the GAL60 mask for all the $200$ different \textit{LiteBIRD} simulated datasets, where $q$ is either the SP-MCNILC foreground residuals map obtained running MCNILC on a single specific random partition $p$ of RP-MCNILC or the mean of foreground residuals up to partition realisation $p$ (red). In the plots where angular power spectra are shown the adopted binning scheme is $\Delta\ell =6$ for visualisation purposes.}
\label{fig:results_nilc_clusters}
\end{figure*}
We perform several tests to assess which pair of \textit{LiteBIRD} frequencies for the ratio leads to the best subtraction of the polarized $B$-mode foregrounds. In the left panel of Fig. \ref{fig:num_and_den},
we compare the trend of the tensor-to-scalar ratio fitted on the foreground residuals average angular power spectrum (see Eq. \ref{eq:like}) when the frequency channel at the numerator in the ratio for the clustering is kept constant to be the \texttt{d1s1} $B$-mode foreground needlet coefficients at $402$ GHz and we vary the frequency of the needlet $B$-mode foreground map at the denominator. In this analysis the CEA partition is adopted and the considered number of clusters ranges between $K=30$ and $K=70$. It is possible to observe that the most effective foreground removal is obtained employing the channel at $119$ GHz. Analogously, fixing this frequency at the denominator and varying the frequency of the simulated foreground $B$-mode map at the numerator permits to state that the channels at the highest frequencies ($337$ and $402$ GHz) are the best ones for the removal of Galactic emission for all the considered range of number of patches (see right panel of Fig. \ref{fig:num_and_den}). \\
\begin{figure}
	\centering
    \includegraphics[width=0.48\textwidth]{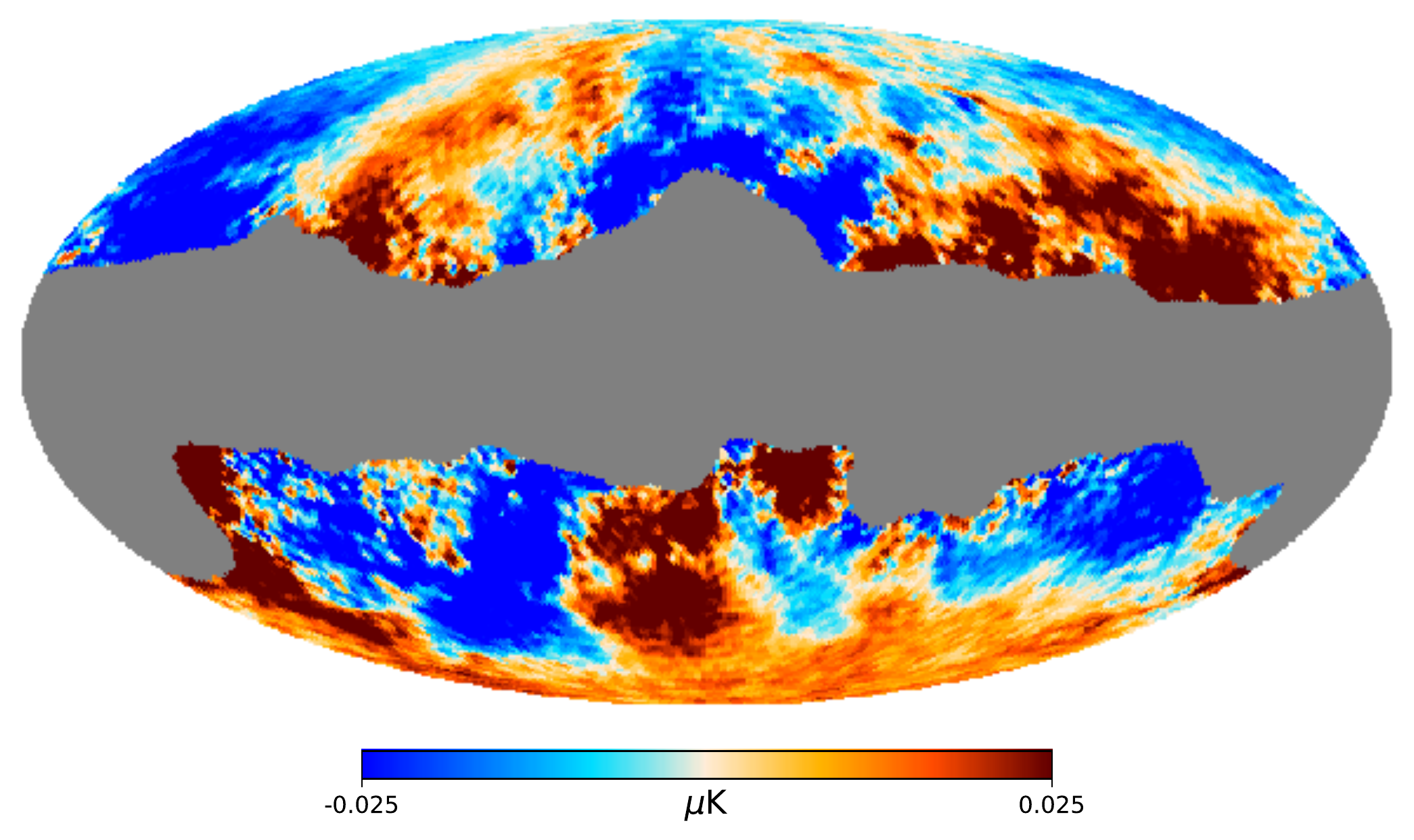} 
    \includegraphics[width=0.48\textwidth]{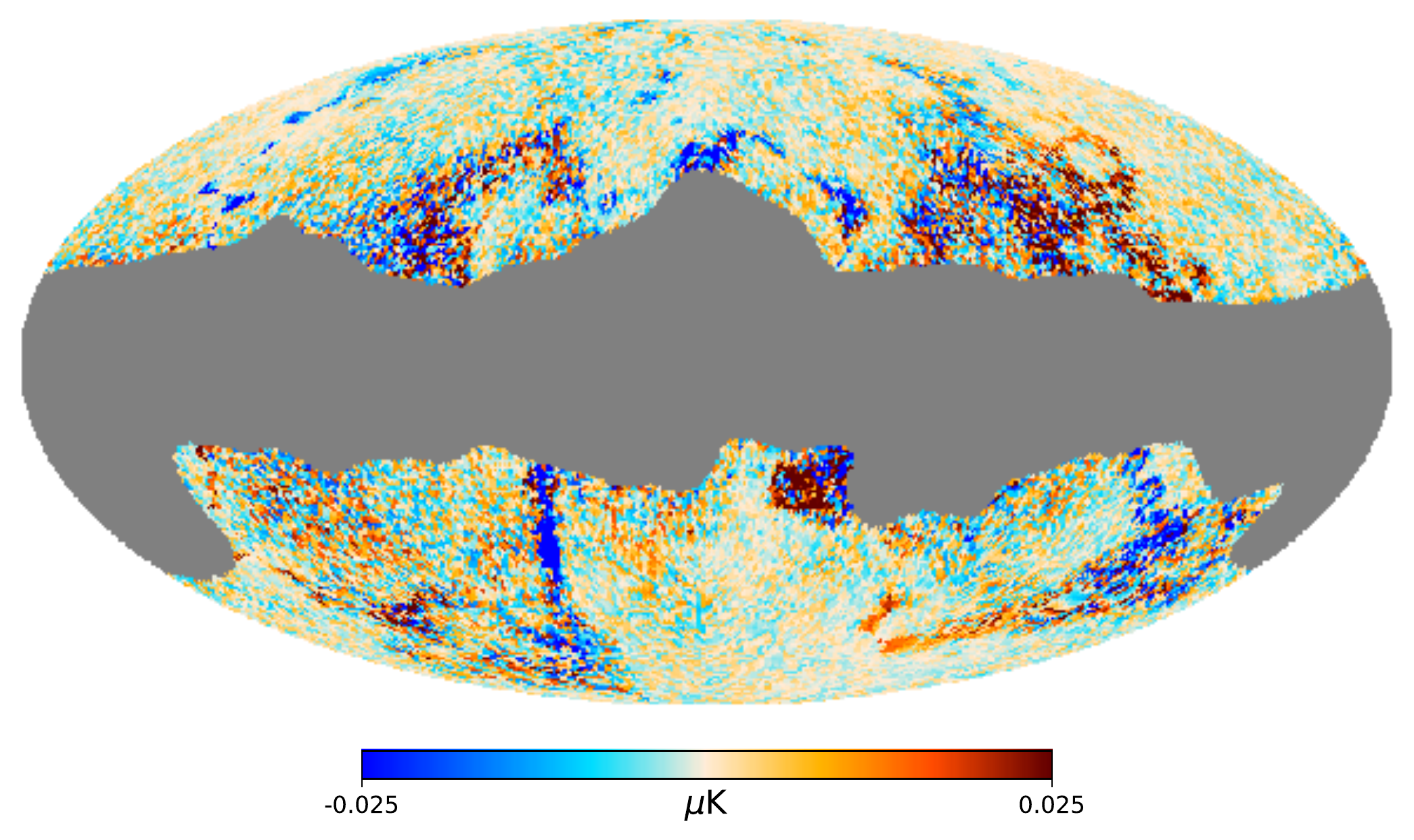} 
    \includegraphics[width=0.48\textwidth]{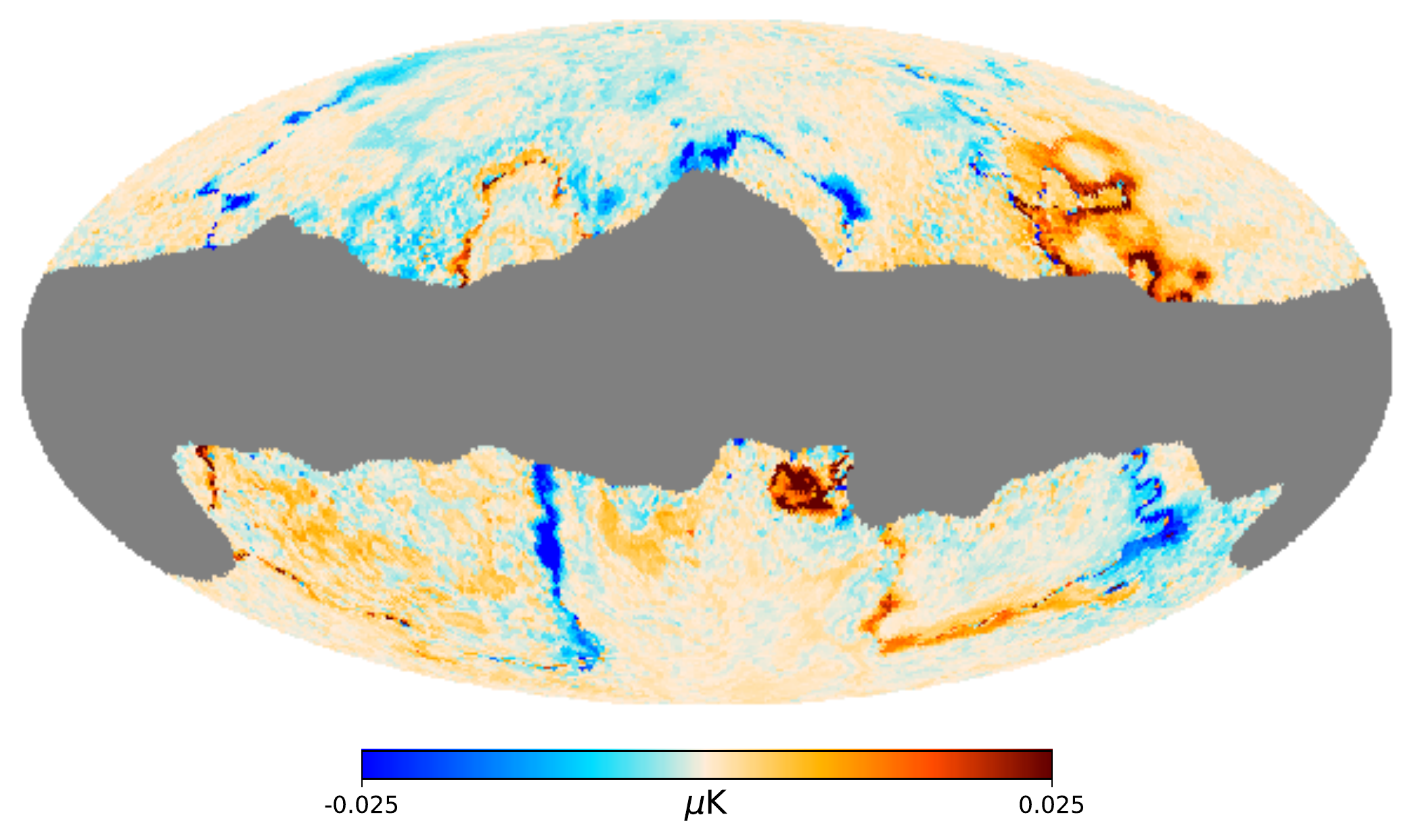} 
	\caption{Foreground residuals maps obtained by applying NILC (top), CEA-MCNILC (middle), and RP-MCNILC (bottom) to a LiteBIRD simulation. For MC-NILC, the sky partitions are obtained in an ideal case by thresholding the ratio of foregrounds-only maps. The Galactic emission is simulated assuming the \texttt{d1s1} model. The Galactic plane is masked with the \textit{Planck} \textit{GAL60} mask  with $f_{\textrm{sky}}=60\,\%$. }
	\label{fig:fgds_res_maps}
\end{figure}
Thus, the optimal choice of frequencies to build the ratio and partition the sky is a high-frequency map ($B_{\textrm{fgds}}^{\textrm{hf}}$) and a CMB channel ($B_{\textrm{fgds}}^{119}$): this configuration allows to trace simultaneously the spatial variations of polarized thermal dust and synchrotron emission in $B$ modes. \\ 
In this \emph{ideal} case, then, the standard MC-NILC configuration consists in applying a NILC minimization in 50 different patches obtained for each needlet scale $j$ from the ratio of needlet filtered foreground maps $B_{\textrm{fgds},j}^{402}/B_{\textrm{fgds},j}^{119}$. 
The partitions of the sky obtained from the ratio of the $B$-mode foreground maps at $402$ and $119$ GHz filtered with the first needlet band with the CEA method and with a single realisation of the RP approach are shown in Fig. \ref{fig:clusters}. In this figure, just for visualisation purposes, a number of patches $K=20$ is adopted. These maps represent, in first approximation, an estimate of the spatial distribution of the spectral properties of Galactic foregrounds in $B$ modes on large and intermediate angular scales. \\
We have studied in detail both the CEA-MCNILC and RP-MCNILC approaches. The results of their application on \textit{LiteBIRD} simulated dataset are shown in Fig. \ref{fig:results_nilc_clusters}. 
We can observe that both CEA-MCNILC and RP-MCNILC lead to lower foreground residuals than NILC at all the angular scales of interest. The improvement introduced by the blind subtraction in clusters is especially evident on the largest scales, where the foreground residuals power is more than two orders of magnitude lower than NILC. Both the reionization and recombination bumps for $r\sim 10^{-3}$ would be observable with both MC-NILC approaches. 
In addition to the reduction of the foreground contamination, the MC-NILC solutions are characterised by even lower noise residuals than NILC. This is expected, because the minimization within each patch, where foregrounds have similar properties, is closer to the one performed on \texttt{d0s0} simulated data where foreground spectral properties do not vary at all across the sky. In that case, indeed, the NILC foreground and noise contamination, as shown in Fig. \ref{fig:NILC_litebird}, is much lower than the one obtained with the \texttt{d1s1} Galactic model. 
Additionally, applying the MC-NILC algorithm on different random partitions of the sky and averaging the obtained CMB solutions (RP-MCNILC) permits to lower the foreground residuals at all scales, except for the largest ones ($\ell \lesssim 15$), without any noise penalty, if compared to CEA-MCNILC.  \\
Looking at the posterior distributions of $r_{\textrm{fgds}}$, the application of the CEA- or RP-MCNILC on \texttt{d1s1} \textit{LiteBIRD} simulations with this ideal estimation of the ratio leads to an upper bound at $68\,\%$ CL of $r_{\textrm{fgds}} \approx 1.6\cdot 10^{-4}$, which is equal to the one obtained with the application of NILC on \textit{LiteBIRD} simulations assuming the \texttt{d0s0} sky model. This value is lower with respect to the NILC's one by more than one order of magnitude. \\
The improvement in the removal of Galactic contamination obtained with RP-MCNILC with respect to CEA-MCNILC can be explained by looking at bottom panels of Fig. \ref{fig:results_nilc_clusters}. On the left, the yellow solid lines display the angular power spectra of foreground residuals maps obtained with \emph{Single-Partition MCNILC} (SP-MCNILC) which consists in applying MC-NILC minimization and CMB estimation employing only one specific random partition of RP-MCNILC. These SP-MCNILC residuals are comparable to the average one of CEA-MCNILC. Therefore, the reduction of the $B$-mode foreground contamination observed in most of the angular scales in RP-MCNILC is given by averaging all the MC-NILC CMB solutions obtained with the different partitions. In each pixel, indeed, $B$-mode foreground residuals can have negative or positive values, depending on the used partition. Therefore the power of the mean among all the MC-NILC solutions is lower than the one of the single realisation because foreground residuals of different solutions cancel out. This is further confirmed looking at the bottom right panel of Fig. \ref{fig:results_nilc_clusters}, where the average over the patch observed outside the GAL60 mask of the absolute values of the foreground residuals obtained from a single partition $p$ or from their mean considering realisations up to partition $p$ are compared for $200$ different simulations. The former has a constant trend, remarking that the amount of foreground residuals is comparable for each single considered partition, while the latter decreases when more foreground residuals maps are averaged. This highlights how computing the mean contributes to reduce the intensity of $B$-mode Galactic contamination in each pixel and how, at most of the angular scales, the foreground residuals of CEA-MCNILC are mostly sourced by a sort of 'partition' noise. This result can be intuitively interpreted with the fact that in RP-MCNILC there is a larger probability that pixels with closer values of the ratio would find themselves in the same cluster more often. \\
The improvement in subtracting Galactic emission led by the application of MC-NILC on LiteBIRD simulated dataset can also be visualised by looking at the foreground residuals map in Fig. \ref{fig:fgds_res_maps}. We report the solutions for a single simulation obtained with NILC, CEA-MCNILC, and RP-MCNILC. Even at the map level, we can observe that CEA-MCNILC residuals present brighter small-scale features with respect to RP-MCNILC. Most of these high-amplitude small-scale residuals are mainly concentrated around the border regions where foreground $B$ modes switch from positive to negative values and vice versa. The morphology of these regions seems correlated with that of the patches (see Fig. \ref{fig:clusters}). Therefore, these features may be partly associated with border effects due to the employment of a single partition in the CEA approach. \\ 
As anticipated in Sect. \ref{Sec:NILC} an important aspect to take into account when an Internal Linear Combination (ILC) method is applied is the CMB reconstruction bias. Although, in principle, this bias can be worsened by performing variance minimization in separate patches of the sky, we have verified that our pipeline is actually unbiased thanks to the procedure (introduced in Sect. \ref{sec:clusters}) of computing for each pixel the covariance matrix taking into account all the other pixels in the same patch but excluding the pixel itself and a circular region around it (see Coulton et al. in prep.). \\
\begin{figure*}
	\centering
\includegraphics[width=0.48\textwidth]{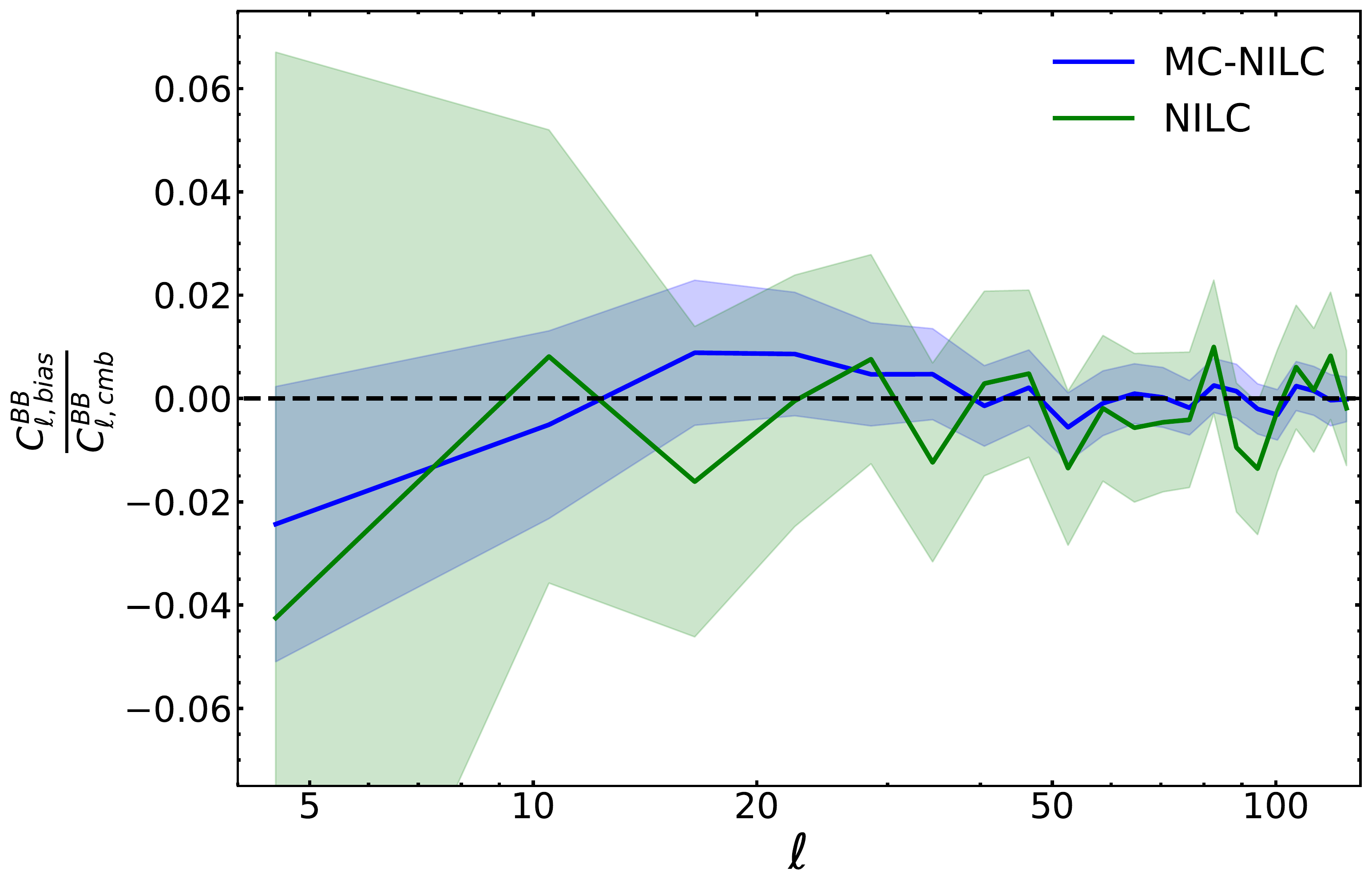} 
\hspace{0.5 cm}
\includegraphics[width=0.48\textwidth]{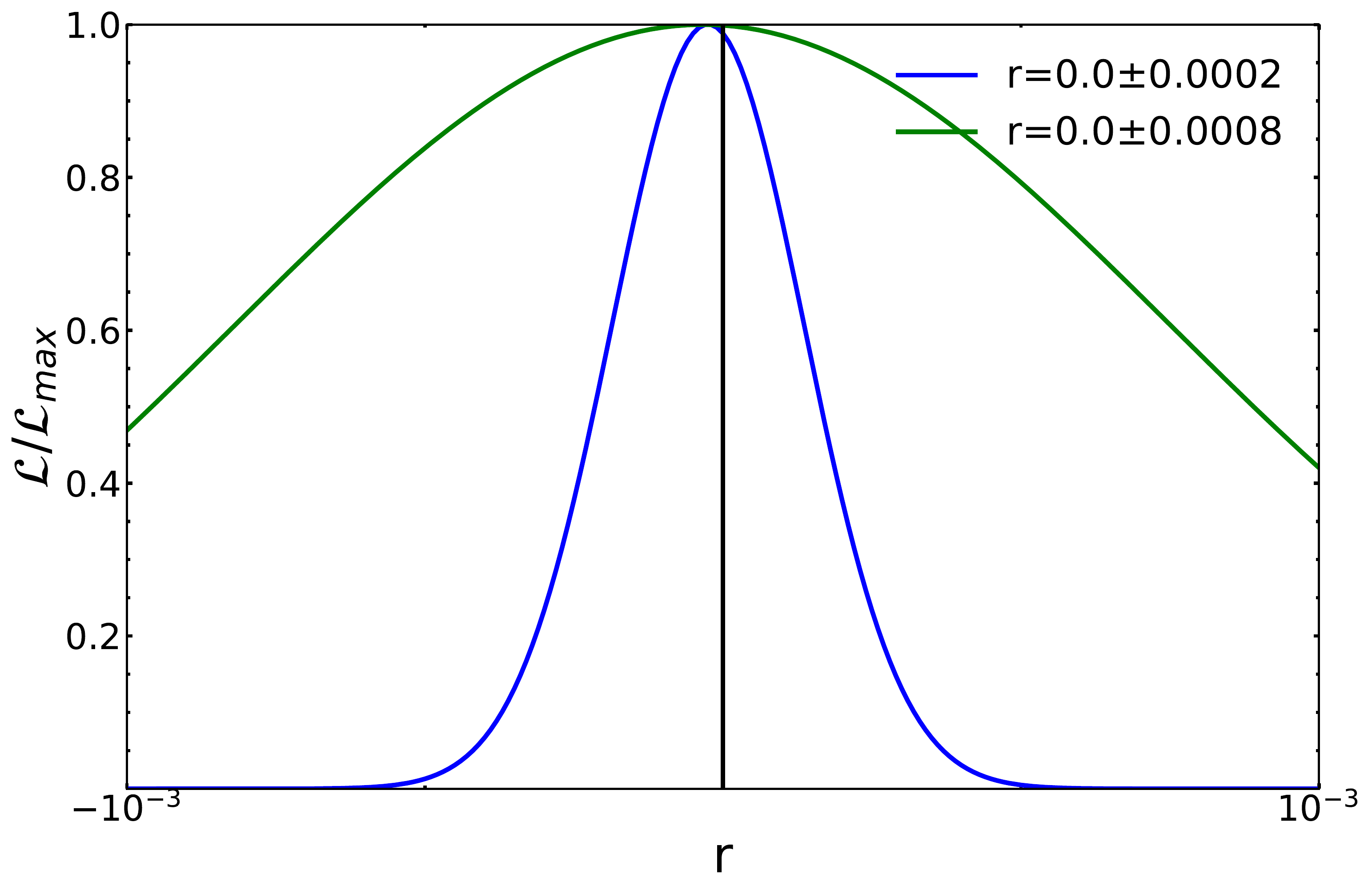}
\caption{On the left: relative bias (solid lines), which is $C_{\ell}^{\textrm{bias}} = C_{\ell}^{\textrm{out}} - C_{\ell}^{\textrm{fgds}} - C_{\ell}^{\textrm{noi}} - C_{\ell}^{\textrm{cmb}}$ divided for the input CMB power spectrum $C_{\ell}^{\textrm{cmb}}$ when MC-NILC (blue) and NILC (green) are applied on \textit{LiteBIRD} data and the Galaxy is modelled with the \texttt{d1s1} sky. Only for MC-NILC, the covariance computation in each pixel neglects the pixel itself and a circular region around it.  The used angular power spectra represent the average among $N_{\textrm{sims}}=500$ different simulations. The angular power spectra are computed employing masks obtained according to the first strategy (\texttt{mask1}) described in Sect. \ref{sec:masks} with $f_{\textrm{sky}}=50\,\%$. The corresponding shaded area represent the uncertainty on the mean at $2\,\sigma$, estimated from the dispersion of the angular power spectra of the simulations divided by $\sqrt{N_{\textrm{sims}}}$. On the right: posterior distribution of an effective tensor-to-scalar ratio fitted on $C_{\ell}^{\textrm{bias}} = C_{\ell}^{\textrm{out}} - C_{\ell}^{\textrm{fgds}} - C_{\ell}^{\textrm{noi}} - C_{\ell}^{\textrm{cmb}}$ according to Eq. \ref{eq:like} where the covariance matrix is estimated considering the CMB lensing signal and the residuals obtained in each method. For the computation of the posteriors a binning scheme of $\Delta\ell =10$ has been used for the angular power spectra (see Sect. \ref{sec:like}). %to make the angular power spectra gaussianly distributed (see \ref{sec:like}). 
The reported confidence intervals refer to the $68\%$ CL. 
In this case we consider even negative values of the effective tensor-to-scalar ratio because $C_{\ell}^{\textrm{bias}}$ can, in principle, present a trend of negative values.}
\label{fig:nilc_clusters_bias}
\end{figure*}
\begin{figure*}
	\centering
	\includegraphics[width=0.48\textwidth]{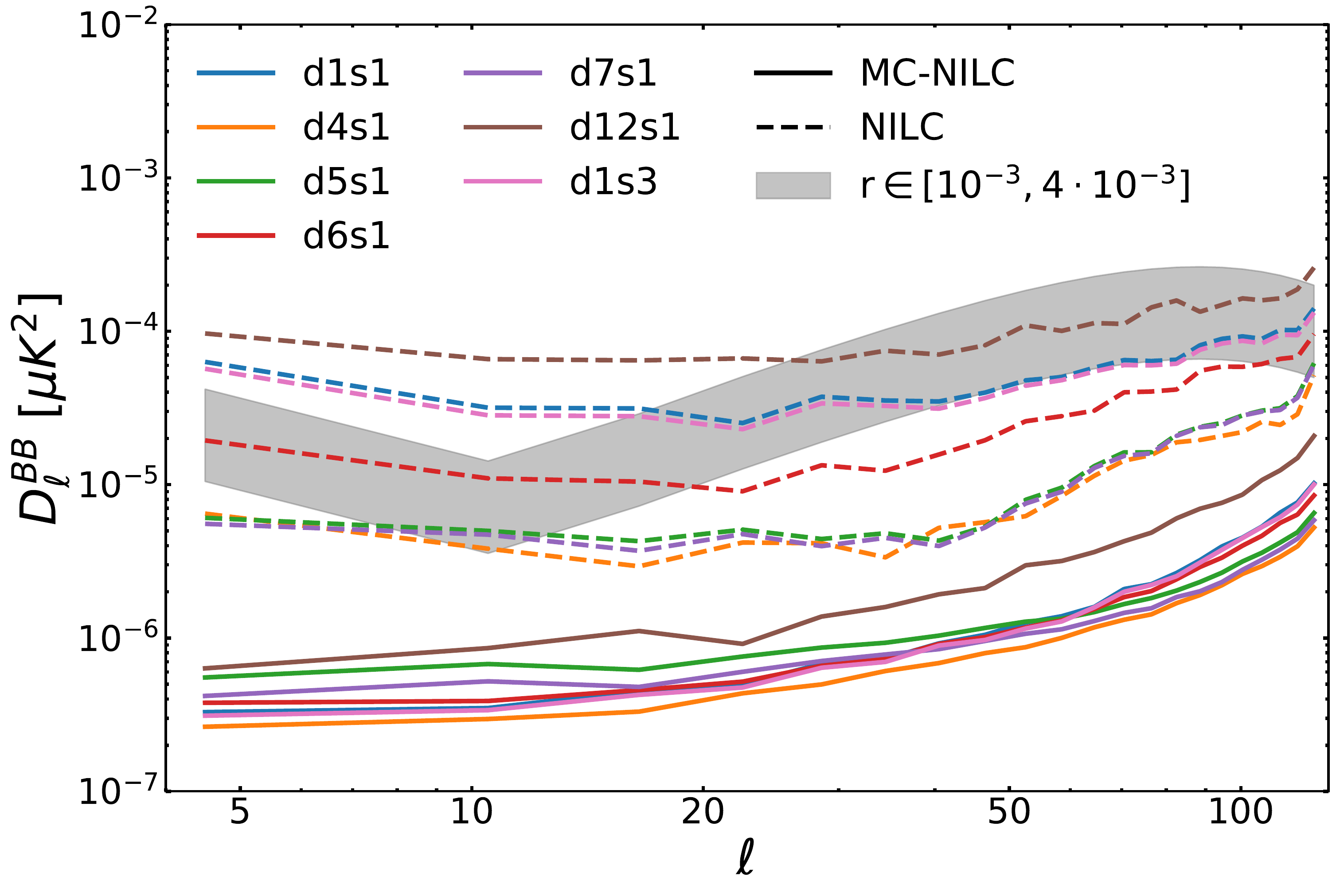}
	\hspace{0.5 cm}
	\includegraphics[width=0.48\textwidth]{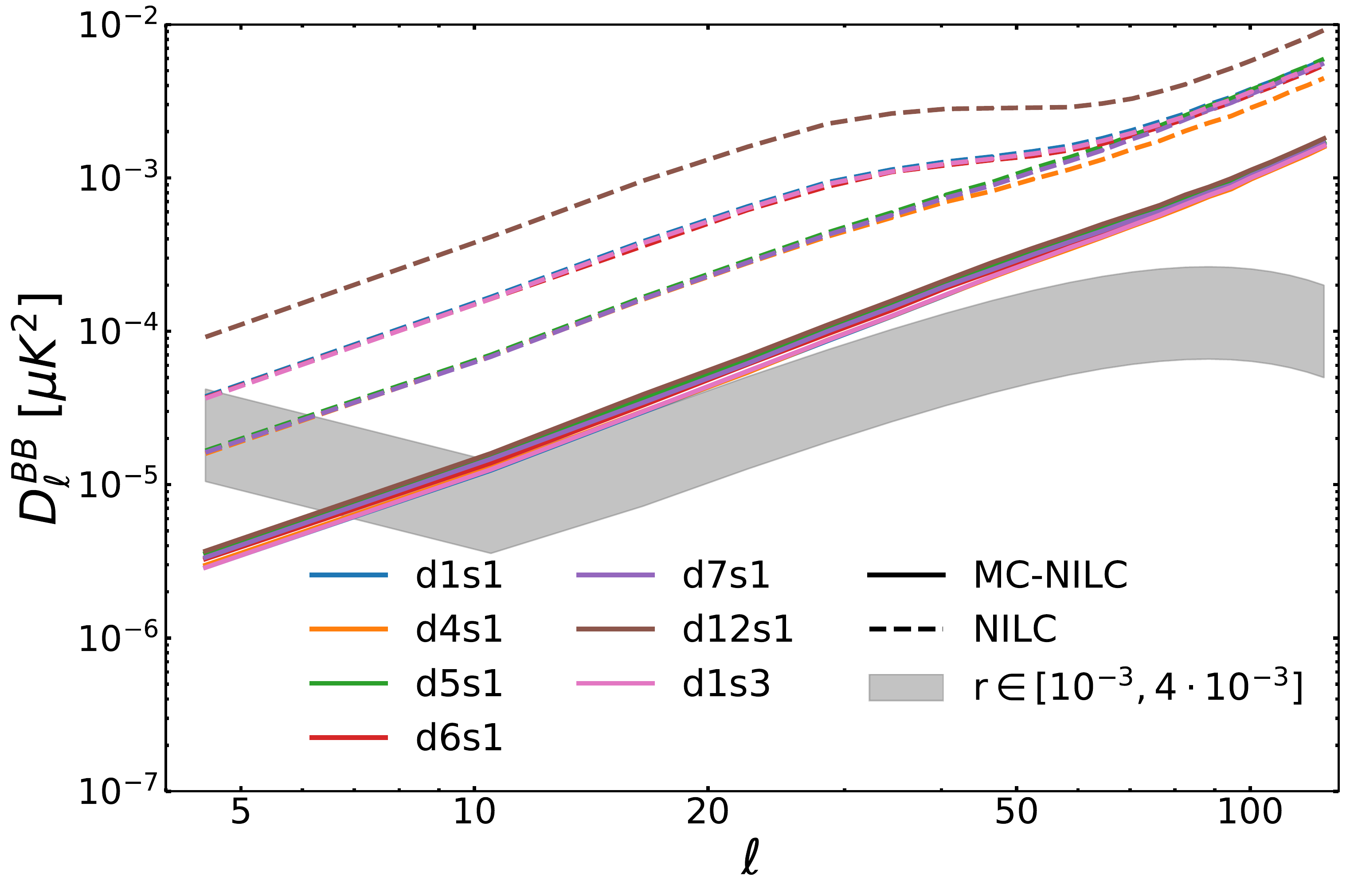}
\caption{Average angular power spectra of foreground (left) and noise residuals (right) when RP-MCNILC (solid lines) or NILC (dashed lines) are applied on $200$ \textit{LiteBIRD} simulated datasets where different Galactic models are assumed. For all cases, the same number of clusters, $K=50$, is adopted. The grey areas highlight the range of amplitudes of the primordial tensor signal targeted by \textit{LiteBIRD}: $r\in [0.001,0.004]$. The angular power spectra are computed employing masks obtained according to the first strategy (\texttt{mask1}) described in \ref{sec:masks} with $f_{\textrm{sky}}=50\,\%$. The adopted binning scheme is $\Delta\ell =6$.}
\label{fig:results_nilc_clusters_diffmodels}
\end{figure*}
In Fig. \ref{fig:nilc_clusters_bias}, we report the relative difference between $\hat{C}_{\ell}^{\textrm{out}}$ of Eq. \ref{eq:bias} and average angular power spectrum of input CMB $C_{\ell}^{\textrm{cmb}}$, which defines the amount of CMB reconstruction bias given by the application of MC-NILC and standard NILC on \textit{LiteBIRD} simulated datasets. For NILC, the covariance estimation in the pixel includes the pixel itself because the problem of the reconstruction bias is tackled by choosing a proper domain $\mathcal{D}$. However, the technique of Coulton et al. in prep. can be easily applied to NILC pipeline further alleviating the bias problem even in this case. We find for both NILC and MC-NILC a bias compatible with zero given the uncertainty on the reconstructed average $C_{\ell}^{\textrm{bias}}$, which is estimated dividing the dispersion of the bias angular power spectra of the different CMB reconstructions by the square root of the number of simulations: $\sigma(C_{\ell}^{\textrm{bias}})/ \sqrt{N_{\textrm{sims}}}$. In this case, $N_{\textrm{sims}}=500$. We can conclude that the input CMB $B$-mode signal is perfectly reconstructed in the MC-NILC and NILC solutions. \\
This result is further confirmed by looking at the posterior distributions of an effective tensor-to-scalar ratio fitted on $C_{\ell}^{\textrm{bias}}$ in the right panel of Fig. \ref{fig:nilc_clusters_bias}. The null value of $r$ is well within $1\,\sigma$ in both cases. In this analysis, we consider even a negative range for $r$, given that $C_{\ell}^{\textrm{bias}}$ can assume negative values. \\
A natural question that might arise is how much the MC-NILC performance depends on the considered foreground model. To answer this question and assess the robustness of the chosen MC-NILC configuration, we apply the RP-MCNILC pipeline to simulated \textit{LiteBIRD} datasets varying the model of the dust and synchrotron emission. We have considered the %\texttt{d2}, 
\texttt{d4}, \texttt{d5}, \texttt{d6}, \texttt{d7}, \texttt{d12}, and \texttt{s3} alternative models presented in Sect. \ref{Sec:sims} combined, respectively, with \texttt{s1} and \texttt{d1} to trace separately the effects of changing the dust or synchrotron models. \\
For all of them, a ratio among the needlet coefficients of $B$-mode foreground emission of each model at $402$ and $119$ GHz and the same number of clusters, $K=50$, is adopted. We apply the clustering on each ratio and then the variance minimization. Figure \ref{fig:results_nilc_clusters_diffmodels} highlights that both MC-NILC foreground and noise residuals are lower with respect to the obtained ones with the application of standard NILC across all the multipoles range of interest and for all models. 
Furthermore, the amount of foreground and noise residuals is comparable for all cases. These results make the choice of $402$ and $119$ as pair of  frequencies in the $B$-mode needlet foregrounds ratio rather robust and independent on what the actual Galactic physical model will be. 
\subsection{Realistic approach}
\label{sec:real_appr}
For all the results previously shown, we assume to perfectly know the $B$-mode emission of the foregrounds in needlet space at two different frequencies. Being this an idealistic case, we evaluate the  performances of MC-NILC approach in a fully realistic framework. 
In order to do that, we should reconstruct clean templates of $B$-mode emission of dust and synchrotron at two different channels, one at high frequency and one at $\sim 119$ GHz. In this realistic case the adopted high frequency is $337$ GHz, whose template turns out to be less contaminated by CMB and instrumental noise with respect to the one at $402$ GHz. \\
As described in Sect. \ref{sec:MCGNILC}, in this realistic application of MC-NILC the ratio is built employing two templates of Galactic emission at $337$ and $119$ GHz obtained from a MC-GNILC run on \textit{LiteBIRD} $B$-mode dataset. These maps include all the independent Galactic modes whose power is larger than the one of CMB and instrumental noise but still include some residual contamination of these components. The MC-GNILC procedure is completely model-independent thus not requiring any prior knowledge about the Galactic emission in $B$ modes. \\
\begin{figure*}
	\centering
	\includegraphics[width=0.48\textwidth]{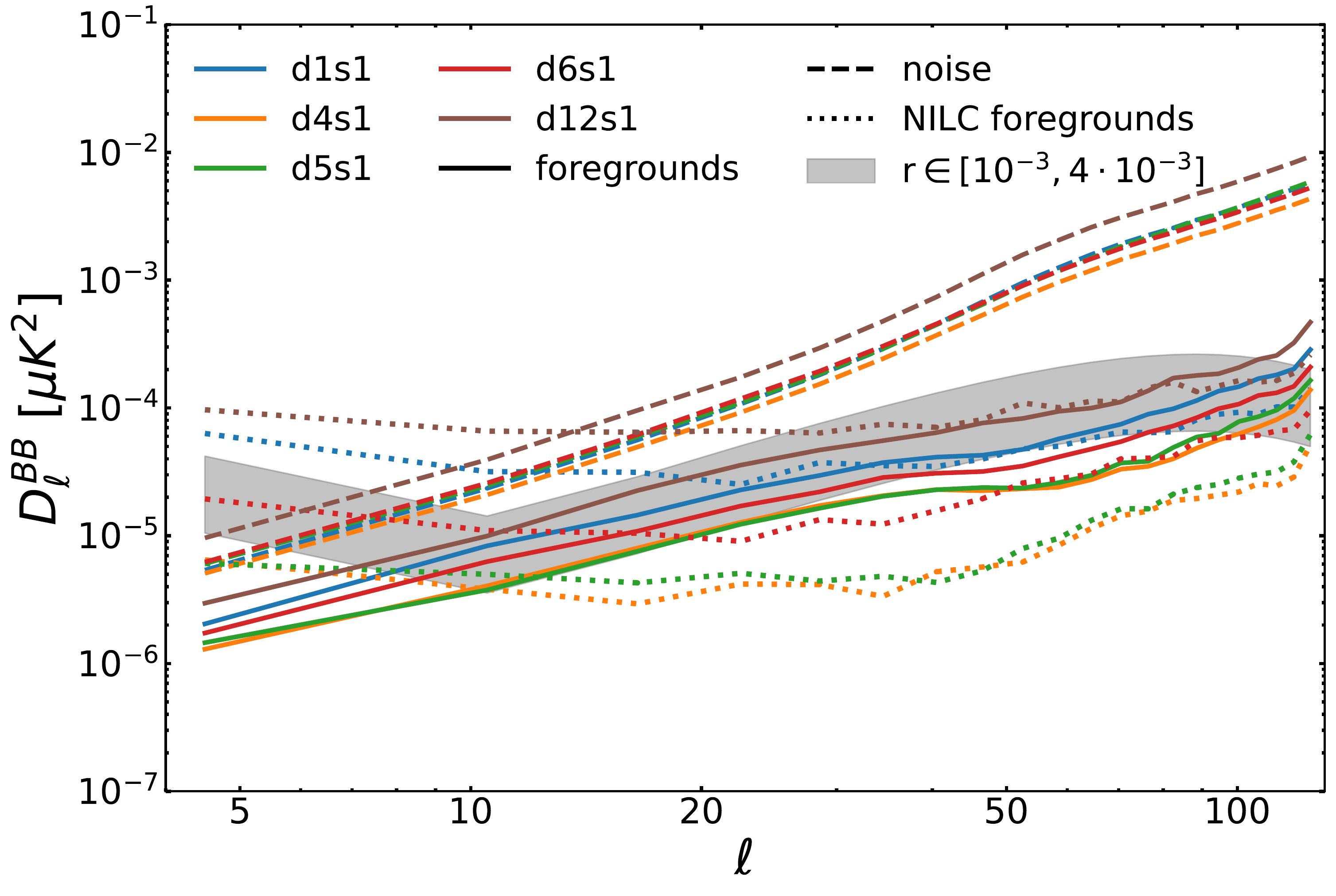}
	\hspace{0.5 cm}
	\includegraphics[width=0.48\textwidth]{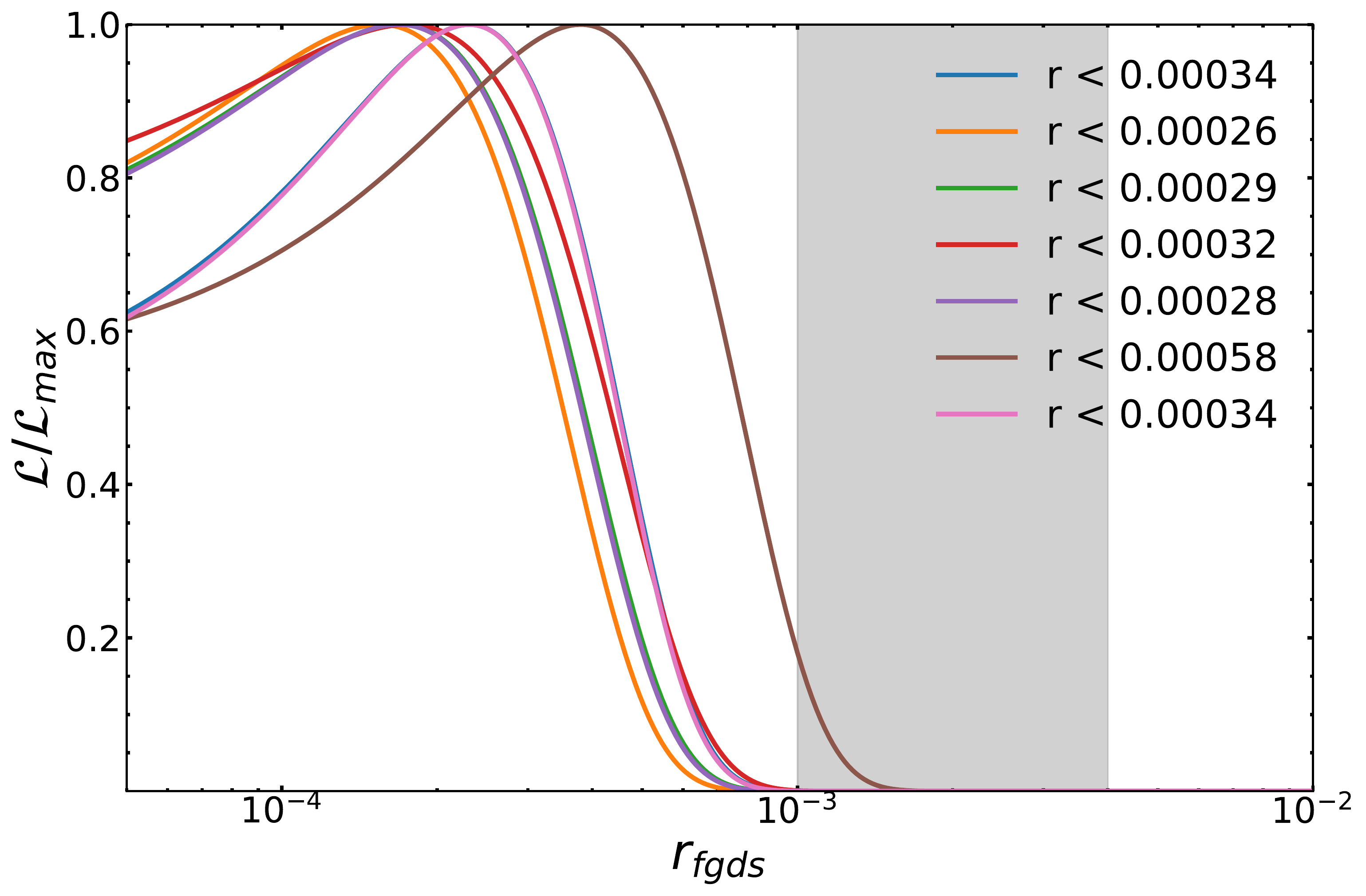} 
\caption{Left: average angular power spectra over $200$ CMB solutions of foreground (solid) and noise (dashed) residuals when CEA-MCNILC is applied in a realistic framework where the ratio is built employing $B$-mode foreground MC-GNILC templates (see text for details). Several different models of the Galactic emission have been considered. The angular power spectra are computed employing masks obtained according to the first strategy (\texttt{mask1}) described in Sect. \ref{sec:masks} with $f_{\textrm{sky}}=50\,\%$. Right: the posterior distribution of an effective tensor-to-scalar ratio fitted on the foreground residuals for the previously mentioned different cases. For the estimation of the posteriors a binning scheme of $\Delta\ell =10$ has been used to make the angular power spectrum gaussianly distributed 
(see Sect. \ref{sec:like}). The reported upper bounds refer to the $68\%$ CL. The grey areas highlight the range of amplitudes of the primordial tensor signal targeted by \textit{LiteBIRD}: $r\in [0.001,0.004]$.}
\label{fig:gnilc_results}
\end{figure*}
In this section we will show the results of this realistic application of MC-NILC to \textit{LiteBIRD} dataset where we adopt a ratio built with the above mentioned MC-GNILC templates. In Sect. \ref{sec:semiblid_res}, instead, we will present an alternative semi-blind approach to obtain $B$-mode Galactic templates and to be applied in case of compelling evidences in favour of a description of thermal dust emission in terms of a MBB given the \textit{LiteBIRD} sensitivity. \\
For this realistic approach, CEA- and RP-MCNILC leads to a similar amount of residuals for $\ell \gtrsim 10$, because Galactic signal de-projection and CMB and noise contamination in the employed templates mis-model the estimation of the ratio thus nullifying the benefits of averaging the CMB solutions obtained from applying MC-NILC on several different random partitions. On large angular scales ($\ell \lesssim 10$), as in the ideal case, CEA-MCNILC better subtracts foreground emission and, thus, it is chosen as the default technique. Furthermore, as already discussed in Sect. \ref{sec:clusters}, we consider for all needlet scales only the ratio where maps are filtered with the first needlet band to avoid larger CMB and noise residuals on small scales. In this analysis, all the realistic Galactic models of Fig. \ref{fig:results_nilc_clusters_diffmodels} and described in Sect. \ref{Sec:sims} have been considered: \texttt{d1s1}, \texttt{d4s1}, \texttt{d5s1}, \texttt{d6s1}, \texttt{d7s1}, \texttt{d12s1}, and \texttt{d1s3}. \\
The average angular power spectra of foreground and noise residuals when MCNILC is applied employing this realistic framework for all different cases are shown in Fig. \ref{fig:gnilc_results} together with the corresponding NILC foreground residuals. In this figure, results for \texttt{d7s1} and \texttt{d1s3} are not reported since they are analogous to those of \texttt{d5s1} and \texttt{d1s1}, respectively. Both foreground and noise residuals power in this realistic MC-NILC framework is larger with respect to the ideal case due to a distortion in the estimation of the tracer caused by the unavoidable de-projection of some Galactic signal and the residual CMB and noise contamination in the MC-GNILC templates, especially in those regions where foreground $B$ modes assume lower values. However, for all sky models, this realistic MC-NILC application leads to much reduced foreground contamination at large angular scales with respect to NILC without any noise penalty. Such low level of foreground and noise residuals at the reionization peak, for all cases, still leads to an unbiased posterior on the effective tensor-to-scalar ratio with upper bounds at $68\%$ CL which would permit to detect a primordial signal with $r \approx 10^{-3}$ (see right panel of Fig. \ref{fig:gnilc_results}). \\
The range of angular scales where MC-NILC Galactic contamination is lower than NILC one depends on the assumed emission law of the foregrounds. For \texttt{d4s1} and \texttt{d5s1}, we have that NILC performs better than MC-NILC at $\ell \gtrsim 10$. These are the cases, where, at intermediate and high multipoles, we observe the best foreground subtraction by NILC and the major differences between NILC and MC-NILC. However, with MC-NILC both the reionization and recombination peaks for values of $r$ targeted by \textit{LiteBIRD} would be observable. Instead, when thermal dust emission is or is close to \texttt{d1} model (\texttt{d1} and \texttt{d6}) or is assumed to be generated by multiple layers along the line of sight (\texttt{d12}), MC-NILC foreground residuals are at a level comparable to the primordial signal targeted by \textit{LiteBIRD} for $\ell \gtrsim 10$ and to NILC only at smaller angular scales. These are the cases where the sky emission is intrinsically more difficult to subtract by blind variance minimization techniques, as shown by the amount of NILC residuals. Moreover, for these models, the improvement in the reduction of foreground contamination at low multipoles led by the realistic MC-NILC application is especially remarkable. 
It would be desirable for this class of models to lower the foreground contamination at $\ell \gtrsim 10$. In the following section, we propose an alternative semi-blind approach to estimate the ratio for those cases where dust emission is or is close to a MBB. \\
\begin{figure}
	\centering
    \includegraphics[width=0.48\textwidth]{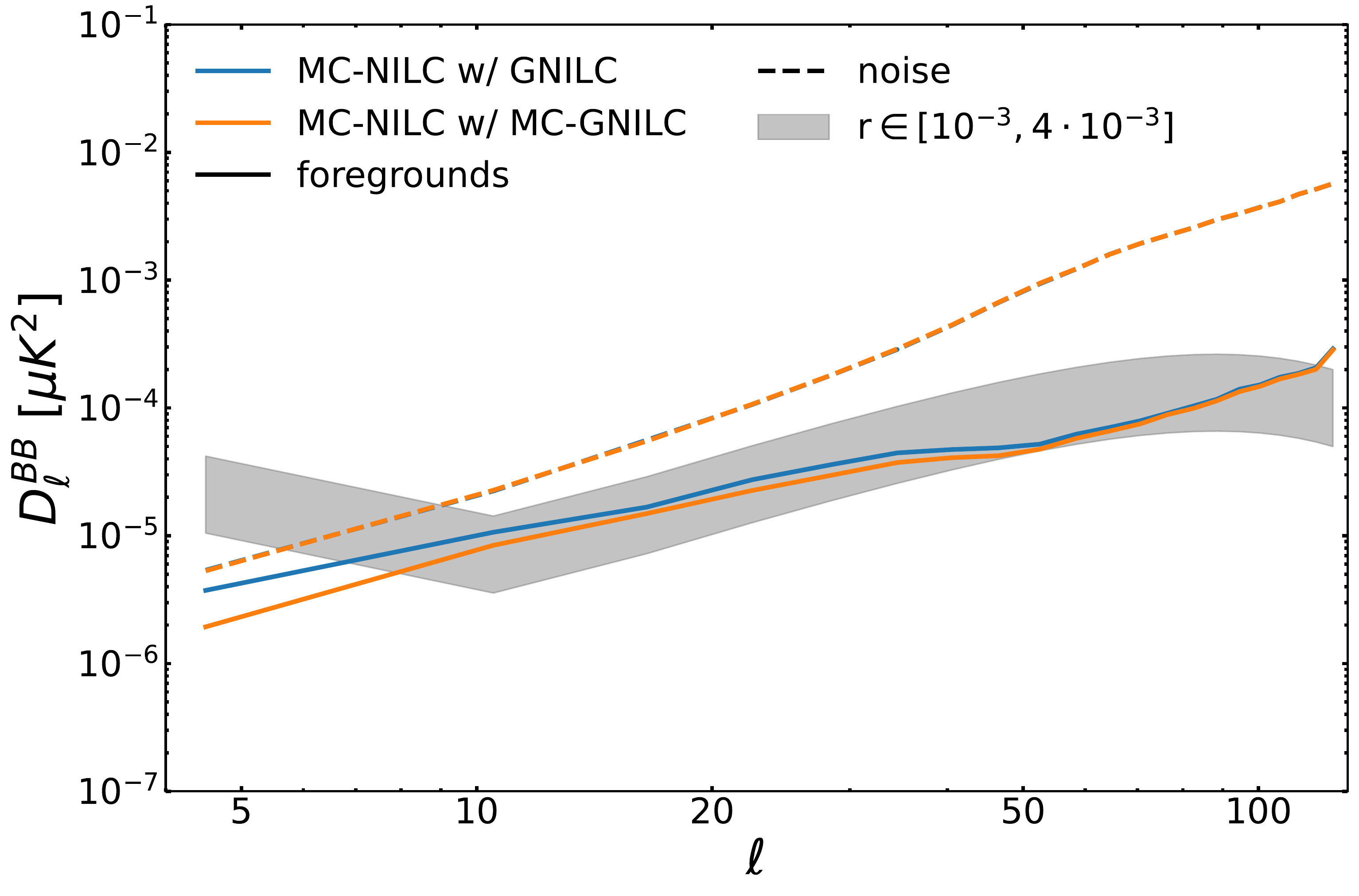} 
	\caption{Average angular power spectra over $200$ CMB solutions of foreground (solid) and noise (dashed) residuals when MC-NILC is applied in a realistic framework where the ratio is built employing $B$-mode foreground templates obtained with the MC-GNILC (blue) and GNILC (orange) pipelines (see text for details). The Galaxy is modelled with the \texttt{d1s1} sky. The angular power spectra are computed employing masks obtained according to the first strategy (\texttt{mask1}) described in Sect. \ref{sec:masks} with $f_{\textrm{sky}}=50\,\%$. The grey area highlights the range of amplitudes of the primordial tensor signal targeted by \textit{LiteBIRD}: $r\in [0.001,0.004]$. The adopted binning scheme is $\Delta\ell =6$.}
	\label{fig:gnilc_vs_mcgnilc}
\end{figure}
\begin{figure}
	\centering
    \includegraphics[width=0.48\textwidth]{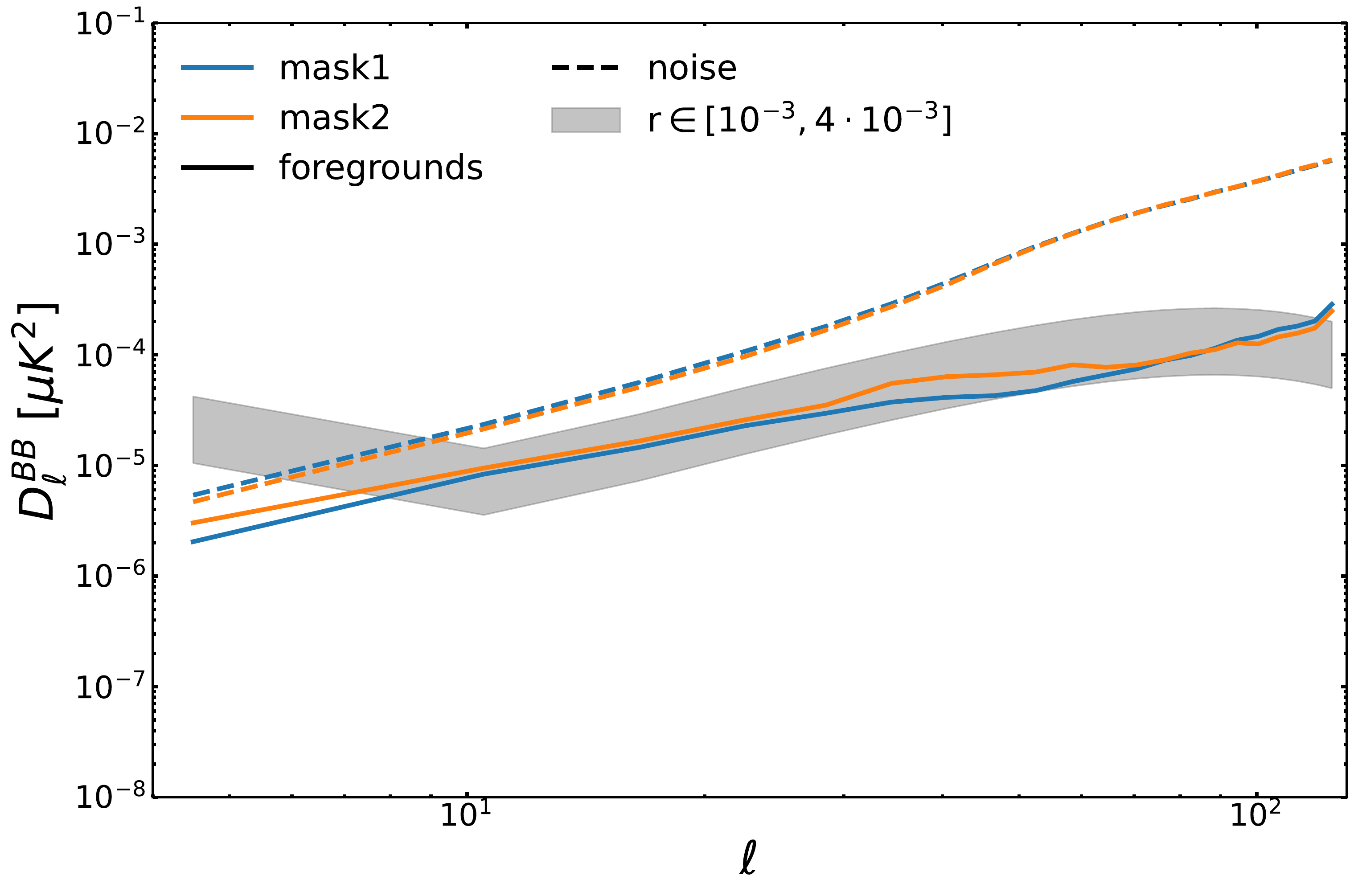} 
	\caption{Average angular power spectra over $200$ CMB solutions of foreground (solid) and noise (dashed) residuals when MC-NILC is applied in a realistic framework and the ratio is built employing $B$-mode foreground templates obtained with the MC-GNILC pipeline. The Galaxy is modelled with the \texttt{d1s1} sky. The angular power spectra are computed employing masks obtained according to the \texttt{mask1} (blue) and \texttt{mask2} (orange) procedures described in Sect. \ref{sec:masks} with, respectively, $f_{\textrm{sky}}=50\,\%$ and $40\,\%$. The grey area highlights the range of amplitudes of the primordial tensor signal targeted by \textit{LiteBIRD}: $r\in [0.001,0.004]$. The adopted binning scheme is $\Delta\ell =6$.}
	\label{fig:cls_diffmasks}
\end{figure}
In Fig. \ref{fig:gnilc_vs_mcgnilc}, we show a comparison of MCNILC residuals when the \texttt{d1s1} sky model is assumed and the ratio is built from Galactic $B$-mode templates obtained either with MC-GNILC or GNILC (average of solutions with $\textit{nl1}$ and $\textit{nl2}$ needlet configurations, see Sect. \ref{sec:MCGNILC}). The former case is characterised by lower foreground residuals, due to the reduced CMB and noise contamination in the MC-GNILC reconstruction. However, Fig. \ref{fig:gnilc_vs_mcgnilc} shows that even the use of GNILC maps to construct the ratio would lead to MC-NILC residuals much lower than the reionization peak of the primordial signal targeted by \textit{LiteBIRD}. \\
All the above results have been obtained adopting the first masking strategy (\texttt{mask1}) described in Sect. \ref{sec:masks} to have a fair comparison with the component separation products reported in \citet{2022arXiv220202773L}. In Fig. \ref{fig:cls_diffmasks}, we compare the foreground reduction performance of this masking approach with a more conservative one (\texttt{mask2}), where we adopt the root mean square map of GNILC templates linearly combined with MC-NILC weights as a tracer of the most contaminated regions by foregrounds (see Sect. \ref{sec:masks} for details). The two masking strategies perform similarly. However, we recall that, for \texttt{mask2}, a more aggressive masking is required and a sky fraction $f_{\textrm{sky}}=40\,\%$ is retained, while we have $f_{\textrm{sky}}=50\,\%$ for \texttt{mask1}. In the shown case, we have assumed the \texttt{d1s1} sky emission, but analogous results have been obtained for the other Galactic models considered in this work.

\subsubsection{Semi-blind ratio}
\label{sec:semiblid_res}
We have shown in Fig. \ref{fig:gnilc_results} that if the actual dust emission was close to a MBB, the application of MC-NILC with the blind ratio built with MC-GNILC templates would lead to significant foreground residuals for $\ell \gtrsim 10$. In order to improve the performance of MC-NILC with a realistic ratio for these considered cases (\texttt{d1s1}, \texttt{d6s1}, and \texttt{d1s3}), we present an alternative semi-blind procedure to obtain templates of the Galaxy in $B$ modes at $337$ and $119$ GHz to be applied in case of compelling evidences in favour of a description of the real polarized dust emission in terms of a MBB.\\
The high-frequency template is the same of the default MC-NILC pipeline, obtained with a MC-GNILC run on the \textit{LiteBIRD} $B$-mode multi-frequency dataset. 
Instead, for the CMB channel foreground template, we exploit some a-priori knowledge of thermal dust \texttt{d1} and synchrotron \texttt{s1} spectral indices and we make use of the Constrained polarization Internal Linear Combination (cPILC, \citealt{2021MNRAS.507.4618A}). 
cPILC represents an extension of the PILC method \citep{2016MNRAS.459..441F}, where the variance minimization is performed on $P^{2}=Q^2 + U^2$ and CMB and moments of the expected SEDs of the foregrounds can be de-projected out to obtain a template of foreground emission at a specific frequency. \\
Specifically, the Galactic $B$-mode emission at $119$ GHz has been re-constructed by implementing the following procedure:
\begin{itemize}
\item Consider the \textit{LiteBIRD} full QU maps (CMB + foregrounds + instrumental noise) at $40$ and $337$ GHz as synchrotron and dust templates, respectively.
\item Use cPILC on the \textit{LiteBIRD}  dataset to lower the noise and CMB contamination in the maps at $40$ GHz fully de-projecting the CMB and the first-order moment of synchrotron emission while preserving its zero-order moment. 
\item Scale these maps to $119$ GHz according to Eqs. \ref{eq:sync}-\ref{eq:dust} and employing the spectral indices of \texttt{d1} and \texttt{s1} accounting for statistical mis-modelling.
\item Add the two obtained QU maps.
\item Compute the corresponding $B$-mode map.
\end{itemize}
\begin{figure*}
	\centering
	\includegraphics[width=0.48\textwidth]{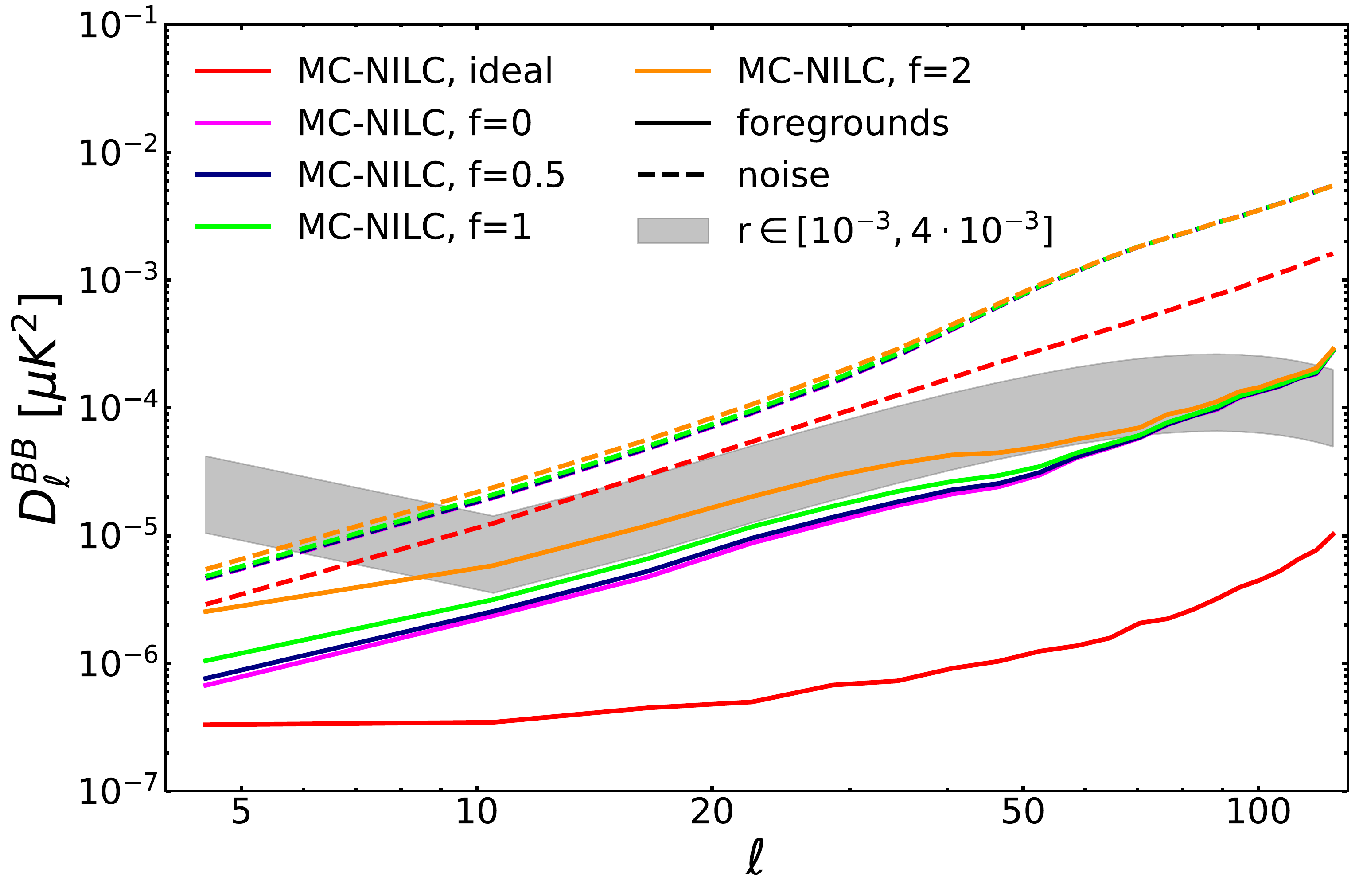}
	\hspace{0.5 cm}
	\includegraphics[width=0.48\textwidth]{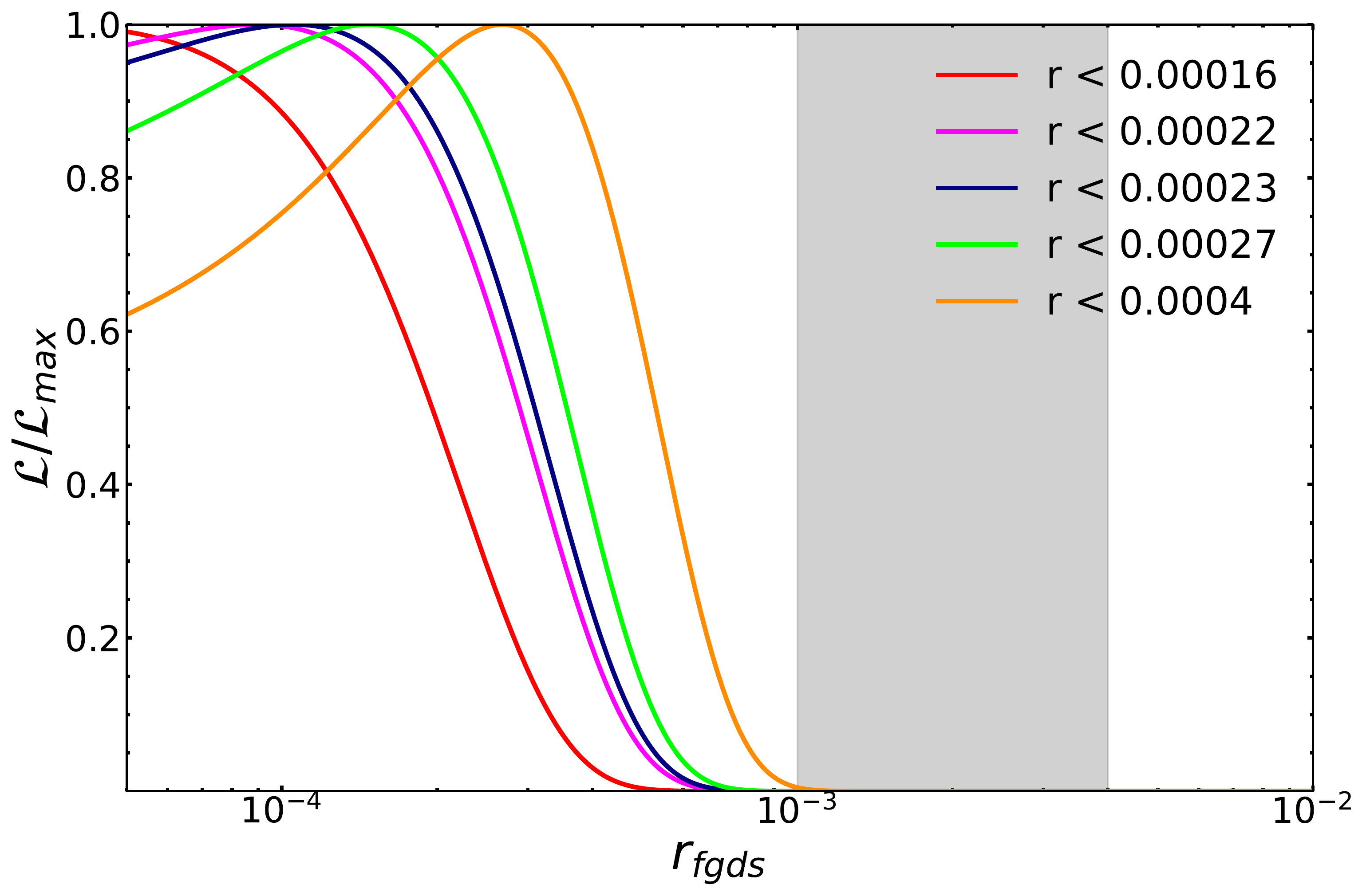}
	\caption{Left panel: average angular power spectra over $200$ CMB solutions of foreground (solid) and noise (dashed) residuals in different cases: in red RP-MCNILC with the ratio of needlet foregrounds-only $B$ modes; in magenta, dark blue, light green, and orange CEA-MCNILC with the realistic semi-blind ratio when a mis-modelling of \texttt{d1s1} spectral indices with, respectively, $\textit{f}=0, 0.5, 1$, and $2$ is considered (see text for details). Here, all the results are obtained adopting the \texttt{d1s1} PySM Galactic model in the simulations. The adopted binning scheme is $\Delta\ell =6$ only for visualisation purposes. The employed mask in each case is obtained following the first procedure (\texttt{mask1}) described in Sect. \ref{sec:masks} with a sky fraction of $50\,\%$. Right: the posterior distributions of an effective tensor-to-scalar ratio fitted on the foreground residuals for the different cases. The binning scheme for the angular power spectra that enter the likelihood is $\Delta\ell =10$  
    (see Sect. \ref{sec:like}). The reported upper bounds refer to the $68\%$ CL. The grey areas highlight the range of amplitudes of the primordial tensor signal targeted by \textit{LiteBIRD}: $r\in [0.001,0.004]$.
    }
\label{fig:results_nilc_clusters_mismodel}
\end{figure*}
This procedure enables to almost perfectly reconstruct the $B$-mode foreground emission at $119$ GHz with very low noise and CMB contamination whose amount at $40$ and $337$ GHz is lowered by scaling them with the synchrotron and dust SEDs, respectively. 
The noise and CMB power is reduced of, respectively, one order and three orders of magnitude with respect to the input one at the same frequency. \\
We remark here that with this procedure we do not require to perfectly know the spectral parameters with which we derive the sky partition. We indeed allow for a certain amount of statistical mis-modelling with respect to the \texttt{d1s1} spectral parameters which is associated to our current uncertainties on them. Indeed, similarly to what \citealt{2022MNRAS.511.2052P} did, we add some noise in all pixels of the map of each spectral index, obtaining mis-modelled distributions of spectral parameters $\hat{X}$:
\begin{equation}
    \hat{X} = X + \textit{f}\cdot N_{\textrm{w}}(\sigma(X)),
\end{equation}
where $X$ is either $\beta_{\textrm{s}}, \beta_{\textrm{d}}$ or $T_{\textrm{d}}$ of Eqs. \ref{eq:sync} and \ref{eq:dust}, $N_{\textrm{w}}(\sigma(X))$ is a random Gaussian realisation with zero mean and standard deviation given by the uncertainty of the parameter in each pixel, and $\textit{f}$ quantifies the degree of mis-modelling we introduce in the spectral indices maps. $\textit{f} = 2$ represents a pessimistic case where the foregrounds
are fully statistically mis-modelled across all the sky given the uncertainty budget in the spectral
parameters, while $\textit{f} = 0$ represents the ideal case where we assume to have a perfect knowledge of the SEDs of the polarized foregrounds. The maps of uncertainties $\sigma(\beta_{\textrm{d}})$ and $\sigma(T_{\textrm{d}})$ are obtained from the posterior distribution on these parameters estimated with the application of the Commander algorithm on a joint dataset of \textit{Planck}, \textit{WMAP}, and $408$-MHz observations \citep{2016A&A...594A..10P}. Instead $\sigma(\beta_{\textrm{s}})$ is estimated in the analysis of \citealt{2008A&A...490.1093M}. \\
Once the foreground $B$-mode emission templates have been obtained with the procedures described above, a ratio is built and CEA-MCNILC variance minimization is performed. 
We report the results only for the \texttt{d1s1} model; analogous outcomes have been obtained in the other cases: \texttt{d6s1} and \texttt{d1s3}. \\ 
The foreground and noise residuals obtained with and without mis-modelling are shown and compared with the best solution (the RP-MCNILC one) in the ideal approach in Fig. \ref{fig:results_nilc_clusters_mismodel}. The increase of the Galactic and noise contamination with respect to the ideal case is caused by the distortion of the sky partition due to a residual presence of noise and CMB especially in the template at the denominator. Figure \ref{fig:results_nilc_clusters_mismodel} shows that in all cases, even with the most pessimistic statistical mis-modelling, it would be possible to detect a primordial $B$-mode signal at the reionization peak, while at the recombination bump the power of residuals is comparable to the primordial tensor signal targeted by \textit{LiteBIRD}. 
Even considering statistical mis-modelling, the performance of MC-NILC in removing polarized foregrounds in $B$-mode data on the largest angular scales is largely better than a standard NILC, while on smaller ones ($\ell \gtrsim 80$) the residuals' power is comparable. \\
Furthermore, the application of MC-NILC with this semi-blind ratio permits to lower the \texttt{d1s1} foreground residuals at large and intermediate angular scales with respect to the case where the ratio is built with both MC-GNILC templates, thanks to a negligible Galactic signal loss and a reduced CMB and noise contamination in the $119$-GHz template which lead to a lower distortion of the clustering. \\
Looking at the posteriors' distribution in the right panel of Fig. \ref{fig:results_nilc_clusters_mismodel}, all the cases that have been considered for the semi-blind approach, with or without mis-modelling, lead to upper bounds at $68\,\%$ CL on $r_{\textrm{fgds}}$ which are lower with respect to the \textit{LiteBIRD} target requirements. These constraints are largely driven by the low level of foreground contamination at large angular scales. \\
We remark the impressiveness of these results, which highlight how \textit{Planck} with its sensitivity could in principle be able to play a major role in the cleaning capability of MC-NILC on \textit{LiteBIRD} polarization data in case we have some knowledge of thermal dust spectral properties. However, this result holds because we are assuming in the sky modelling that foreground emission is described in intensity and polarization by the same set of spectral parameters. If this will be the case even with up-coming more sensitive CMB surveys, the \textit{Planck} temperature analysis would be of great benefit for future polarization CMB experiments. 

\section{Conclusions}
\label{sec:concs}
NILC has proven to be a powerful blind method for the subtraction of Galactic foregrounds in the analysis of \textit{WMAP} and \textit{Planck} CMB temperature anisotropies data. Given its low number of a-priori assumptions, it potentially represents a key alternative to the parametric component separation techniques also for the processing of $B$-mode data from future CMB experiments. We have, however, compelling evidence that the spectral energy distribution of the polarized Galactic emission is characterised by a complex spatial variability across the sky, which complicates the NILC estimation and mitigation of the foreground contamination.
This fact is particularly relevant in the analysis of $B$-mode data given the weakness of the primordial signal with respect to the Galactic foregrounds. For example, the application of NILC to simulated \textit{LiteBIRD} $B$-mode data where Galactic emission is assumed to have anisotropic spectral properties leads to significant foreground residuals, especially on large angular scales ($\ell \lesssim 30$, see Fig. \ref{fig:NILC_litebird}). Such a contamination would bias the estimation of the tensor-to-scalar ratio, a result also pointed out by \citet{2021MNRAS.503.2478R}. \\
In order to reach the ambitious targets of future CMB experiments, we therefore propose a novel blind technique, MC-NILC, which performs variance minimization at the various needlet scales in different regions of the sky that present similar properties of the $B$-mode foregrounds. 
Here we highlight the main methodological developments required to build this component separation pipeline: 
\begin{enumerate}[label=\alph*)]
    \item Performing ILC variance minimization on portions of the sky induces a bias in the reconstruction of the CMB signal (see Sec. \ref{Sec:NILC}). To limit this issue we build an appropriate configuration of needlet bands which limits mode-loss and we implement an estimator of the covariance matrix in each pixel which excludes the pixel itself and a circular region around it 
    \item We identify a tracer of the spatial distribution of spectral properties of $B$-mode foregrounds which requires minimal modelling assumptions: the ratio of $B$-mode maps at two separate frequencies.
    \item To limit the impact of CMB and instrumental noise when building the ratio, we present an optimised method to construct clean templates of $B$-mode foreground emission from input data: MC-GNILC (see Sect. \ref{sec:MCGNILC}).
    \item We explore different clustering techniques to identify regions of the sky with similar foreground properties, namely CEA (Clusters of Equal Area) and RP (Random Partitions): 
    \begin{itemize}
       \item in \emph{CEA-MCNILC} a unique partition of the sky is identified with all patches composed of an equal number of pixels and the variance is minimized within each cluster independently for each needlet scale;
       \item in \emph{RP-MCNILC} we generate several different partitions of the sky where each cluster collects a random amount of pixels, we then apply MC-NILC on each partition and, finally, compute the mean of all the CMB solutions corresponding to the different partitions.
    \end{itemize}
    \item We introduce a partition metric to estimate the optimal number of clusters %($50$) 
    taking into account both the amount of foreground residuals and the bias in the  power spectrum of the recovered CMB (see Fig. \ref{fig:fgds_vs_bias}).
\end{enumerate}
In this work we have studied in detail the application of MC-NILC to simulations of the \textit{LiteBIRD} CMB satellite. We have obtained the following main results:
\begin{itemize}
    \item The optimal pair of frequencies to build the ratio used as tracer of the spatial distribution of Galactic emission spectral properties results to be $402$ (or $337$) and $119$ GHz. The choice of these frequencies does not depend on the assumed Galactic model of emission as we find consistent results for several models with varying foreground complexity implemented in the PySM simulation package (see Fig. \ref{fig:results_nilc_clusters_diffmodels})
    \item the selection of the domains where GNILC is independently applied has been optimised by accounting for the spatial variability of the spectral properties of the $B$-mode foregrounds; such optimisation (MC-GNILC) leads to templates of Galactic emission with lower contamination of CMB and noise components (see Sect. \ref{sec:MCGNILC})
    \item In an ideal case where the ratio is built with $B$-mode foregrounds-only maps, RP-MCNILC results to be the best configuration for $\ell \gtrsim 15$ because in this case there is larger probability that pixels with closer values of the ratio would find themselves in the same cluster more often; both MC-NILC implementations reduce sensibly the amount of foreground and noise residuals at all angular scales with respect to NILC and it allows to reach the sensitivity goal on $r$.
    \item In the case of a realistic approach, the ratio is built with $B$-mode foreground templates that also include CMB and instrumental noise contamination and are estimated with a MC-GNILC run on the \textit{LiteBIRD} simulated data. In this case, foreground residuals in the recovered MC-NILC CMB solution are lower at the reionization peak with respect to a primordial tensor signal with $r=0.001$ for all considered Galactic models, while at the recombination bump they are comparable if thermal dust emission is assumed to follow a MBB or to be generated by multiple layers (see left panel of Fig. \ref{fig:gnilc_results}).
    \item To lower the Galactic contamination at $\ell \gtrsim 10$ for those models where thermal dust emission is or is close to a MBB, we implement a semi-blind approach where the foreground template at $119$ GHz is obtained exploiting some a-priori knowledge of the spectral parameters even accounting for their statistical mis-modelling (see Sect. \ref{sec:semiblid_res}). With such approach, we are able to lower the foreground residuals at all angular scales with respect to the case where the default MC-NILC realistic estimation of the ratio is employed, thanks to a negligible Galactic signal loss and a reduced CMB and noise contamination in the $119$ GHz template which lead to a lower distortion of the clustering (see Fig. \ref{fig:results_nilc_clusters_mismodel}).
\end{itemize}
These results highlight that the proposed method overcomes standard NILC limitations when complex foreground emission skies are considered and leads to a bias on $r$ associated to the foreground residuals which is below the targeted sensitivity of future CMB missions. \\ %\\
To our knowledge, this is the first time that a minimum variance technique is applied on CMB polarization data separately on different regions of the sky according to the spatial variability of the Galactic foregrounds both to reconstruct the CMB and the Galactic signal. 
Although, in this work, the MC-NILC pipeline had been tested and applied on \textit{LiteBIRD} simulated dataset, it represents a promising approach for the analysis of $B$-mode data of any future satellite or sub-orbital CMB experiment, such as Simons Observatory \citep{2019JCAP...02..056A}, CMB-S4 \citep{2022arXiv220308024A} or \textit{PICO} \citep{2019arXiv190210541H}.

\section*{Acknowledgements}
      We thank Mathieu Remazeilles, David Alonso and Jonathan Aumont for helpful comments. We acknowledge support by ASI/COSMOS grant n. 2016-24-H.0, ASI/LiteBIRD grant n. 2020-9-HH.0, and from the MIUR Excellence Project awarded to the Department of Mathematics, Università di Roma Tor Vergata, CUP E83C18000100006. This research used resources of the National Energy Research Scientific Computing Center (NERSC), a U.S. Department of Energy Office of Science User Facility located at Lawrence Berkeley National Laboratory. Part of this work was also supported by the InDark and LiteBIRD INFN projects. The Italian \textit{LiteBIRD} phase A contribution is supported by the Italian Space Agency (ASI Grants No. 2020-9-HH.0 and 2016-24-H.1-2018), the National Institute for Nuclear Physics (INFN), and the National Institute for Astrophysics (INAF).

\section*{Data Availability}
The data underlying this article will be shared on reasonable request to the corresponding author.

\bibliographystyle{mnras}
\bibliography{biblio}

\clearpage
\appendix
\section{Choice of the tracer of $B$-mode foreground spectral properties}
\label{app:tracers}
\begin{figure*}
\centering
\includegraphics[width=0.48\textwidth]{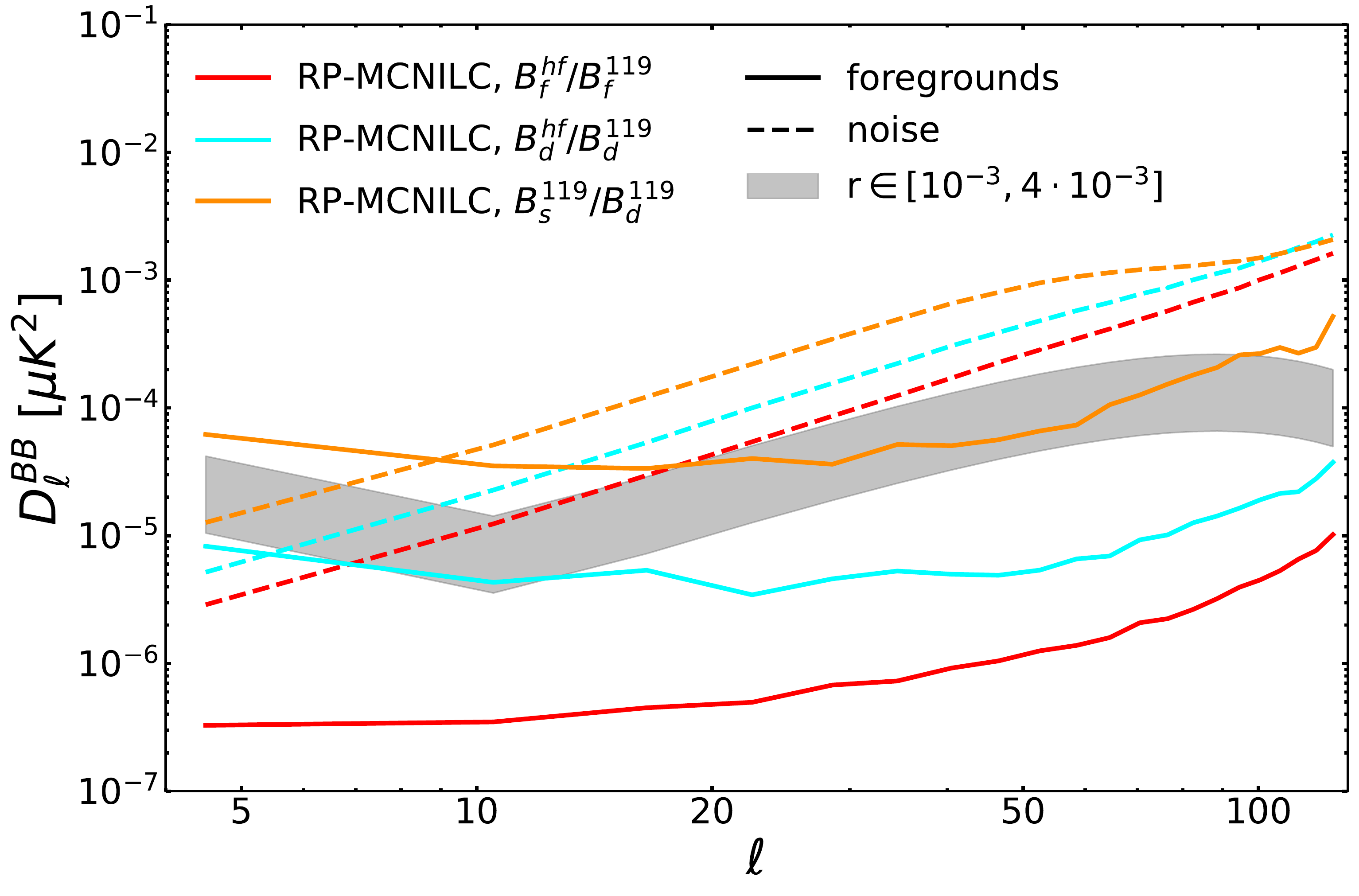}
\hspace{0.5 cm}
\includegraphics[width=0.48\textwidth]{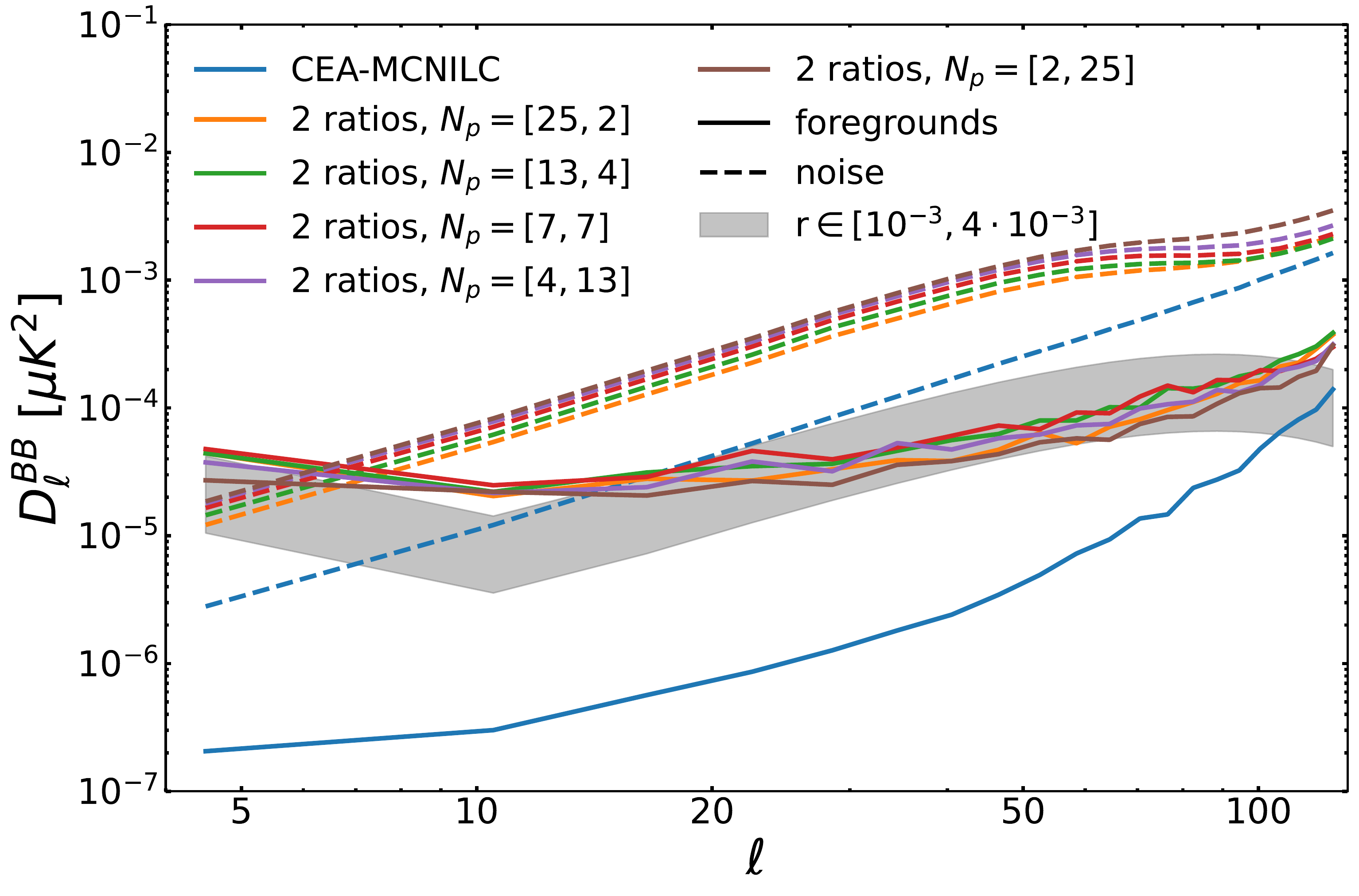}
\caption{Average angular power spectra of foreground (solid) and noise (dashed) residuals among $200$ different MC-NILC CMB solutions. The sky emission is simulated assuming the \texttt{d1s1} model. Left: results when RP-MCNILC is applied and the sky is partitioned using ratios built with: needlet $B$-mode foregrounds at $402$ and $119$ GHz (red), needlet $B$-mode dust maps at $402$ and $119$ GHz (cyan), needlet $B$-mode synchrotron and dust maps at $119$ GHz (orange). Right: comparison between the MC-NILC residuals when the sky partition is obtained by either thresholding the single ratio of Eq. \ref{eq:ratio_comps} (blue) or combining two separate partitions from a ratio at high frequencies and one at low frequencies. In all cases, the CEA approach is employed to threshold the histograms of the ratios (see Sect. \ref{sec:clustering}). We report results for different combinations of the number of patches for the high- and low-frequency ratios. The grey areas highlight the range of amplitudes of the primordial tensor signal targeted by \textit{LiteBIRD}: $r\in [0.001,0.004]$. We mask $50\,\%$ of the sky adopting the first strategy (\texttt{mask1}) in Sect. \ref{sec:masks}. The adopted binning scheme is $\Delta\ell =6$.}
\label{fig:results_nilc_clusters_dust}
\end{figure*}

In Sect. \ref{sec:GNILC&cPILC}, we present the ratio of $B$-mode foreground templates at two separate frequencies as a tracer of the distribution of the spectral properties of the Galactic emission in $B$ modes across the sky. Such choice allows us to preserve the blindness of the NILC methodology and results to be effective for the $B$-mode foreground subtraction, as proved in Sects. \ref{sec:ideal_app} and \ref{sec:real_appr}. 
In this Appendix, we will show what this ratio is effectively tracing and some comparisons with an alternative tracer. \\
Given the linearity of the $B$-mode decomposition and the fact that $\textit{Q}$ and $\textit{U}$ parameters are additive for multiple components, a $B$-mode foreground map at a frequency $\nu$ is given by:
\begin{equation*}
    B_{\textit{fgds}}^{\nu} = B_{\textit{dust}}^{\nu} + B_{\textit{synch}}^{\nu},
\end{equation*}
where $B_{\textit{dust}}^{\nu}$ and $B_{\textit{synch}}^{\nu}$ are, respectively, the dust and synchrotron total contribution to the $B$-mode map. Therefore, a ratio of foreground templates in $B$ modes at two distinct frequencies (introduced in Sect. \ref{sec:GNILC&cPILC}) can be re-written as:
\begin{equation}
    \frac{B_{\textit{fgds}}^{\textit{hf}}}{B_{\textit{fgds}}^{119}} = \frac{B_{\textit{dust}}^{\textit{hf}} + B_{\textit{synch}}^{\textit{hf}}}{B_{\textit{dust}}^{119} + B_{\textit{synch}}^{119}} =  \frac{B_{\textit{dust}}^{\textit{hf}}}{B_{\textit{dust}}^{119}} \cdot \frac{1 + \frac{B_{\textit{synch}}^{\textit{hf}}}{B_{\textit{dust}}^{\textit{hf}}}}{1 + \frac{B_{\textit{synch}}^{119}}{B_{\textit{dust}}^{119}}} \simeq \frac{B_{\textit{dust}}^{\textit{hf}}}{B_{\textit{dust}}^{119}} \cdot \frac{1}{1 + \frac{B_{\textit{synch}}^{119}}{B_{\textit{dust}}^{119}}},
\label{eq:ratio_comps}
\end{equation}
where in the last step we confidently neglect the term $B_{\textit{synch}}^{\textit{hf}}/B_{\textit{dust}}^{\textit{hf}}$, since synchrotron component is totally subdominant at high frequencies with respect to thermal dust. \\
Eq. \ref{eq:ratio_comps} shows that with a single ratio, we are sampling at the same time an effective spectral index of the thermal dust $B$-mode emission through the first term and an emission ratio between the two polarized components at a 'CMB' frequency channel, where both are supposed to contribute to the total observed emission. To assess the relevance of the two terms in Eq. \ref{eq:ratio_comps}, we have applied RP-MCNILC to the LiteBIRD dataset by adopting sky partitions obtained by thresholding: i) the ratio of input $B$-mode foreground needlet maps at $402$ and $119$ GHz, ii) the ratio of input $B$-mode thermal dust needlet maps at $402$ and $119$ GHz, iii) the emission ratio of dust and synchrotron input needlet maps at $119$ GHz. The comparison of the angular power spectra of foreground and noise residuals between the three different cases is shown in the left panel of Fig. \ref{fig:results_nilc_clusters_dust}. We can observe that neglecting one of the two terms in the ratio leads to brighter foreground residuals in the output MC-NILC solution. The increase of the Galactic contamination is particularly relevant when the $B$-mode thermal dust spectral index is ignored. These results highlight the importance of considering both terms when constructing the ratio. \\
In this work, we have also studied the possibility to adopt two distinct ratios to separately trace the spectral properties of thermal dust and synchrotron components. Specifically, we have considered the ratio $B_{\textit{fgds}}^{50}/B_{\textit{fgds}}^{40}$ of B-modes foreground maps at $50$ and $40$ GHz as a tracer of synchrotron spectral index and $B_{\textit{fgds}}^{402}/B_{\textit{fgds}}^{280}$ for thermal dust. We generate two distinct partitions of the sky by thresholding the histograms of the two ratios with the CEA approach, as described in Sect. \ref{sec:clustering}. The final partition is then generated grouping together all those pixels belonging to the same patch in both synchrotron and dust partitions. In order to have, in such framework, approximately the same number of patches of the partition obtained from a single ratio, we set the following combination of number of patches to threshold the synchrotron (first number) and dust (second number) ratios: $[2-25]$, $[4-13]$, $[7-7]$, $[13-4]$, and $[25-2]$, leading, respectively, to $50$, $52$, $49$, $52$, and $50$ final patches. MC-NILC is then applied considering these new partitions. The obtained results in terms of foreground and noise residuals are compared with those of CEA-MCNILC with a single ratio in the right panel of Fig. \ref{fig:results_nilc_clusters_dust}. It is possible to observe that, in none of the cases, we can reach a foreground subtraction comparable to that obtained by thresholding the single ratio. Therefore, considering two distinct ratios for the two polarized Galactic components is less effective than a unique partition obtained by thresholding the single ratio of Eq. \ref{eq:ratio_comps}, which simultaneously traces both components in the way explained above.

\label{lastpage}	

\end{document}